\documentclass[prd,showpacs,altaffilletter,twocolumn,
superscriptaddress,floatfix,nofootinbib,preprintnumbers]{revtex4-1} 

\pdfoutput=1
\usepackage{amsfonts,amsmath,graphicx,bm} %,siunitx}
\usepackage{dcolumn}
\usepackage[pdftex,
       colorlinks=true,
       urlcolor=blue,       % \href{...}{...} external (URL)
       linkcolor=red,       % \ref{...} and \pageref{...}
       pdfpagemode=None,
       bookmarksopen=true]{hyperref}
\usepackage[capitalise]{cleveref}
\usepackage{ifpdf}
\usepackage{multirow}
\usepackage[colorlinks=true]{hyperref}
\usepackage{url}

\newcommand{\glasgow}{SUPA, School of Physics and Astronomy,
  University of Glasgow, Glasgow, G12 8QQ, UK}

\def\today{\number\day\space\ifcase\month\or
January\or February\or March\or April\or May\or June\or
July\or August\or September\or October\or November\or December\fi
\space\number\year}
\newcount\mins \newcount\hours
\def\now{\hours=\time \mins=\time
	\divide\hours by60 \multiply\hours by60 \advance\mins by-\hours
	\divide\hours by60 
	\number\hours:\ifnum\mins<10 0\fi\number\mins }
\newcommand{\GAMMAmuTOTAL}{2.06(16)_\mathrm{latt}(2)_\mathrm{EM}\times 10^{13} ~\mathrm{s}^{-1}}
\newcommand{\GAMMAmuGEV}{13.6(1.1)_\mathrm{latt}(0.1)_\mathrm{EM}\times 10^{-12} ~\mathrm{GeV}}

\newcommand{\GAMMAeTOTAL}{2.07(17)_\mathrm{latt}(2)_\mathrm{EM}\times 10^{13} ~\mathrm{s}^{-1}}
\newcommand{\GAMMAeGEV}{13.6(1.1)_\mathrm{latt}(0.1)_\mathrm{EM}\times 10^{-12} ~\mathrm{GeV}}

\newcommand{\GAMMAtauTOTAL}{5.14(37)_\mathrm{latt}(5)_\mathrm{EM}\times 10^{12} ~\mathrm{s}^{-1}}
\newcommand{\GAMMAtauGEV}{3.38(24)_\mathrm{latt}(3)_\mathrm{EM}\times 10^{-12} ~\mathrm{GeV}}

\newcommand{\RDsstar}{0.2490(60)_\mathrm{latt}(35)_\mathrm{EM}}

\newcommand{\RDsstarIMP}{0.3324(31)_\mathrm{latt}(47)_\mathrm{EM}}
\begin{document}

\title{$B_s \rightarrow D_s^*$ Form Factors for the full $q^2$ range from Lattice QCD}
%The dispersion relation of lattice non-relativistic QCD to $\mathcal{O}(p^6)$ with kinetic couplings corrected through $\mathcal{O}(\alpha_s p^4)$} %\bm

\author{Judd \surname{Harrison}}
\email[]{judd.harrison@glasgow.ac.uk}
\affiliation{\glasgow}

\author{Christine~T.~H.~\surname{Davies}} 
\email[]{christine.davies@glasgow.ac.uk}
\affiliation{\glasgow}

\collaboration{HPQCD Collaboration}
\email[]{http://www.physics.gla.ac.uk/HPQCD}

%\preprint{DAMTP-2015-xx}

\begin{abstract}
We compute the Standard Model semileptonic vector and axial-vector form factors for $B_s\rightarrow D_s^*$ decay across the full $q^2$ range using lattice QCD. We use the Highly Improved Staggered Quark (HISQ) action for all valence quarks, enabling us to normalise weak currents nonperturbatively. Working on second generation MILC ensembles of gluon field configurations which include $u$, $d$, $s$ and $c$ HISQ sea quarks and HISQ heavy quarks with masses from that of $c$ mass up to that of the $b$ on our finest lattices, allows us to map out the heavy quark mass dependence of the form factors, and to constrain the associated discretisation effects. We can then determine the physical form factors at the $b$ mass. We use these to construct the differential and total rates for $\Gamma\left(B_s^0\to D_s^{*-}\ell^+{\nu}_\ell\right)$ and find $\Gamma_{\ell=e}/|\eta_\mathrm{EW}V_{cb}|^2=\GAMMAeTOTAL$, $\Gamma_{\ell=\mu}/|\eta_\mathrm{EW}V_{cb}|^2=\GAMMAmuTOTAL$ and $\Gamma_{\ell=\tau}/|\eta_\mathrm{EW}V_{cb}|^2=\GAMMAtauTOTAL$, where $\eta_\mathrm{EW}$ contains the short distance electroweak correction to $G_F$, the first uncertainty is from our lattice calculation, and the second allows for long-distance QED effects.The ratio  $R(D_s^{*-})\equiv \Gamma_{\ell=\tau}/\Gamma_{\ell=\mu}=\RDsstar$. We also obtain a value for the ratio of decay rates $\Gamma_{\ell=\mu}(B_s\to D_s)/\Gamma_{\ell=\mu}(B_s\to D_s^*)=0.443(40)_\mathrm{latt}(4)_\mathrm{EM}$, which agrees well with recent LHCb results. We can determine $V_{cb}$ by combining our lattice results across the full kinematic range of the decay with experimental results from LHCb and obtain $|V_{cb}|=42.2 (1.5)_\mathrm{latt}(1.7)_\mathrm{exp}(0.4)_\mathrm{EM} \times 10^{-3}$. A comparison of our lattice results for the shape of the differential decay rate to the binned, normalised differential decay rate from LHCb shows good agreement. We also test the impact of new physics couplings on angular observables and ratios which are sensitive to lepton flavor universality violation.
\end{abstract}

\maketitle

%Version compiled on \today~at \now.

%===========================================================================8

\section{Introduction}
\label{sec:intro}
The determination of the Cabibbo-Kobayashi-Maskawa (CKM) matrix elements requires precise theoretical calculations within the Standard Model (SM) and experimental measurements of quark flavor-changing decay processes. Meson semileptonic decay rates, to a meson in the final state, are parameterised by form factors that are related to matrix elements of the relevant quark flavor-changing weak current between the initial and final meson states. Lattice QCD has become the method of choice for the calculation of these matrix elements and continuing efforts are being made to systematically improve the precision with which form factors are known, in line with the projected reductions in experimental uncertainty.

Here we focus on semileptonic decays mediated by the quark level weak transition $b\rightarrow c \ell^-\overline{\nu}_\ell$, which have seen many recent theoretical developments.  
The most precise determinations of the corresponding CKM matrix element, $V_{cb}$, make use of measurements of $B\rightarrow D^*$ and $B\rightarrow D$~\cite{Amhis:2019ckw} semileptonic decay, emphasising the former due to favourable kinematic factors which do not suppress the differential decay rate as strongly near zero recoil. In these determinations the experimental data for $B\rightarrow D^*$ is extrapolated using a parameterisation scheme to zero recoil, where only a single form factor is needed, and matched to lattice calculations (e.g.\cite{Bailey:2014tva,Harrison:2017fmw}). Until recently determinations of $V_{cb}$ done in this way were in tension with the alternative inclusive determination, in which all charmed final state mesons are considered, and it has since become apparent that the systematic uncertainties associated with the underlying model dependence of extrapolating to zero recoil were being underestimated, with more general parameterisations going some way in resolving the tension~\cite{BIGI2017441,Bordone:2019vic}. It is clear then that an improved comparision between theory and experiment is needed, and that this must be done across the full physical kinematic range in order to remove any possible dependence upon the choice of parameterisation scheme. Such a comparison requires an accurate calculation of the SM form factors using lattice QCD, which is made challenging by the presence of the light spectator quark accompanying the $b$ and $c$ quarks in the $B$ and $D^*$ meson states respectively.

Recent work by LHCb~\cite{Aaij:2020hsi} has also provided a complementary determination of $|V_{cb}|$ using the related $B_s\rightarrow D_s^{(*)}$ decay. While $B_s\rightarrow D_s^{(*)}$ decay has not been measured as precisely as $B\rightarrow D^{(*)}$, it is expected that the experimental uncertainty entering this determination will be decreased in future measurements. LHCb has also measured the shape of the normalised differential decay rate with respect to $q^2$, the squared four-momentum transfer, for $B_s\rightarrow D_s^{(*)}$, allowing a direct comparison between theory and experiment.

%A lattice calculation of the $B_s\rightarrow D_s^{(*)}$ form factors will then allow for a complementary, model-independent determination of $|V_{cb}|$, another channel in which to probe lepton flavor universality and a test of the underlying theory, as well as forming a stepping stone to a lattice computation of the $B\rightarrow D^*$ form factors.

Decays such as $B_s\rightarrow D_s^{(*)}$ and $B\rightarrow D^{(*)}$ involving a $b\rightarrow c \ell^-\overline{\nu}_\ell$ weak decay also allow us to probe lepton flavor universality. This may be done most straightforwardly by comparing the theoretical ratio in the SM of branching fractions for decays to a $\tau$ final state lepton to those to a $\mu$ or $e$ to the experimentally measured ratio. The corresponding ratio for $B\rightarrow D^{(*)}$, $R(D^{(*)})$, has been a source of tension with the SM for some time, e.g.~\cite{Kim:2016yth,Gambino:2019sif}, though recent measurements by Belle show consistency with the SM~\cite{Belle:2019rba}. The sensitivity of other observables for $B\rightarrow D^{(*)}$ decay to lepton flavor universality violation~(LFUV) has also recently been investigated~\cite{Becirevic:2019tpx}.
The ratio for the related decay $B_c\rightarrow J/\psi\ell^-\bar{\nu}_\ell$, $R(J/\psi)$, has also recently been measured for the first time as part of the experimental programme at LHCb~\cite{PhysRevLett.120.121801}. Although this result currently has a large uncertainty, it is expected that this will be reduced significantly in future runs~\cite{Bediaga:2018lhg} to provide a further test of lepton flavor universality in that channel.

Pseudoscalar to pseudoscalar semileptonic decays, which are described in the SM by 2 form factors, have been more extensively studied using lattice QCD than pseudoscalar to vector semileptonic decays, which are described by 4 form factors in the SM. The $B\rightarrow D$ form factors have been computed away from zero recoil~\cite{Lattice:2015rga}, with work to compute $B \rightarrow D^*$ away from zero recoil currently underway~\cite{Vaquero:2019ary,Na:2015kha}. The related $B_s\rightarrow D_s$ semileptonic form factors have recently been computed across the full physical $q^2$ range~\cite{EuanBsDs}, while the relevant form factor for $B_s\rightarrow D_s^*$ has been computed only at zero recoil~\cite{EuanBsDsstar,Harrison:2017fmw}. 
The form factors for $B_c\rightarrow J/\psi$ decay are less computationally expensive to calculate using lattice QCD than the form factors for $B\rightarrow D^*$ or $B_s\rightarrow D_s^*$, owing to the fact that all of the valence quarks are heavy. The $J/\psi$ is also very narrow and far from strong decay thresholds (unlike the $D^*$), making $B_c \to J/\psi$ decay an ideal starting point for lattice QCD calculations of pseudoscalar to vector form factors. The corresponding $B_c\rightarrow J/\psi$ form factors, computed in lattice QCD across the full $q^2$ range, recently became available~\cite{Harrison:2020gvo}, together with a high precision theoretical value for the ratio $R(J/\psi)$~\cite{Harrison:2020nrv}, together with other relevant LFUV-sensitive observables from~\cite{Becirevic:2019tpx}.

A lattice QCD calculation of the $B_s\rightarrow D_s^*$ form factors is then well motivated: it will form an important stepping stone between the recent calculation of $B_c\rightarrow J/\psi$ form factors and a future lattice QCD calculation of the $B\rightarrow D^*$ form factors, as well as allowing for a complementary, model-independent determination of $|V_{cb}|$ when combined with LHCb results and a further channel to probe lepton 
flavour universality. $B_s \rightarrow D_s^*$ has an advantage over 
$B \rightarrow D^*$ from a lattice QCD perspective; there are no valence $u/d$ quarks and the $D_s^*$ has no Zweig-allowed strong two-body decay mode and so has a very narrow 
width~\cite{Donald:2013sra} that allows us to treat it as a stable meson. 

Lattice QCD calculations involving a $b$ quark have historically relied upon QCD discretisations which make use of the nonrelativistic nature of the $b$, or on the large mass of the $b$, to avoid the trade-off between numerical expense and large discretisation effects associated with placing a relativistic $b$ quark on the lattice (e.g.~\cite{Bailey:2014tva,Harrison:2017fmw}). Recently however it has become computationally viable to use lattices with sufficiently small lattice spacings to simulate relativistic heavy quarks with masses very close to the physical $b$ mass~\cite{EuanBsDs,EuanBsDsstar,Harrison:2020gvo}\footnote{This builds on the approach developed by HPQCD for heavy meson decay constants that has proved very successful~\cite{McNeile:2011ng, Bazavov:2017lyh, Hatton:2021dvg}.}. These calculations carry with them the advantage that the same fully relativistic action is used for both $b$ and $c$ quarks and so the current renormalisation factors may be computed non-perturbatively and with a high precision. This is not typically possible for calculations using nonrelativistic actions where the current must instead be renormalised perturbatively, with the resulting truncation of the perturbation series contributing a potentially large systematic uncertainty (e.g.~\cite{Harrison:2017fmw}).

In this paper we apply the method of heavy-HISQ, which has seen much recent application in studies of the decays of $b$ quarks (e.g.~\cite{EuanBsDsstar,EuanBsDs,Harrison:2020gvo}), to the study of ${B}_s^0\rightarrow D_s^{*-}(\rightarrow D_s^-\gamma)\ell^+{\nu}_\ell$ decay across the full $q^2$ range.

\begin{figure}
\includegraphics[scale=0.4]{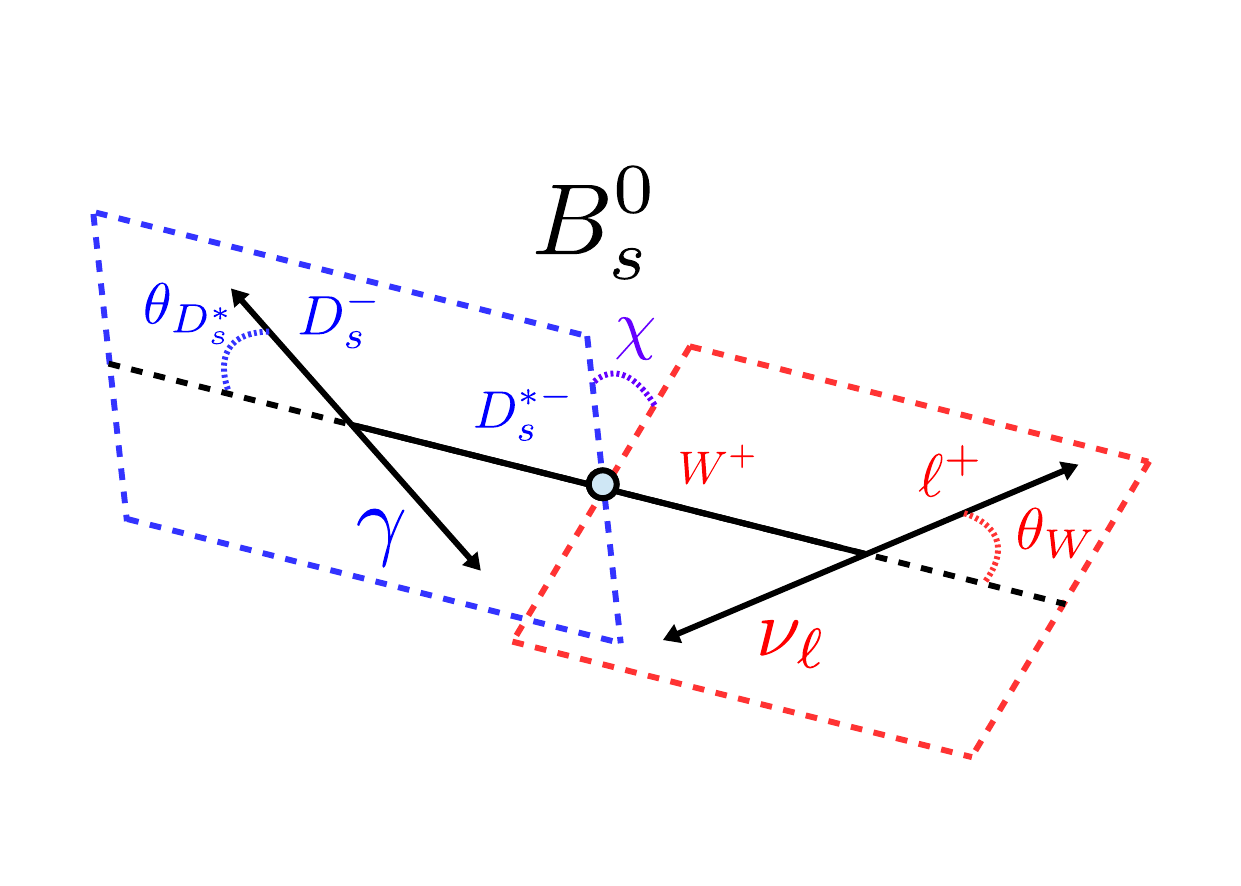}
\caption{\label{BsDsstarangles}Conventions for the angular variables entering the differential decay rate.}
\end{figure}

The subsequent sections are arranged as follows:
\begin{itemize}
\item In Section~\ref{sec:theory} we give expressions for the differential decay rates, helicity amplitudes and form factors relevant for ${B}^0_s \rightarrow D_s^{*-}\ell^+{\nu}_\ell$.
\item Section~\ref{sec:lattcalc} gives the technical details of the lattice calculation.
\item In Section~\ref{sec:results} we give the results for form factors from the lattice and discuss how to determine the form factors in the physical continuum limit. We also provide several tests of the stability of the analysis.
\item In Section~\ref{sec:discussion} we use the physical continuum form factors to make predictions for the differential decay rates for ${B}_s^0\rightarrow D_s^{*-}(\rightarrow D_s^-\gamma)\ell^+{\nu}_\ell$ and related quantities as well as providing a breakdown of sources of uncertainty. 
\item In Section~\ref{sec:lhcbcomp} we compare our results to recent experimental results from LHCb for the shape of the differential rate, provide a determination of $|V_{cb}|$ using our results across the full kinematic range.
\item Finally in Section~\ref{sec:lfuv} we investigate the effects of new physics couplings on lepton flavor universality violating ratios.
\end{itemize}

\section{Theoretical Background}
\label{sec:theory}
${B}_s^0\rightarrow D_s^{*-}(\rightarrow D_s^-\gamma)\ell^+{\nu}_\ell$ decay has the same angular structure as the decay of $B_c^-\rightarrow J/\psi(\rightarrow \mu^+\mu^-)\ell^-\overline{\nu}_\ell$ in the limit that the $\mu^+\mu^-$ pair are massless, and hence pure helicity states. The latter case was studied in~\cite{Harrison:2020gvo}. We adopt the same angular definitions, given in Figure~\ref{BsDsstarangles}, as for the $B_c\to J/\psi$ case. The differential rate is given by:
\begin{align}
\label{eq:diffrate}
\frac{d^4\Gamma({B}_s^0\rightarrow D_s^{*-}(\rightarrow D_s^-\gamma)\ell^+{\nu}_\ell)}{d\cos(\theta_{D_s^*})d\cos(\theta_{W})d\chi dq^2}&=\nonumber\\
 \mathcal{B}(D_s^*\rightarrow D_s\gamma)\mathcal{N}\sum_ik_i&(\theta_W,\theta_{D_s^*},\chi)\mathcal{H}_i(q^2) 
\end{align}
where
\begin{equation}\label{normfacdiff}
\mathcal{N}=\frac{G_F^2}{(4\pi)^4}(1+\delta_\mathrm{EM})|\eta_\mathrm{EW}|^2|V_{cb}|^2\frac{ 3(q^2-m_\ell^2)^2|{\vec{p}}~'| }{8M_{B_s}^2 q^2}
\end{equation}
Here $|{\vec{p}}~'|$ is the magnitude of 
the $D_s^*$ spatial momentum in the $B_s$ rest frame
and $\eta_\mathrm{EW}$ is a process-independent electroweak correction coming from box diagrams~\cite{Sirlin:1981ie}. We include the factor $(1+\delta_\mathrm{EM})$, which we take as a $q^2$-independent uncertainty, to allow for the effects of QED long-distance radiation (we expect this to be dominated by final-state interactions between the electrically charged lepton and $D_s^*$). Following~\cite{Chakraborty:2021qav} we include this as a $1\%$ uncertainty, which we take as the same for $e$ and $\mu$ final states, but independent for the $\tau$ final state.  Note that this choice is conservative, since in practice we expect there to be some amount of correlation between these effects for $\mu$ and $\tau$ final states, which we are neglecting here. We include this uncertainty seperately in quoted results so that it may be adjusted in light of future calculations.
The $k_i$ and $\mathcal{H}_i$ are given in Table~\ref{tab:diffterms}. 
\begin{table}
\centering
\caption{The helicity amplitude combinations and coefficients for them that 
appear in the differential rate, Eq.~(\ref{eq:diffrate}). Note that $k_i$ for terms 1 and 2 have been swapped, as well as terms 4 and 5, compared to~\cite{Harrison:2020nrv} since here we 
work with the conjugate mode ${B}_s^0\rightarrow D_s^{*-}(\rightarrow D_s^-\gamma)\ell^+{\nu}_\ell$. \label{tab:diffterms}}
\begin{tabular}{ c c | c  }
\hline
$i$ & $\mathcal{H}_i$ & $k_i(\theta_W,\theta_{D_s^*},\chi)$\\
\hline
1 & $|H_+(q^2)|^2$ & $\frac{1}{2}(1+\cos(\theta_W))^2(1+\cos^2(\theta_{D_s^*}))$\\
2 & $|H_-(q^2)|^2$ & $\frac{1}{2}(1-\cos(\theta_W))^2(1+\cos^2(\theta_{D_s^*}))$\\
3 & $|H_0|^2$& $2\sin^2(\theta_W)\sin^2(\theta_{D_s^*})$\\
4 & $\mathrm{Re}(H_+H_0^*)$&$-\sin(\theta_W)\sin(2\theta_{D_s^*})\cos(\chi)(1+\cos(\theta_W))$\\
5 & $\mathrm{Re}(H_-H_0^*)$&$\sin(\theta_W)\sin(2\theta_{D_s^*})\cos(\chi)(1-\cos(\theta_W))$\\
6 & $\mathrm{Re}(H_+H_-^*)$& $\sin^2(\theta_W)\sin^2(\theta_{D_s^*})\cos(2\chi)$\\
7 & $\frac{m_\ell^2}{q^2}|H_+(q^2)|^2$ & $\frac{1}{2}(1-\cos^2(\theta_W))(1+\cos^2(\theta_{D_s^*}))$\\
8 & $\frac{m_\ell^2}{q^2}|H_-(q^2)|^2$ & $\frac{1}{2}(1-\cos^2(\theta_W))(1+\cos^2(\theta_{D_s^*}))$\\
9 & $\frac{m_\ell^2}{q^2}|H_0|^2$& $2\cos^2(\theta_W)\sin^2(\theta_{D_s^*})$\\
10 & $\frac{m_\ell^2}{q^2}|H_t(q^2)|^2$ & $2\sin^2(\theta_{D_s^*})$\\
11 & $\frac{m_\ell^2}{q^2}\mathrm{Re}(H_+H_0^*)$&$\sin(\theta_W)\sin(2\theta_{D_s^*})\cos(\chi)\cos(\theta_W)$\\
12 & $\frac{m_\ell^2}{q^2}\mathrm{Re}(H_-H_0^*)$&$\sin(\theta_W)\sin(2\theta_{D_s^*})\cos(\chi)\cos(\theta_W)$\\
13 & $\frac{m_\ell^2}{q^2}\mathrm{Re}(H_+H_-^*)$& $-\sin^2(\theta_W)\sin^2(\theta_{D_s^*})\cos(2\chi)$\\
14 & $\frac{m_\ell^2}{q^2}\mathrm{Re}(H_tH_0^*)$& $-4\sin^2(\theta_{D_s^*})\cos(\theta_W)$\\
15 & $\frac{m_\ell^2}{q^2}\mathrm{Re}(H_+H_t^*)$& $-\sin(\theta_W)\sin(2\theta_{D_s^*})\cos(\chi)$\\
16 & $\frac{m_\ell^2}{q^2}\mathrm{Re}(H_-H_t^*)$& $-\sin(\theta_W)\sin(2\theta_{D_s^*})\cos(\chi)$
\end{tabular}
\end{table}
Integrating over angles, the differential rate in $q^2$ is then given by
\begin{align}\label{dgammadq2}
&\frac{d\Gamma}{dq^2} = \mathcal{N}\times\frac{64\pi}{9}\Big[\left({H_-}^2+{H_0}^2+{H_+}^2\right) \nonumber\\
+&\frac{{m_\ell^2}}{2q^2}{ \left({H_-}^2+{H_0}^2+{H_+}^2+3 {H_t}^2\right)}\Big],
\end{align}
The helicity amplitudes are defined in terms of standard Lorentz-invariant form 
factors~\cite{RevModPhys.67.893} as
\begin{align}
H_\pm(q^2) =& (M_{B_s}+M_{D_s^*})A_1(q^2) \mp \frac{2M_{B_s}|{\vec{p}}~'|}{M_{B_s}+M_{D_s^*}}V(q^2),\nonumber\\
H_0(q^2) =& \frac{1}{2M_{D_s^*} \sqrt{q^2}} \Big(-4\frac{M_{B_s}^2{|{\vec{p}}~'|}^2}{M_{B_s}+M_{D_s^*}}A_2(q^2)\nonumber\\
&  +  (M_{B_s}+M_{D_s^*})(M_{B_s}^2 - M_{D_s^*}^2 - q^2)A_1(q^2) \Big),\nonumber\\
H_t(q^2) =& \frac{2M_{B_s}|{\vec{p}}~'|}{\sqrt{q^2}}A_0(q^2)\label{helicityamplitudes}
\end{align}
and correspond to the nonzero values of $\overline{\epsilon}^*_\mu(q,\lambda^\prime)\langle D_s^*(p^\prime,\lambda)|\bar{c}\gamma^\mu (1-\gamma^5)  b|{B}_s^0(p)\rangle$ for the different combinations of the $D_s^*$ and $W^-$ polarisations $\lambda$ and $\lambda^\prime$ respectively. The form factors in Eq.~(\ref{helicityamplitudes}) are the standard Lorentz invariant ones; their relations to the matrix elements are given by \cite{RevModPhys.67.893}
\begin{align}
& \langle  D_s^*(p',\lambda)|\bar{c}\gamma^\mu  b|{B}_s^0(p)\rangle=\nonumber\\
& \frac{2i V(q^2)}{M_{B_s} + M_{D_s^*}} \varepsilon^{\mu\nu\rho\sigma}\epsilon^*_\nu(p',\lambda) p'_\rho p_\sigma \nonumber \\
&\langle  D_s^*(p',\lambda)|\bar{c}\gamma^\mu \gamma^5 b|{B}_s^0(p)\rangle = \nonumber\\
& 2M_{D_s^*}A_0(q^2)\frac{\epsilon^*(p',\lambda)\cdot q}{q^2} q^\mu \nonumber\\
&  +(M_{B_s}+M_{D_s^*})A_1(q^2)\Big[ \epsilon^{*\mu}(p',\lambda) - \frac{\epsilon^*(p',\lambda)\cdot q}{q^2} q^\mu \Big]\nonumber \\
&-A_2(q^2)\frac{\epsilon^*(p',\lambda)\cdot q}{M_{B_s}+M_{D_s^*}}\Big[ p^\mu + p'^\mu - \frac{M_{B_s}^2-M_{D_s^*}^2}{q^2}q^\mu \Big]. \label{formfactors}
\end{align}
We also have
\begin{equation}
\langle 0 |\bar{s}\gamma^\nu c|D_s^*(p',\lambda)\rangle = N_{D_s^*}\epsilon^\nu(p',\lambda),
\end{equation}
\begin{equation}
\langle {B}_s^0(p) |\bar{b}\gamma^5 c|0)\rangle = N_{B_s},
\end{equation}
and
\begin{equation}\label{spinsumident}
\sum_{\lambda} \epsilon_\nu(p',\lambda)\epsilon_{\mu}^{*}(p',\lambda) = -g_{\nu\mu}+\frac{p'_\nu p'_\mu}{M^2}
\end{equation}
where $N_{D_s^*}$ and $N_{B_s}$ are amplitudes proportional to the decay constant of the corresponding meson and $\epsilon$ is the $D_s^*$ polarisation vector. We use these when we come to extract the form factors in Eq.~(\ref{formfactors}) from our lattice correlation functions. 
\section{Lattice Calculation}
\label{sec:lattcalc}
\begin{figure}
\includegraphics[scale=0.4]{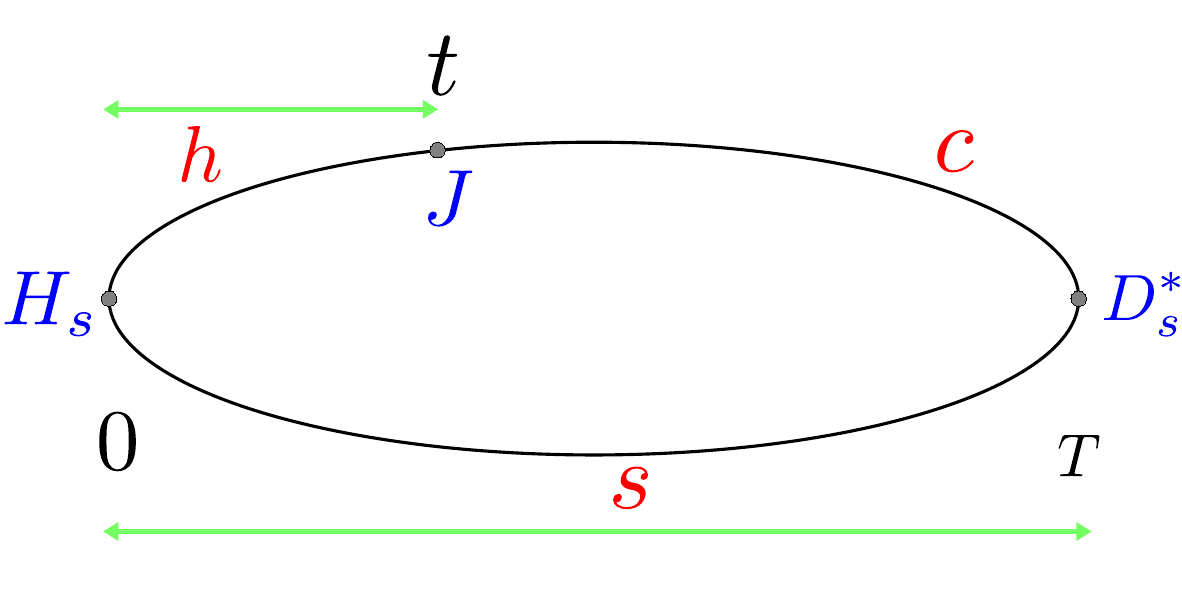}
\caption{\label{3ptfig}Arrangement of propagators in the three point function; we refer to $c$ as the `active' charm quark, $h$ as the `extended' heavy quark and $s$ as the `spectator' strange quark. $J$ represents the insertion of either a vector or axial-vector current and $H_s$ and $D_s^*$ represent the insertion of the corresponding meson interpolating operators.}
\end{figure}

As with previous heavy-HISQ calculations of semileptonic form factors involving $b\rightarrow c$ transitions, e.g.~\cite{Harrison:2020gvo,EuanBsDs}, we work with the heavy meson at rest and give momentum to the charm quark. We use a range of heavy quark masses $m_h$ between the charm and physical bottom quark mass, on ensembles with a range of lattice spacings between $0.09~\mathrm{fm}$ and $0.045~\mathrm{fm}$. We use the the second generation $N_f=2+1+1$ MILC ensembles which include light, strange and charm HISQ sea quarks \cite{PhysRevD.87.054505,PhysRevD.82.074501}. The details of these ensembles, together with the number of configurations we use, are given in Table~\ref{lattdets}. 
\begin{table}
\caption{Details of the gauge field configurations used in our calculation \cite{PhysRevD.87.054505,PhysRevD.82.074501}. We use the Wilson flow parameter~\cite{Borsanyi:2012zs}, $w_0$, to fix 
the lattice spacing given in column 2. The physical value of $w_0$ was determined in \cite{PhysRevD.88.074504} to be 0.1715(9)fm and the values of $w_0/a$, 
which are used together with $w_0$ to compute $a$, were taken from \cite{PhysRevD.96.034516,PhysRevD.91.054508,EuanBsDs}. Set 1 with $w_0/a=1.9006(20)$ is referred to as `fine', set 2 with $w_0/a=2.896(6)$ as `superfine', set 3 with $w_0/a=3.892(12)$ as `ultrafine' and set 4 with $w_0/a=1.9518(7)$ as `physical fine'. $n_\mathrm{cfg}$ and $n_\mathrm{t}$ give the number of configurations and the number of time sources respectively. $am_{l0}$, $am_{s0}$ and $am_{c0}$ are the masses of the sea up/down, strange and charm quarks in lattice units. We also include the approximate mass of the Goldstone pion, computed in~\cite{Bazavov:2017lyh}.\label{lattdets}}
\begin{tabular}{c c c c c c c c c}\hline
 Set &$a$ & $N_x\times N_t$ &$am_{l0}$&$am_{s0}$& $am_{c0}$ & $M_\pi$ &$n_\mathrm{cfg}\times n_\mathrm{t}$ \\ 
  & $(\mathrm{fm})$&  &&&  & $(\mathrm{MeV})$ & \\ \hline
1 & $0.0884$   & $32\times96 $    &$0.0074$ &$0.037$  & $0.440$ & $316$ & $980\times 16$\\
2 & $0.0592$   & $48\times144  $    &$0.0048$ &$0.024$  & $0.286$ & $329$ & $489\times 4$\\
3 & $0.0441$   &$ 64\times192  $    &$0.00316$ &$0.0158$  & $0.188$ & $315$ & $374\times 4$\\
4 & $0.08787$  &$ 64\times96  $    &$0.0012$ &$0.0363$  & $0.432$ & $129$ & $300\times 8$\\\hline
\end{tabular}
\end{table}
On the finest lattices with $a\approx 0.045\mathrm{fm}$ we are able to reach very near to the physical $b$ mass. The heavy quark masses we use, together with the charm and strange quark valence masses, are given in Table~\ref{masses}.
\begin{table}
\centering
\caption{Details of the strange, charm and heavy valence masses. \label{masses}}
\begin{tabular}{c c c c }\hline
 Set &$am_h^\mathrm{val}$ & $am_c^\mathrm{val} $ & $am_s^\mathrm{val} $\\ \hline
1 & $0.65,0.725,0.8$  & $0.449$  &   $0.0376$   \\
2 & $0.427,0.525,0.65,0.8$  & $0.274$ &  $0.0234$    \\
3 & $0.5,0.65,0.8$  & $0.194$ &   $0.0165$   \\
4 & $0.5,0.65,0.8$  & $0.433$ &   $0.036$  \\ \hline
\end{tabular}
\end{table}
Note that in this section, for notational simplicity, we consider the matrix elements in terms of continuum current operators. The nonperturbative renormalisation of our lattice current operators is discussed in Section~\ref{sec:renorm}, where we use the values computed in~\cite{EuanBsDs} and~\cite{EuanBsDsstar}. We calculate, for general choices of current operator $J=\bar{c}\Gamma h$ and $D_s^*$ interpolating operator $\bar{c}\gamma^\nu s$, the correlation functions
\begin{align}
C_\mathrm{2pt}^{D_s^*}(t,0) =& \langle  0|\bar{s}\gamma^\nu c(t) \left(\bar{s}\gamma^\nu c(0)\right)^\dagger| 0 \rangle,\nonumber\\
C_\mathrm{2pt}^{H_s}(t,0) =& \langle  0|\left(\bar{h}\gamma^5 s(t)\right)^\dagger\bar{h}\gamma^5 s(0) | 0 \rangle,\nonumber\\
C_\mathrm{3pt}(T,t,0) =& \langle  0|\bar{s}\gamma^\nu c(T) ~ \bar{c}\Gamma h(t) ~ \bar{h}\gamma^5 s(0)| 0 \rangle. \label{threepointcorr}
\end{align}
In order to improve statistics we work with a random wall source placed at multiple origin times $T_0$. From each of these sources we compute charm propagators with momenta inserted using twisted boundary conditions~\cite{Sachrajda:2004mi, Guadagnoli:2005be}, as well as zero momentum strange propagators and zero momentum heavy quark propagators. Correlation functions with different $T_0$ on a single configuration are binned. We tie the strange and charm propagators together (with the appropriate operators and conjugation) at time $T_0+t$ to construct $C_\mathrm{2pt}^{D_s^*}(t,0)$, and tie the strange and heavy propagators together at time $T_0+t$ to construct $C_\mathrm{2pt}^{H_s}(t,0)$. To construct the three-point correlation functions we use the strange propagator at time $T_0-T$ as a source for the heavy quark propagator which we tie together with the charm quark at time $T_0+t-T$. The arrangement of propagators in $C_\mathrm{3pt}$ is shown in figure~\ref{3ptfig}, shifted so that the $H_s$ operator is at time $0$.
We use twisted boundary conditions in the $(1,1,0)$ direction with the twist chosen such that for the largest value of $am_h$ on a given lattice the points span the physical $q^2$ range evenly, where we estimate the maximum value of $q^2$ using the measured values of $M_{H_s}$ from \cite{EuanBsDs}. The twists we use are given in Table~\ref{twists}, in units of $\pi/L_x$, together with the values of $T$ used in the three point functions in eq.~(\ref{threepointcorr}).
\begin{table}
\centering
\caption{Values of twists, $\theta$, in units of $\pi/L_x$ together with values of $T$ used in the three point functions in eq.~(\ref{threepointcorr}).\label{twists}}
\begin{tabular}{c c c }\hline
Set &$\theta$ & $T$\\ \hline
1 & 0,0.3656,0.7312,1.097,1.462,1.828 & 14,17,20\\
2 & 0,0.8019,1.604,2.406,3.208,4.009 &  22,25,28\\
3 & 0,1.193,2.387,3.580,4.773,5.967 &   31,36,41\\
4 & 0,0.7268,1.454,2.180,2.907,3.634 &  14,17,20\\\hline
\end{tabular}
\end{table}

The correlation functions in Eq.~(\ref{threepointcorr}) are fit to the forms derived by considering the insertion of complete sets of states:
\begin{align}\label{twopointfit}
{C}_\mathrm{2pt}^{D_s^*}(t,0) =&\sum_{n}\Big((A^n)^2{e^{-tE^{n}}}\nonumber+(-1)^{t}(A^n_o)^2{e^{-tE^n_o}}\Big),\\
{C}_\mathrm{2pt}^{H_s}(t,0) =&\sum_{n}\Big((B^n)^2{e^{-tM^{n}}}+(-1)^{t}(B^n_o)^2{e^{-tM^n_o}}\Big)
\end{align}
and
\begin{align}\label{threepointfit}
{C}_\mathrm{3pt}(T,t,0) &=\sum_{n,m}\Big({  A^n B^m J^{nm} e^{-(T-t)E^{n} - tM^{m}} }\nonumber\\
+&{(-1)^{T+t}}  A^n_o B^m J^{nm}_{oe} e^{-(T-t)E^n_o - tM^{m}} \nonumber\\
+&{(-1)^{t}}  A^n B^m_o J^{nm}_{eo} e^{-(T-t)E^{n} - tM^m_o} \nonumber\\
+&{(-1)^{T}}  A^n_o B^m_o J^{nm}_{oo} e^{-(T-t)E^n_o - tM^m_o} \Big) .
\end{align}
Here $n$ and $m$ are integers corresponding to on shell particle states of increasing energies, $A^n$ and $B^m$ are the amplitudes (together with relativistic normalisation factors) of the $D_s^*$ and $H_s$ operators respectively and $E_{n}$ and $M_{n}$ are their energies and masses. The time oscillating terms are a consequence of the use of staggered quarks, and the amplitudes and energies with an $o$ subscript denote those quantities corresponding to time-doubled states. Note that for the matrix elements of $J$ involving time-doubled states, we include an additional subscript $e$ to make clear which of the $D_s^*$ or $H_s$ states is time-doubled. $J^{nm}$ is then related to the matrix element of the current $\bar{c}\Gamma h(t)$ in Eq.~(\ref{threepointcorr}) between the states labelled $n$ and $m$. $J^{nm}_{eo}$ then corresponds to the matrix element between the $A^n$ state and the time-doubled $B^m_o$ state, $J^{nm}_{oe}$ to the matrix element between the time-doubled $A^n_o$ state and $B^m$ state and $J^{nm}_{oo}$ to the matrix element between the time-doubled $A^n_o$ state and the time-doubled $B^m_o$ state. The ground state parameters are related to matrix elements as:
\begin{align}
A^0=&\frac{N_{D_s^*} }{\sqrt{2E_{D_s^*}}}\left(1+\frac{{\vec{p}}_{\nu}^{~\prime 2}}{M_{D_s^*}^2}\right)^{1/2},\nonumber\\
B^0=&\frac{N_{H_s}}{\sqrt{2M_{H_s}}}
\end{align}
and
\begin{equation}
J^{00}_{(\nu,\Gamma)} = \sum_{\lambda}\frac{\epsilon^\nu(p',\lambda) \langle  D_s^*(p',\lambda ) |\bar{c}\Gamma b |H_s\rangle}{\sqrt{2E_{D_s^*}2M_{H_s}\left(1+{\vec{p}}_{\nu}^{~\prime 2}/M_{D_s^*}^2\right)}}\label{relnorm}
\end{equation}
where ${\vec{p}}~'_{\nu}$ is the $\nu$ component of the $D_s^*$ spatial momentum, with $\nu$ corresponding to the choice of polarisation in Eq.~(\ref{threepointcorr}), with current $\overline{c}\Gamma h$. We also compute $\gamma_5 \otimes \gamma_5$ pseudoscalar $\eta_h$ and $\eta_c$ correlation functions, which we will use to parameterise the physical $m_h$ dependence of our form factors as well as to determine the valence and sea charm quark mass mistunings. Note that for the $\eta$ correlation functions we neglect disconnected contributions.
%In the subsequent subsections we discuss the combinations of $\nu$ and $\Gamma$ for which we must compute correlation functions in order to extract the full set of correlation functions defined in Eq.~(\ref{formfactors}).
%#######################################
%The combinations of spin-taste operators we use here to access the form factors, and the methods used to extract the form factors from the matrix elements, are the same as in~\cite{Harrison:2020gvo}. We repeat these combinations here in Table\ref{spintastetable} for reference. 
%#######################################
\subsection{Extraction of Form Factors}
The combinations of spin-taste operators we use here to access the form factors, and the methods used to extract the form factors from the matrix elements, are the same as in~\cite{Harrison:2020gvo}. We repeat these methods below for reference. The combinations of spin taste operator are given in Table~\ref{spintastetable}. 

We give the $D_s^*$ spatial momentum $\vec{p'} = (k,k,0)$.
In order to isolate all the form factors we need one component of 
$\vec{p'}$ to be zero. Keeping both of the others non-zero minmises the discretisation 
errors for a given magnitude of $p'$.

\subsubsection{Extracting $V(q^2)$} 
Here we choose $\mu=3$ and $\nu=1$ and find 
\begin{align}
V(q^2)=&\Phi(k)\frac{M_{H_s} + M_{D_s^*}}{2i k M_{H_s} }J^{00}_{(1,\gamma^3)}
\end{align}
where we have defined the relativistic normalisation
\begin{equation}
\Phi(k) = \sqrt{2E_{D_s^*}2M_{H_s}\left(1+k^{2}/M_{D_s^*}^2\right)}
\end{equation}
with $k$ the $\nu$ component of $p^{\prime}$. 
\subsubsection{Extracting $A_0(q^2)$}
In order to isolate $A_0(q^2)$, following \cite{PhysRevD.90.074506}, we make use of the partially conserved axial-vector current (PCAC) relation $\langle \partial A \rangle = (m_c+m_h) \langle P \rangle$ where $A=\overline{c}\gamma^5 \gamma^\nu h$ and $P=\overline{c}\gamma^5 h$. From Eq.~(\ref{formfactors}) we have
\begin{align}
\sum_{\lambda}\epsilon^\nu(p',\lambda)& q_\mu\langle  D_s^*(p',\lambda)|\bar{c}\gamma^\mu \gamma^5 h|H_s\rangle \nonumber\\
=& \sum_{\lambda}\epsilon^\nu(p',\lambda)2M_{D_s^*}A_0(q^2){\epsilon^*(p',\lambda)\cdot q}\nonumber\\
=& \frac{2kE_{D_s^*}M_{H_s}}{M_{D_s^*}}A_0(q^2).
\end{align}
Taking $\Gamma = \gamma^5$ and $\nu = 1$ in Eq.~(\ref{relnorm}) and multiplying by $m_c+m_b$ we then have
\begin{equation}
A_0(q^2) = \Phi(k)\frac{(m_c+m_b)M_{D_s^*}}{2kE_{D_s^*}M_{H_s}} J^{00}_{(1,{\gamma^5})}
\end{equation}
\subsubsection{Extracting $A_1(q^2)$}
In order to isolate $A_1$ we use the axial-vector current $\Gamma = \bar{c}\gamma^\mu \gamma^5 h$ and $D_s^*$ operator $\bar{s}\gamma^\nu c$ and choose $\mu =\nu= 3$ along the spatial direction with zero $D_s^*$ momentum. Using Eq.~(\ref{formfactors}) this gives 
\begin{align}
\sum_{\lambda}\epsilon^3(p',\lambda)\langle  D_s^*(p',\lambda)|&\bar{c}\gamma^3 \gamma^5 h|H_s\rangle =\nonumber\\
&(M_{D_s^*}+M_{H_s})A_1(q^2)
\end{align}
which gives, in terms of $J^{00}$ 
\begin{equation}
A_1(q^2) = \Phi(0) \frac{J^{00}_{(3,{\gamma^3\gamma^5})}}{M_{D_s^*}+M_{H_s}}.
\end{equation}
\subsubsection{Extracting $A_2(q^2)$}
The extraction of $A_2$ is more complicated than the extraction of the other form factors since no trivial choice of directions in axial-vector and $D_s^*$ operators isolates the contribution of $A_2$ relative to $A_1$ or $A_0$. We use axial-vector current operator $J = \bar{c}\gamma^1 \gamma^5 h$ and $D_s^*$ operator $\bar{s}\gamma^1 c$. This yields contributions from each form factor, 
\begin{align}
&\Phi(k)J^{00}_{(1,\gamma^1 \gamma^5)}=\sum_{\lambda}\epsilon^1(p',\lambda)\langle  D_s^*(p',\lambda)|\bar{c}\gamma^1 \gamma^5 b|H_s\rangle= \nonumber\\
& -\frac{2k^2E_{D_s^*}M_{H_s}}{q^2M_{D_s^*}}A_0(q^2)\nonumber\\
&+(M_{H_s} + M_{D_s^*})\left(1+\frac{k^2}{M_{D_s^*}^2}+\frac{E_{D_s^*}M_{H_s}k^2}{M_{D_s^*}^2q^2}\right)A_1(q^2)\nonumber\\
&-A_2(q^2)\frac{k^2E_{D_s^*}M_{H_s}}{M_{D_s^*}^2(M_{H_s}+M_{D_s^*})}\left(1+\frac{M_{H_s}^2-M_{D_s^*}^2}{q^2}\right).\label{A2subbit}
\end{align}
We must then subtract the $A_0$ and $A_1$ contributions to obtain $A_2$. 
%#######################################
\begin{table}
\centering
\caption{Spin-taste operators used to isolate form factors. The first column is the operator used for the $H_s$, the second for the $D_s^*$ and the third column is the operator used at the current. \label{spintastetable}}
\begin{tabular}{c | c c c }\hline
 &$\mathcal{O}_{H_s}$ & $\mathcal{O}_{D_s^*}$ & $\mathcal{O}_J$   \\
\hline
$V$ & $\gamma_0\gamma_5\otimes \gamma_0\gamma_5$ & $\gamma_1\otimes \gamma_1\gamma_2$ & $\gamma_3\otimes \gamma_3$  \\
$A_0$ & $\gamma_5\otimes \gamma_5$ & $\gamma_1\otimes 1$ & $\gamma_5\otimes \gamma_5$  \\
$A_1$ & $\gamma_5\otimes \gamma_5$ & $\gamma_3\otimes \gamma_3$ & $\gamma_3\gamma_5\otimes \gamma_3\gamma_5$ \\
$A_2$ & $\gamma_5\otimes \gamma_5$ & $\gamma_1\otimes \gamma_1$ & $\gamma_1\gamma_5\otimes \gamma_1\gamma_5$ \\ \hline    
\end{tabular}
\end{table}

\subsection{Fit Parameters}
The correlator fits to Eq.~(\ref{twopointfit}) and Eq.~(\ref{threepointfit}) were done using the \textbf{corrfitter} python package \cite{corrfitter}. These were done simultaneously for all correlation functions on each ensemble, taking all correlations into account. For ground state priors we take $E^{D_s^*}_0=2.1(0.6)\mathrm{GeV}$ and $M^{H_s}_0=M^{H_s}_\mathrm{max} (am_h/0.8)^{1/2}\times 1(0.3) \mathrm{GeV}$ where $M^{H_s}_\mathrm{max}$ is the value of $M_{H_s}$ from \cite{EuanBsDs} corresponding to the largest value of $am_h$. The $m_h$ dependence of the prior for $M^{H_s}_0$ was chosen heuristically to give prior values approximately following the observed $H_s$ masses on each set while remaining suitably loose so as not to constrain the fit results. Our priors for the $\eta_c$, $\eta_h$ and lowest oscillating state energies, as well as amplitudes, are given in table \ref{priortable}, while for the matrix elements of $J$ we take priors $0(1)$.

\begin{table}
\centering
\caption{Correlator fit priors. We take $\Delta E^{(o)}_i=\Lambda_\mathrm{QCD}\times 1.0(0.75)$ where $\Delta E^{(o)}_i = E^{(o)}_{i+1}-E^{(o)}_{i},~i\geq 0$ and here for our correlator fits we take $\Lambda_\mathrm{QCD}=0.75\mathrm{GeV}$, $\Omega_{D_s^*} =(2.1^2+p'^2)^\frac{1}{2}$ and $\Omega_{H_s}=M^\mathrm{max}_{H_s}\left(\frac{am_h}{0.8}\right)^\frac{1}{2}$ where $M^\mathrm{max}_{H_s}$ is the value of $M_{H_s}$ corresponding to the largest $am_h$, taken from \cite{EuanBsDsstar}. While $\Omega_{D_s^*}$ was chosen to follow the relativistic dispersion relation, $\Omega_{H_s}$ was chosen heuristically to give prior values approximately following the observed $H_s$ masses on each set while remaining suitably loose so as not to constrain the fit results. \label{priortable}}
\begin{tabular}{ c c c c c }\hline
 Prior & $\eta_h$ & $\eta_c$ & $D_s^*(p')$ & $H_s$\\\hline
$E_0/\mathrm{GeV}$	&$m_h\times 2.5(0.5)$	&$3.0(0.9)$  	&$\Omega_{D_s^*} 1.0(0.3)$ 	&$\Omega_{H_s}1.0(0.3)$ 	\\
$E^o_0/\mathrm{GeV}$	&$-$		&$-$		&$\Omega_{D_s^*} 1.2(0.5)$	&$\Omega_{H_s}1.2(0.5)$		\\
$A(B)_{(o)}^n$		&$0.1(5.0)$	&$0.1(5.0)$	&$0.1(5.0)$			&$0.1(5.0)$\\\hline
\end{tabular}
\end{table}

In order to reduce excited state contamination and to improve the stability and convergence of the fits we exclude data for $t<t_\mathrm{min}$ and for $t>t_\mathrm{max}$. We also specify an SVD cut using the tools available in~\cite{corrfitter}. 
%For very large fits such as these involving a large number of fit parameters, methods of estimating suitable choices of $t_\mathrm{min}$, $t_\mathrm{max}$ and SVD cut are impractical. Instead,
We use several different choices of $t_\mathrm{min}$, $t_\mathrm{max}$ and SVD cut and investigate the stability of our subsequent analysis with respect to taking different combinations of fit parameters. These fit parameters are given in Table~\ref{fitparams}, and the stability of our analysis will be discussed later in Section~\ref{sec:results}.

\begin{table}
\centering
\caption{Details of fit parameters, together with variations used in section \ref{stabsec} to check stability. $\Delta T$ indicates the number of data points at the extremities of correlation functions not included in the fit.\label{fitparams} Bold values are those used to produce our final results. $\chi^2/\mathrm{dof}$ is estimated by introducing svd and prior noise as in \cite{corrfitter}. We do not compute $\chi^2$ values including prior and svd noise for those fits with $n_\mathrm{exp}=4$.}
\begin{tabular}{ c c c c c c c c }\hline
Set & $n_\mathrm{exp}$ & $\Delta T_\mathrm{3pt}$ & $\Delta T_\mathrm{2pt}^{D_s^*}$& $\Delta T^{H_s}_\mathrm{2pt}$ & SVD cut & $\chi^2/\mathrm{dof}$ & $\delta$ \\ \hline
1  & \textbf{3} & \textbf{2} & \textbf{4} & \textbf{4} & \textbf{0.005} & \textbf{1.06} & 0\\
  & 3 & 2 & 4 & 4 & 0.01 & 1.09 & 1\\
  & 3 & 3 & 6 & 6 & 0.005 & 0.96 & 2 \\
  & 4 & 2 & 4 & 4 & 0.005 & $-$ & 3\\\hline
2  & \textbf{3} & \textbf{2} & \textbf{4} & \textbf{4} & \textbf{0.025} & \textbf{1.04}& 0\\
  & 3 & 2 & 4 & 4 & 0.05 & 1.00& 1\\
  & 3 & 3 & 7 & 7 & 0.025 & 0.98& 2\\
  & 4 & 2 & 4 & 4 & 0.025 & $-$ & 3\\\hline
3  & \textbf{3} & \textbf{3} & \textbf{6} & \textbf{6} & \textbf{0.005} & \textbf{1.02}& 0\\
  & 3 &  3 & 6 & 6 & 0.01 & 1.01& 1\\
  & 3 &  4 & 8 & 8 & 0.005 & 0.99& 2\\
  & 4 &  3 & 6 & 6 & 0.005 & $-$ & 3\\\hline
4  & \textbf{3} &  \textbf{2} & \textbf{5} & \textbf{5} & \textbf{0.01} & \textbf{1.03}& 0\\
  & 3 &  2 & 5 & 5 & 0.025 & 1.04& 1\\
  & 3 &  3 & 7 & 7 & 0.01 & 1.05 & 2\\
  & 4 &  2 & 5 & 5 & 0.01 & $-$ & 3\\\hline
\end{tabular}
\end{table}

\subsection{Non-Perturbative Current Renormalisation}
\label{sec:renorm}
The renormalisation factors relating the HISQ lattice currents in Table~\ref{spintastetable} to the continuum currents considered in Section~\ref{sec:theory} are the same as those used in~\cite{Harrison:2020gvo}. These were computed previously in~\cite{EuanBsDs} for the local vector current and in~\cite{EuanBsDsstar} for the axial current using the partially conserved vector curent~(PCVC) and axial-vector current~(PCAC) relations respectively. As in the previous $B_c \to J/\psi$ form factor calculation we include some values of the heavy quark masses for which the $Z$ factors were not computed. We interpolate between the values computed in~\cite{EuanBsDs} and~\cite{EuanBsDsstar} on sets 1 and 4 for masses $am_h=0.725$ and $am_h=0.65$ respectively, setting the uncertainties of the interpolated factors equal to the largest uncertainty of the other values. We tabulate these renormalisation factors in Table~\ref{Zfactors}, where we include the $am$-dependent discretisation correction terms, $Z^\mathrm{disc}$, for the HISQ-quark tree level wavefunction renormalisation computed in~\cite{ronrenormdisc}. These receive contributions beginning at $\mathcal{O}((am_h)^4)$ for HISQ and as such are close to $1$. The total renormalisation factor for an (axial-)vector current is then given by $Z^{V(A)}Z^\mathrm{disc}$. Note however that no renormalisation factor is required for $A_0$ since we determine it using the absolutely normalised $\gamma_5 \otimes \gamma_5$ pseudoscalar current together with the PCAC relation.

We neglect correlations between the renormalisation factors $Z$ and our lattice data, since the $Z$ factors have uncertainties which are typically at least an order of 
magnitude smaller than the corresponding form factors.

\begin{table}
\centering
\caption{ $Z$ factors from \cite{EuanBsDsstar} and \cite{EuanBsDs} for the axial-vector and vector operators used in this work, together with the discretisation corrections. $Z^A$ and $Z^V$ values for $am_h=0.725$ on set 1 and $am_h=0.65$ on set 4 were obtained by interpolation from the other values for those sets. The uncertainties of the interpolated factors are set equal to the largest uncertainty of the other values. \label{Zfactors}}
\begin{tabular}{ c c c c c }
\hline
Set & $am_h$ & $Z^A$& $Z^V$& $Z^\mathrm{disc}$ \\\hline
1 & 0.65 & 1.03740(58)& 1.0254(35) & 0.99635 \\
 & 0.725 & 1.04030(58)& 1.0309(35) & 0.99491 \\
 & 0.8 & 1.04367(56)& 1.0372(32) & 0.99306 \\
\hline
2 & 0.427 & 1.0141(12)& 1.0025(31) & 0.99931 \\
 & 0.525 & 1.0172(12)& 1.0059(33) & 0.99859 \\
 & 0.65 & 1.0214(12)& 1.0116(37) & 0.99697 \\
 & 0.8 & 1.0275(12)& 1.0204(46) & 0.99367 \\
\hline
3 & 0.5 & 1.00896(44)& 1.0029(38) & 0.99889 \\
 & 0.65 & 1.01363(49)& 1.0081(43) & 0.99704 \\
 & 0.8 & 1.01968(55)& 1.0150(49) & 0.99375 \\
\hline
4 & 0.5 & 1.03184(47)& 1.0134(24) & 0.99829 \\
 & 0.65 & 1.03717(47)& 1.0229(29) & 0.99645 \\
 & 0.8 & 1.04390(39)& 1.0348(29) & 0.99315 \\
\hline
\end{tabular}
\end{table}

\section{Results}
\label{sec:results}
Here we give the lattice results for the form factors, which were extracted from the matrix elements in Eq.~(\ref{relnorm}) resulting from the fits discussed in Section~\ref{sec:lattcalc}. For $V$, $A_1$ and $A_2$ these include the renormalisation factors given in Table~\ref{Zfactors}. The values of the form factors are tabulated in Appendix~\ref{app:lattice_results} in Tables~\ref{set1},~\ref{set2},~\ref{set3}, and ~\ref{set4} together with the value of the momentum component in lattice units, $ak$, of the $D_s^*$ in the $x$ and $y$ directions.

Our results for the $D_s^*$ and $\eta_c$ masses on each set are given in lattice units in Table~\ref{charmMasses} together with the corresponding spin-taste operators. Our results for the $H_s$ and $\eta_h$ masses are given in lattice units in Table~\ref{Massesheavy} together with the corresponding spin-taste operators.

We fit the heavy mass and lattice spacing dependence of our lattice form factor results in order to determine the physical continuum form factors, following the method in~\cite{Harrison:2020gvo}. We repeat the details of this fit here, as well as performing similar tests of stability.
\begin{table}
\centering
\caption{Results for the $D_s^*$ masses for the local spin-taste operator $\gamma_1\otimes \gamma_1$ and $1-$link operators $\gamma_1\otimes 1$ and $\gamma_1\otimes \gamma_1\gamma_2$ used in our calculation, see table \ref{spintastetable}. Here we also include values for the local $\gamma_5\otimes \gamma_5$ $\eta_c$ mass.\label{charmMasses}}
\begin{tabular}{ c c c c | c }
\hline
& $aM_{D_s^*}$ & & & $aM_{\eta_c}$\\\hline
Set & $\gamma_1\otimes \gamma_1$&  $\gamma_1\otimes 1$ & $\gamma_1\otimes \gamma_1\gamma_2$& $\gamma_5\otimes \gamma_5$\\\hline
1&0.96451(48)	&0.96493(56)	&0.96422(67)	&1.364940(48)	\\
\hline
2&0.63482(80)	&0.6350(12)	&0.6347(12)	&0.896802(67)	\\
\hline
3&0.47327(58)	&0.47284(99)	&0.4727(11)	&0.66721(12)	\\
\hline
4&0.93975(48)	&0.93957(63)	&0.93928(68)	&1.329310(45)	\\
\hline
\end{tabular}
\end{table}

\begin{table}
\centering
\caption{Results for the $\eta_h$ masses and $H_s$ masses for the local spin-taste operators $\gamma_5\otimes \gamma_5$ and $\gamma_0\gamma_5\otimes \gamma_0\gamma_5$ that we use in our calculation, see table \ref{spintastetable}. \label{Massesheavy}}
\begin{tabular}{ c c c c c }
\hline
Set & $am_h$ & $aM_{H_s}(\gamma_5\otimes \gamma_5)$& $aM_{H_s}(\gamma_0\gamma_5\otimes \gamma_0\gamma_5)$& $aM_{\eta_h}$\\\hline
1&	0.65	&1.12498(16)	&1.12550(27)	&1.775201(42)	\\
&	0.725	&1.20419(17)	&1.20467(30)	&1.921510(40)	\\
&	0.8	&1.28122(19)	&1.28166(33)	&2.064184(39)	\\
\hline
2&	0.427	&0.77431(28)	&0.77456(59)	&1.233625(58)	\\
&	0.525	&0.88460(35)	&0.88496(73)	&1.439573(54)	\\
&	0.65	&1.01969(45)	&1.02019(92)	&1.693959(49)	\\
&	0.8	&1.17464(56)	&1.1752(12)	&1.987607(46)	\\
\hline
3&	0.5	&0.80339(34)	&0.8020(12)	&1.343315(81)	\\
&	0.65	&0.96484(40)	&0.9634(14)	&1.650857(69)	\\
&	0.8	&1.11894(45)	&1.1174(15)	&1.946422(60)	\\
\hline
4&	0.5	&0.95452(14)	&0.95487(26)	&1.470108(45)	\\
&	0.65	&1.11976(19)	&1.11987(34)	&1.773763(42)	\\
&	0.8	&1.27577(25)	&1.27571(44)	&2.062919(42)	\\
\hline
\end{tabular}
\end{table}

\subsection{Extrapolation to the Physical Point}
\label{sec:physextrap}

\begin{figure*}
\centering
\includegraphics[scale=0.225]{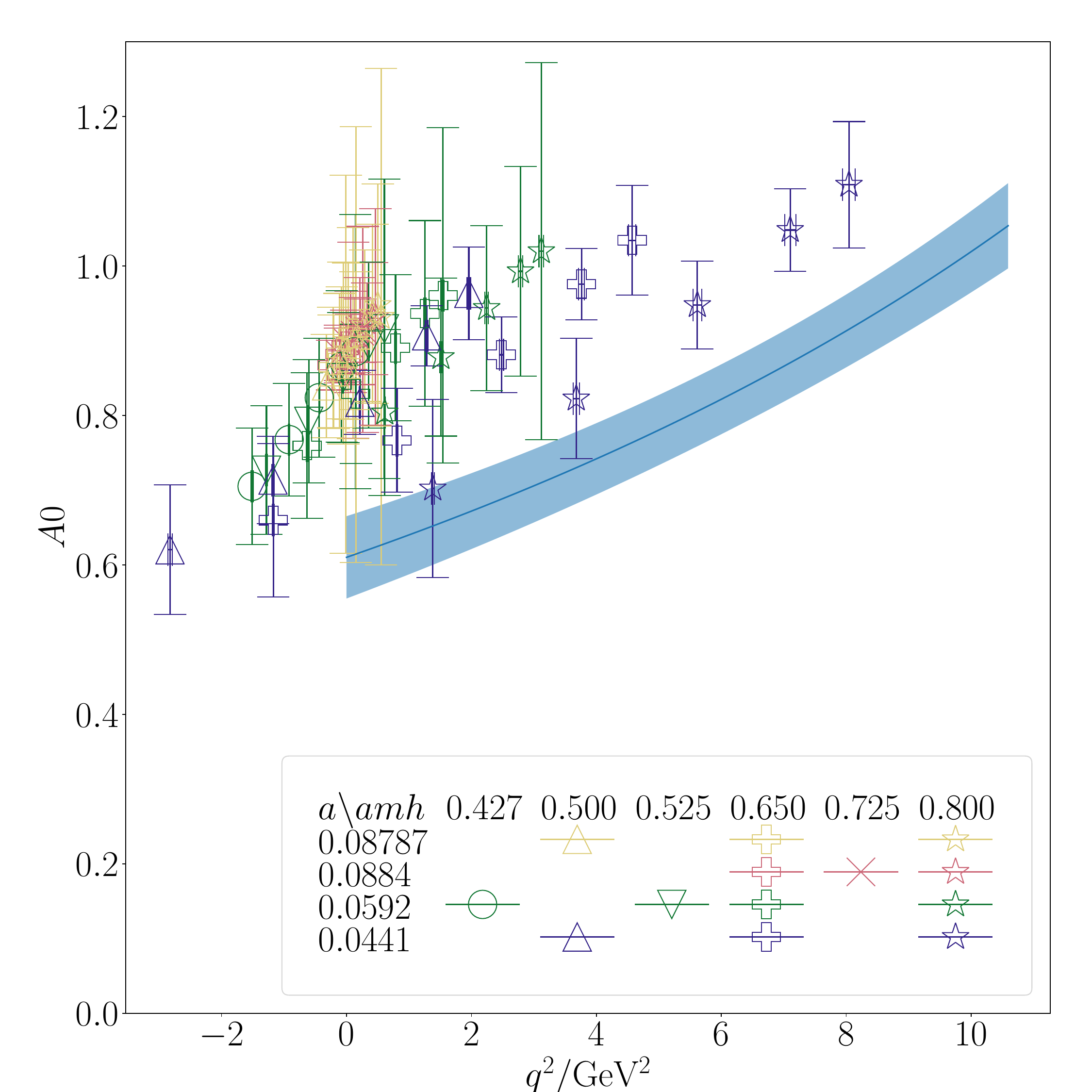}
\includegraphics[scale=0.225]{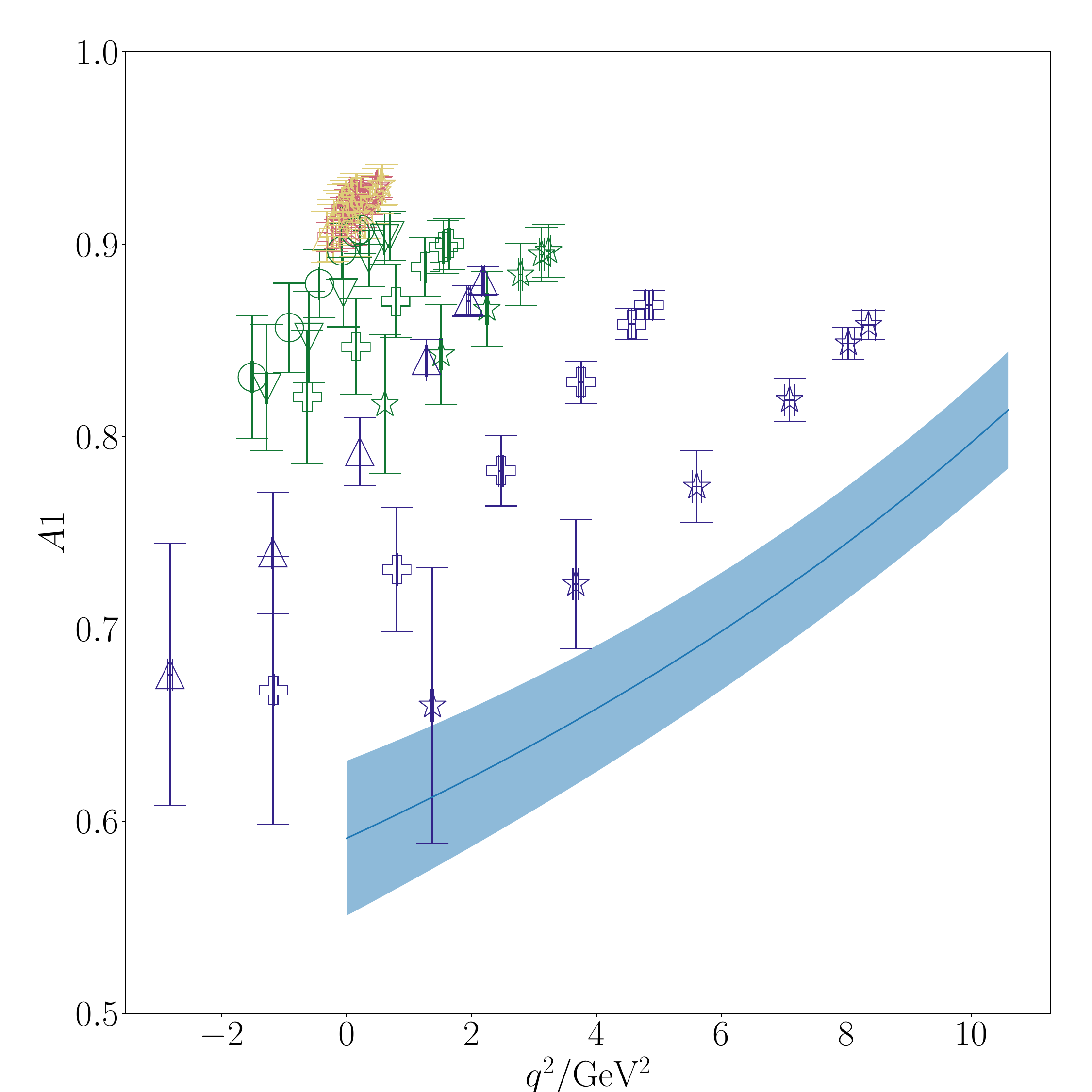}
\includegraphics[scale=0.225]{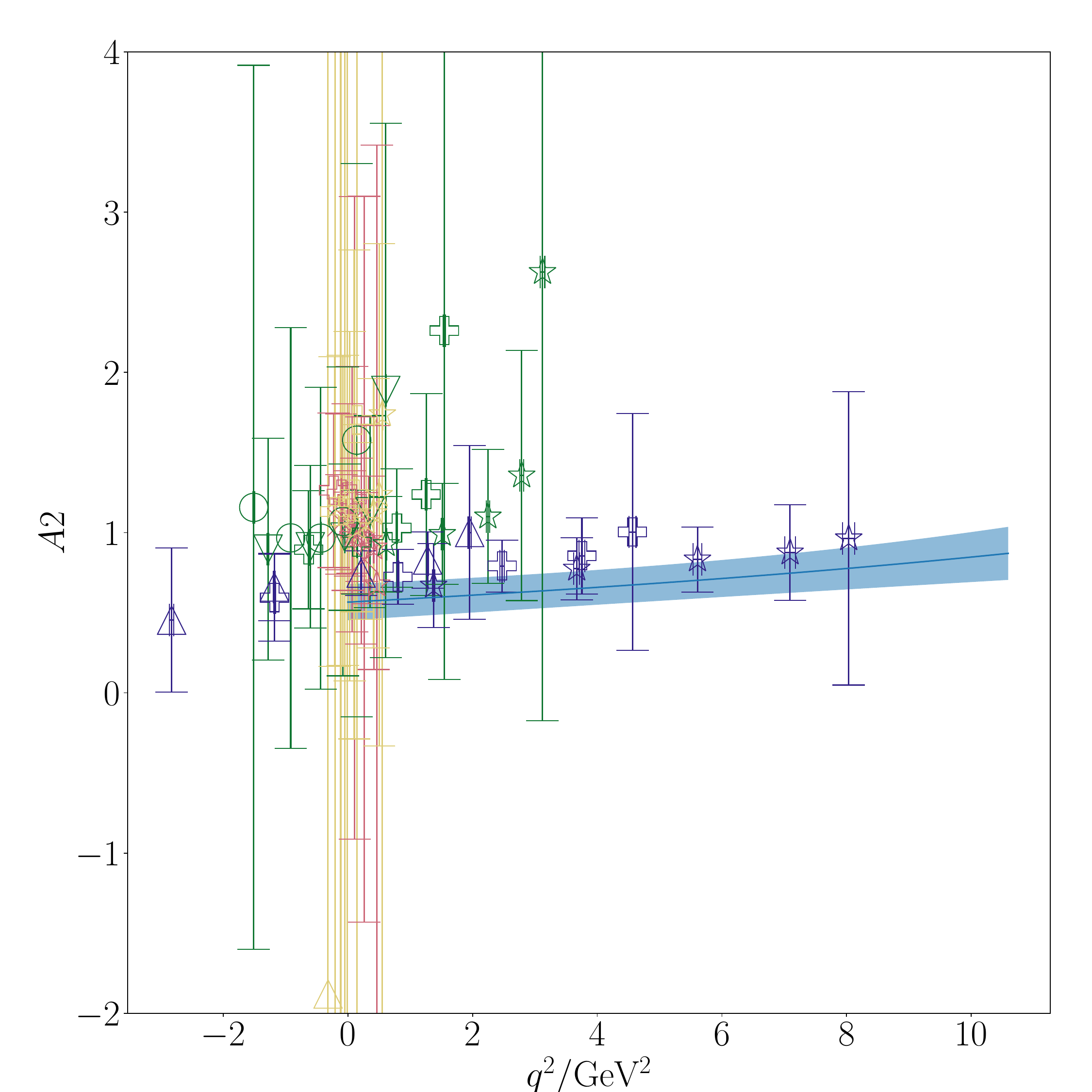}
\includegraphics[scale=0.225]{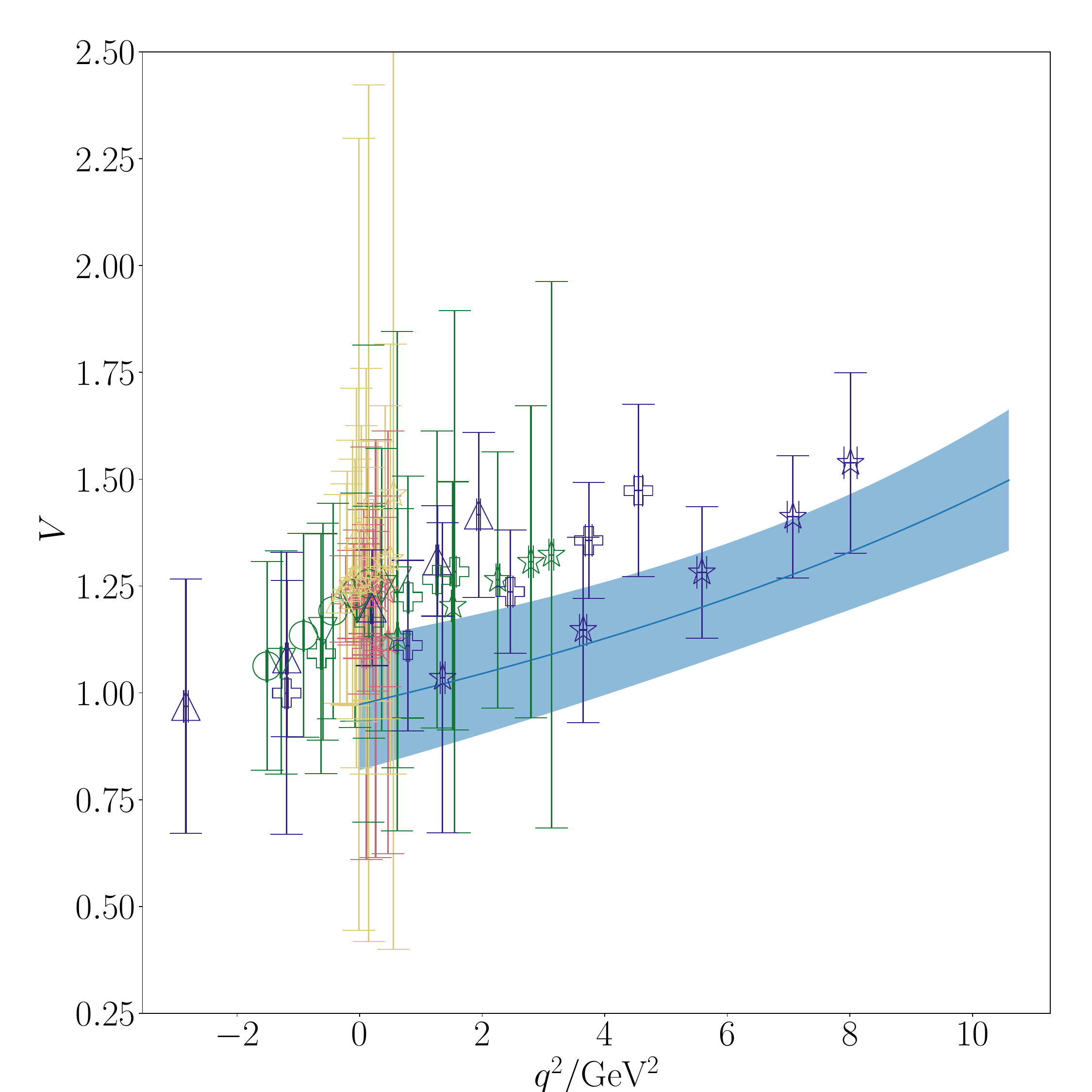}
\caption{\label{FFqsq} The points show our lattice QCD results for each 
form factor as given in Tables~\ref{set1},~\ref{set2},~\ref{set3} 
and~\ref{set4} as a function of squared four-momentum transfer, $q^2$. 
The legend gives the mapping between symbol colour and shape and the 
set of gluon field configurations used, as given by the lattice 
spacing, and the heavy quark mass in lattice units (see Tables~\ref{lattdets} and~\ref{masses}). 
The blue curve with error band is the result of our fit in the continuum 
limit and with the physical $b$ quark mass. 
}
\end{figure*}

\begin{figure*}
\centering
\includegraphics[scale=0.225]{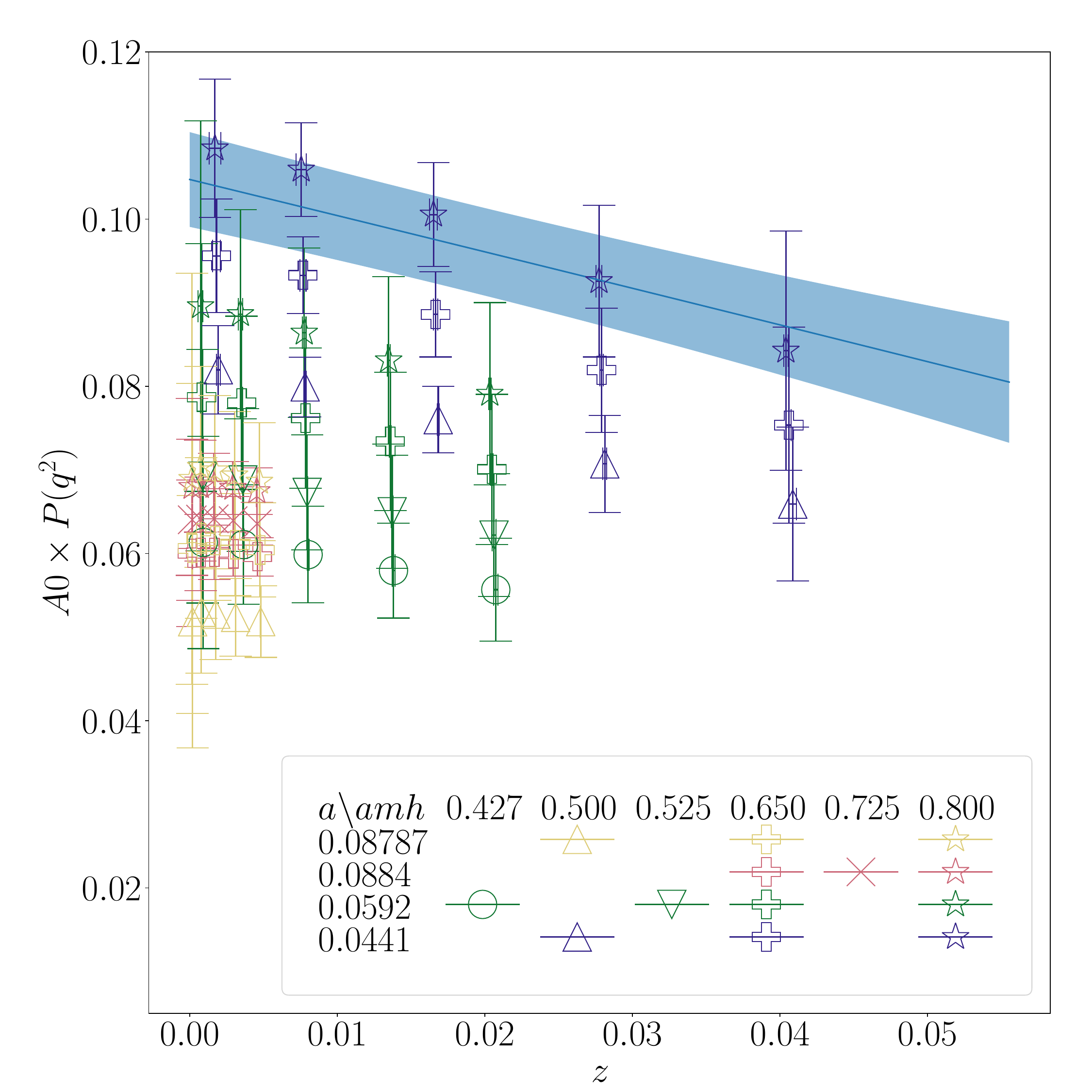}
\includegraphics[scale=0.225]{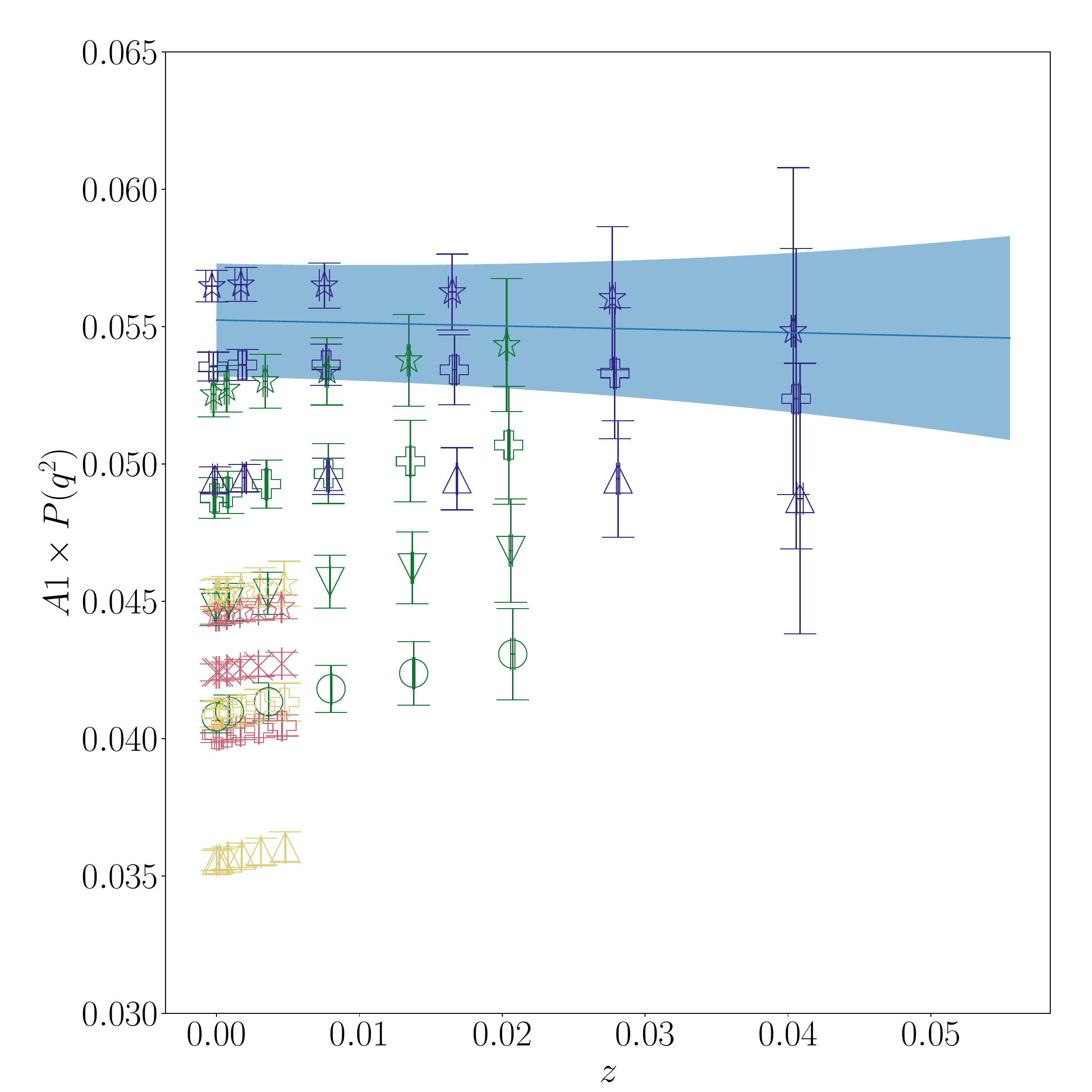}
\includegraphics[scale=0.225]{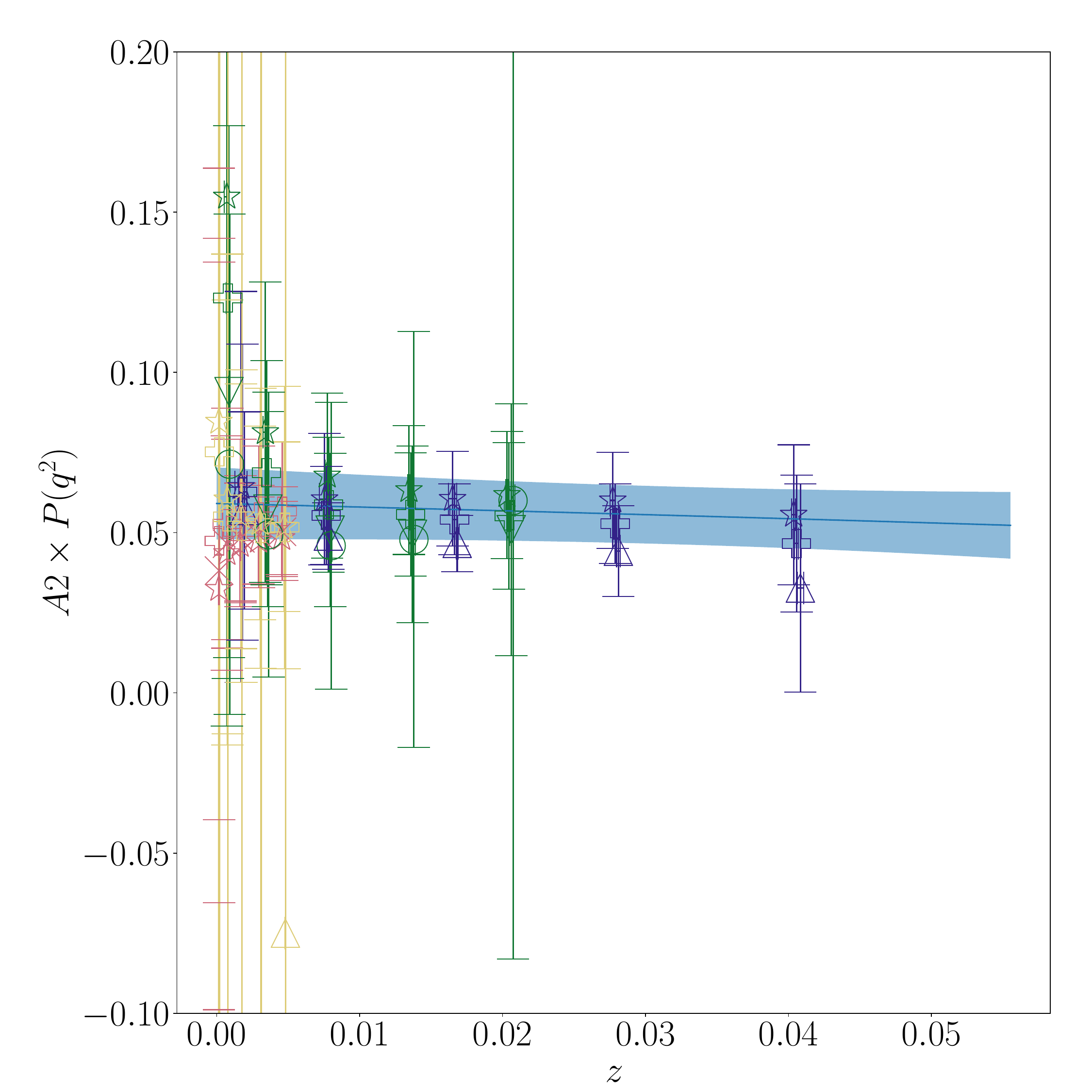}
\includegraphics[scale=0.225]{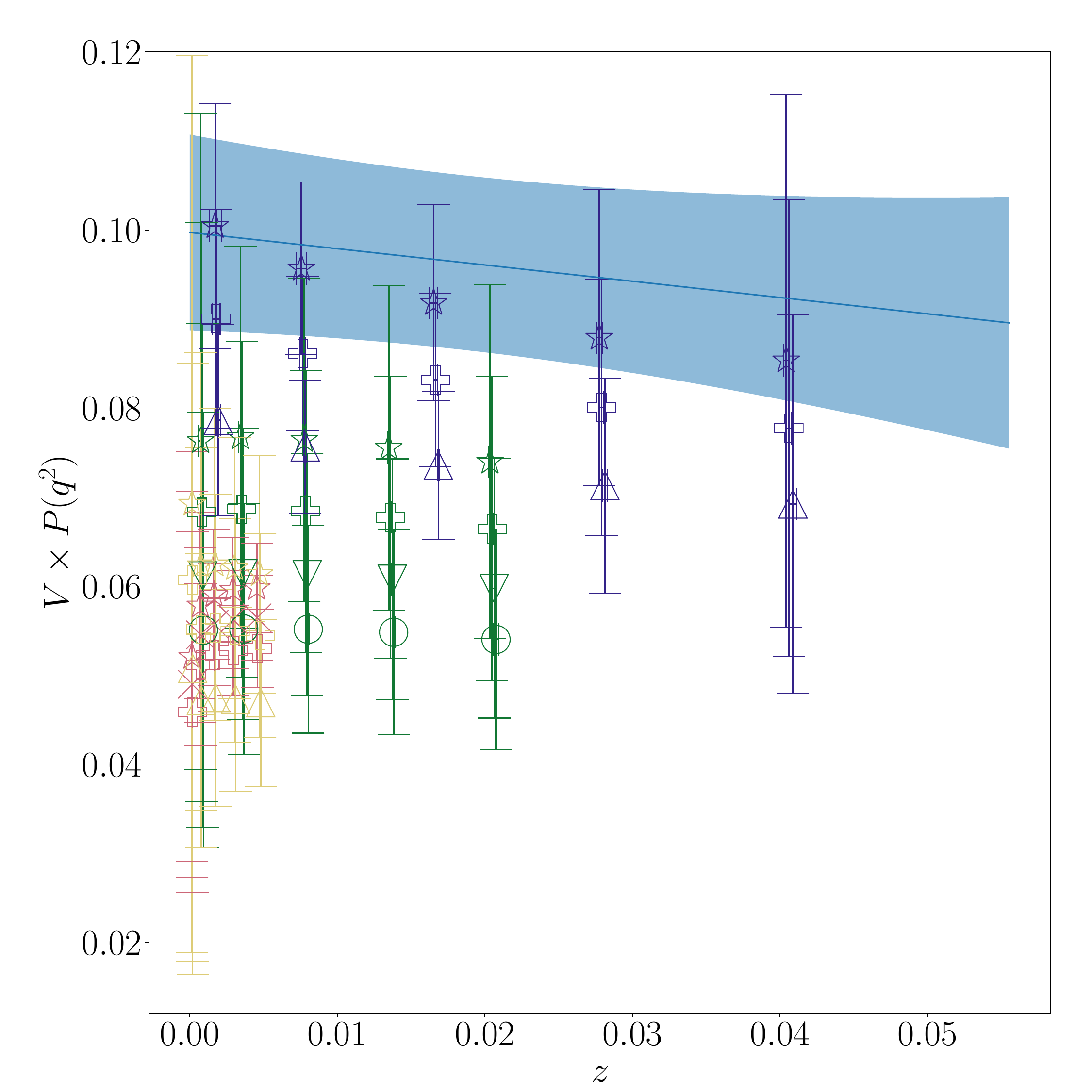}
\caption{\label{zspacepoleremoved} The points show our lattice QCD results for each 
form factor as given in Tables~\ref{set1},~\ref{set2},~\ref{set3} 
and~\ref{set4} multiplied by the pole function of Eq.~(\ref{poleformeq}) and plotted 
in $z$-space. 
The legend gives the mapping between symbol colour and shape and the 
set of gluon field configurations used, as given by the lattice 
spacing, and the heavy quark in lattice units (see Tables~\ref{lattdets} and~\ref{masses}). 
The blue curve with error band is the result of our polynomial fit in $z$ with lattice 
spacing and heavy quark mass dependence (Eq.~(\ref{fitfunctionequation})), evaluating the result in the continuum 
limit and for the $b$ quark mass, to give the physical form factor for $B_s \to D_s^*$. 
}
\end{figure*}

We parameterise the $q^2$ dependence using the $z$-expansion~\cite{Boyd:1997kz, Caprini:1997mu, poleform}. We first change variables from $q^2$ to $z(q^2,t_0)$, with
\begin{equation}
z(q^2,t_0)=\frac{\sqrt{t_+-q^2}-\sqrt{t_+-t_0}}{\sqrt{t_+-q^2}+\sqrt{t_+-t_0}}.
\end{equation}
We take $t_0$ equal to the maximum physical value of $q^2$,
\begin{equation}
t_0=(M_{H_s}-M_{D_s^*})^2,
\end{equation}
and
\begin{equation}
t_+=(M_{H}+M_{D^*})^2.
\end{equation}
Since we do not have direct access to $M_H$, the mass of the $h\bar{u}$ pseudoscalar meson, we instead use $M_H^\mathrm{latt}=M_{H_s}-(M_{B_s}^\mathrm{phys}-M_B^\mathrm{phys})$. We use the physical value of $M_{D^*}$ from experiment, since our valence charm masses are tuned to the physical value, and the light quark mass mistuning is accounted for elsewhere in the fit function.  The numerical values of the physical $B$, ${B_s}$ and ${D^*}$ masses are given in Table~\ref{physmasses}.
The form factors include poles resulting from $b\bar{c}$ states, with masses below the pair production threshold $t_+$, with the same quantum numbers as the corresponding current. We include these poles, which are the same for $B_s\to D_s^*$ as for $B_c\to J/\psi$ and $B\to D^*$, using the same form as in~\cite{Harrison:2020gvo}, which was taken from~\cite{Boyd:1997kz, poleform}:
\begin{equation}\label{poleformeq}
P(q^2)=\prod_{M_\mathrm{pole}}z(q^2,M_\mathrm{pole}^2).
\end{equation}
We approximate the heavy mass dependence of the pole masses using $M_\mathrm{pole}=M_{H_s} +M_\mathrm{pole}^\mathrm{phys} - M_{B_s}^\mathrm{phys}$, which ensures that in the physical continuum limit the correct physical pole masses are recovered. The physical pole masses used here are listed in Table~\ref{poletab}.

Our fit function then takes the form 
\begin{equation}
F(q^2) = \frac{1}{P(q^2)}\sum_{n=0}^3 a_n z^n \mathcal{N}_n \label{fitfunctionequation},
\end{equation}
where $P(q^2)$ is the appropriate pole form for that form 
factor (constructed using $1^-$ states for $V(q^2)$, $1^+$ states for $A_1(q^2)$ and $A_2(q^2)$ or $0^-$ states for $A_0(q^2)$) 
as in Eq.~(\ref{poleformeq}). 
The remainder of the fit function is a polynomial in $z$ with 
separate coefficients, $a_n$, for each form factor that take the form
\begin{equation}
\label{eq:anfitform}
a_n = \sum_{j,k,l=0}^3 b_n^{jkl}\Delta_{h}^{(j)} \left(\frac{am_c^\mathrm{val}}{\pi}\right)^{2k} \left(\frac{am_h^\mathrm{val}}{\pi}\right)^{2l}.
\end{equation}
The $\Delta_{h}^{(j)}$ allow for the dependence on the heavy quark 
mass using the $\eta_h$ mass as a physical proxy for this. 
We have $\Delta_{h}^{(0)}=1$ and
\begin{equation}
\Delta_{h}^{(j\neq 0)}=\left(\frac{2\Lambda}{M_{\eta_h}}\right)^j-\left(\frac{2\Lambda}{M_{\eta_b}^\mathrm{phys}}\right)^j.
\end{equation}
The physical value of the $\eta_b$ mass is given in Table~\ref{physmasses} and we take $\Lambda$ = 0.5 GeV. 
The remainder of Eq.~(\ref{fitfunctionequation}), $\mathcal{N}_n$, takes into account 
the effect of mistuning the valence and sea quark masses for each form factor, where 
\begin{equation}
\mathcal{N}_n = 1 + A_n \delta_{m_c}^\mathrm{val}+ B_n \delta_{m_c}^\mathrm{sea}+ C_n \delta_{m_s}^\mathrm{val}+ D_n \delta_{m_s}^\mathrm{sea} + E_n \delta_{m_l}^\mathrm{sea}
\end{equation}
with
\begin{align}
\delta_{m_c}^\mathrm{val} = (am_c^\mathrm{val}-am_c^\mathrm{tuned})/am_c^\mathrm{tuned},\nonumber\\
\delta_{m_c}^\mathrm{sea} = (am_{c}^\mathrm{sea}-am_c^\mathrm{tuned})/am_c^\mathrm{tuned},\nonumber\\
\delta_{m_s}^\mathrm{val} = (am_{s}^\mathrm{val} - am_{s}^\mathrm{tuned})/(10am_{s}^\mathrm{tuned}),\nonumber\\
\delta_{m_{s(l)}}^\mathrm{sea} = (am_{s(l)}^\mathrm{sea} - am_{s(l)}^\mathrm{tuned})/(10am_{s}^\mathrm{tuned}).\label{deltatermseq}
\end{align}
Using a ratio of lattice quark masses to $10am_s^\mathrm{tuned}$ is a convenient proxy for the usual chiral expansion parameter which is a ratio of squared meson masses to $\Lambda_\chi^2$ where $\Lambda_\chi=4\pi f_\pi$. The tuned values of the quark masses are given by
\begin{equation}
am_c^\mathrm{tuned} = am_c^\mathrm{val}\frac{M_{\eta_c}^\mathrm{phys}}{M_{\eta_c}},
\end{equation}
and
\begin{equation}
am_s^\mathrm{tuned} = am_s^\mathrm{val}\left(\frac{M_{\eta_s}^\mathrm{phys}}{M_{\eta_s}}\right)^2
\end{equation}
$M_{\eta_c}$ on each set is given in lattice units in Table~\ref{charmMasses} and we use the values of $M_{\eta_s}$ given in~\cite{EuanBsDsstar} 
which used the same values of $am_s^\mathrm{val}$. To determine the mistuning of the $u/d = l$ quark mass in the sea we take
\begin{equation}
am_{l}^\mathrm{tuned}=am_{s}^\mathrm{tuned}/{[m_s/m_l]^\mathrm{phys}},
\end{equation}
with ${[m_s/m_l]^\mathrm{phys}} = 27.18(10)$ from~\cite{Bazavov:2017lyh}. We take priors of $0(1)$ for each $b_n$, multiplying terms of order $\mathcal{O}(a^2)$ by $0.5$ in line with the tree level $a^2$ improvement of the HISQ action~\cite{PhysRevD.75.054502}. We also use priors of $0.0(0.1)$ for $B_n$, motivated by the results of the analysis 
of $m_c^\mathrm{sea}$ effects on $w_0$ in \cite{PhysRevD.91.054508}. We take priors of $0.0(0.5)$ for $D_n$ and $E_n$ for each form factor, since sea quark mistuning effects enter at 1-loop. All of the remaining priors are taken as $0(1)$, motivated by the analysis done in~\cite{Harrison:2020gvo} using the empirical Bayes criterion which showed that for the $B_c\to J/\psi$ form factors this choice was conservative, which we expect to be the case here also. The physical masses used for the $\eta_c$ and $\eta_s$ are given in Table~\ref{physmasses}.

We impose the kinematical constraint 
\begin{equation}
\label{eq:contconstraint}
2M_{D_s^*}A_0(0) = (M_{D_s^*}+M_{H_s})A_1(0) + (M_{D_s^*}-M_{H_s})A_2(0).
\end{equation}
We do this, using our lattice meson masses at each value of $am_h$ 
and allowing for discretisation and quark mass 
mistuning effects, by requiring 
\begin{eqnarray}
\label{eq:ourconstraint}
A_0(0) - (aM_{D_s^*}+aM_{H_s})/(2aM_{D_s^*})A_1(0) &+& \\
(aM_{D_s^*}-aM_{H_s})/(2aM_{D_s^*})A_2(0) &=& \Delta_\mathrm{kin}. \nonumber 
\end{eqnarray}
$\Delta_\mathrm{kin}$ here is a nuisance term made up of leading order 
discretisation and mistuning effects to account for the use of lattice 
masses rather than values in the physical continuum limit. 
We take 
\begin{eqnarray}
\label{eq:deltadef}
\Delta_\mathrm{kin} &=& \sum_{i=1}^3\alpha_{c,i} (am^\mathrm{val}_c/\pi)^{2i} + \alpha_{h,i}(am_h/\pi)^{2i} \nonumber \\
+ \beta_c \delta_{m_c}^\mathrm{val} &+& \beta'_c \delta_{m_c}^\mathrm{sea} + \beta'_s\delta_{m_{s}}^\mathrm{val} + \beta_s\delta_{m_{s}}^\mathrm{sea}+ \beta_l\delta_{m_{l}}^\mathrm{sea}
\end{eqnarray} 
where $\alpha$ and $\beta$ are priors taken as $0(1)$. 
We find that the fit returns values for $\alpha$ and $\beta$ well 
within their prior widths. 

\begin{table}
\centering
\caption{\label{physmasses}Values used in our fits for the physical masses 
of relevant mesons, in GeV. These are from the Particle Data 
Group~\cite{Tanabashi:2018oca} 
except for the unphysical $\eta_s$ which we take from lattice 
calculations of the mass of the pion and kaon~\cite{Dowdall:2013rya}. The 
$\eta_s$ and $\eta_c$ masses are used to set mass mistuning terms in our 
fit and so include an uncertainty. The other masses are used as kinematic 
parameters in setting up our fit in $z$-space and used without uncertainties. In light of the results of~\cite{Hatton:2020qhk} we have checked that using a value of $M_{\eta_c}^\mathrm{phys}$ $10~\mathrm{MeV}$ lower than that given here, allowing for the effects of QED and $c\bar{c}$ annihilation, has only a very small effect on our results at the level of $0.05\sigma$.}
\begin{tabular}{ c | c }\hline
meson & $M^\mathrm{phys}$[GeV]\\\hline
${\eta_b}$&9.3889\\
${B_s}$   & 5.3669\\
${B}$     &5.27964 \\
$\eta_c$ & 2.9863(27)\\
${D_s^*}$& 2.112\\
${D^*}$   &2.010 \\
${\eta_s}$   &0.6885(22)\\\hline
\end{tabular}
\end{table}

\begin{table}
\centering
\caption{\label{poletab} Expected $B_c$ pseudoscalar, 
vector and axial vector 
masses below $BD^*$ threshold that we use in our pole factor, Eq.~(\ref{poleformeq}). 
Pseudoscalar values for the ground-state and first radial excitation 
are taken from experiment~\cite{Aaij:2016qlz, Sirunyan:2019osb, Aaij:2019ldo, Tanabashi:2018oca}; the other values are taken from~\cite{Harrison:2017fmw} and are derived from 
lattice QCD calculations~\cite{Dowdall:2012ab} and model 
estimates~\cite{Eichten:1994gt,Godfrey:2004ya,Devlani:2014nda}.
}
\begin{tabular}{ c c c }\hline
$0^-/$GeV & ${1^-}/$GeV & ${1^+}/$GeV \\\hline
6.275 & 6.335 & 6.745\\
6.872 & 6.926 & 6.75\\
7.25 & 7.02 & 7.15\\
& 7.28 & 7.15\\\hline
\end{tabular}
\end{table}

The physical continuum form factors are given by setting $a_n = b_n^{000}$ and $\mathcal{N}_n=1$ in Eq.~(\ref{fitfunctionequation}), to give
\begin{equation}\label{contphysfitfunc}
F^\mathrm{phys}(q^2) = \frac{1}{P(q^2)}\sum_{n=0}^3 a_n z^n,
\end{equation}
where $P(q^2)$ is computed using the physical masses given in Table~\ref{poletab}.
These $F^\mathrm{phys}$ are plotted in Figure~\ref{FFqsq} together with our lattice data. 
The continuum values of the $z$-expansion coefficients $a_n = b_n^{000}$ are given in Table~\ref{zexpcoefficients}. The correlation matrices between these parameters are 
given in Appendix~\ref{anfullcov}.

\begin{table}
\caption{\label{zexpcoefficients} Physical $z$-expansion coefficients for the pseudoscalar, axial-vector and vector form factors for $B_s \rightarrow D_s^*$ decay. The full correlation matrices for these coefficients are given in Appendix~\ref{anfullcov}.}
\begin{tabular}{ c | c c c c }
\hline
& $a_0$	& $a_1$	& $a_2$	& $a_3$	\\\hline
${A0}$&	0.1047(57)&	-0.43(13)&	-0.10(96)&	-0.03(1.00)\\
${A1}$&	0.0552(21)&	-0.010(54)&	-0.03(77)&	0.06(99)\\
${A2}$&	0.059(11)&	-0.11(22)&	-0.25(79)&	-0.05(1.00)\\
${V}$&	0.100(11)&	-0.18(27)&	-0.006(0.998)&	0.0(1.0)\\
\hline
\end{tabular}
\end{table}

In Figure~\ref{zspacepoleremoved} we plot the form factor data, together with the extrapolated physical continuum form factors, multiplied by the pole function Eq.~(\ref{poleformeq}) against $z(q^2,t_0)$. There we see that the fit to the polynomial part of Eq.~(\ref{fitfunctionequation}) is straightforward, with only very simple dependence on $z(q^2,t_0)$ and only mild heavy quark mass dependence. Note that here the $A_1$ form factor data and extrapolated curve are flatter than was found for $B_c\to J/\psi$ in~\cite{Harrison:2020gvo}.

\subsection{Heavy Mass Dependence}
\label{metabdependence}

\begin{figure*}
\centering
\includegraphics[scale=0.225]{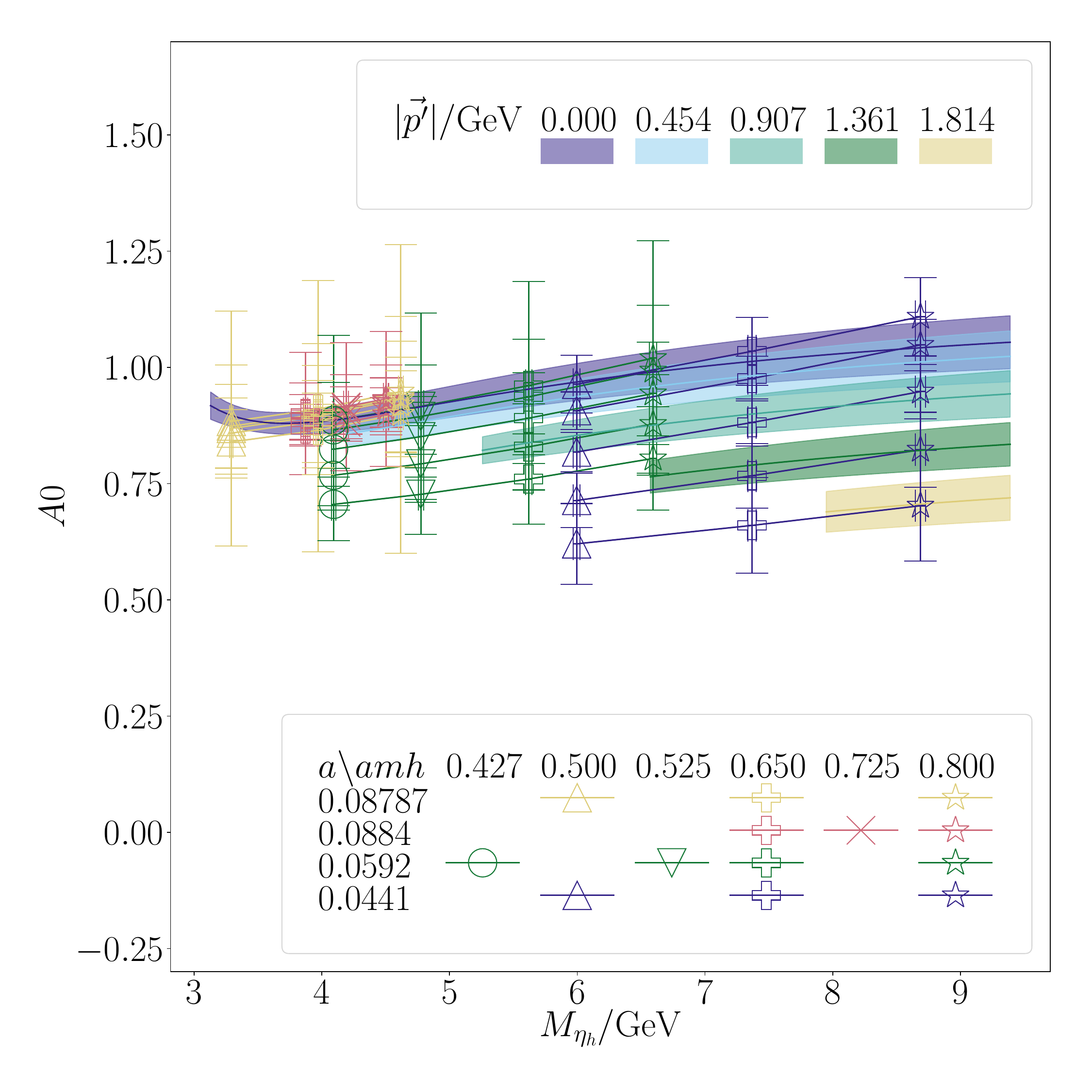}
\includegraphics[scale=0.225]{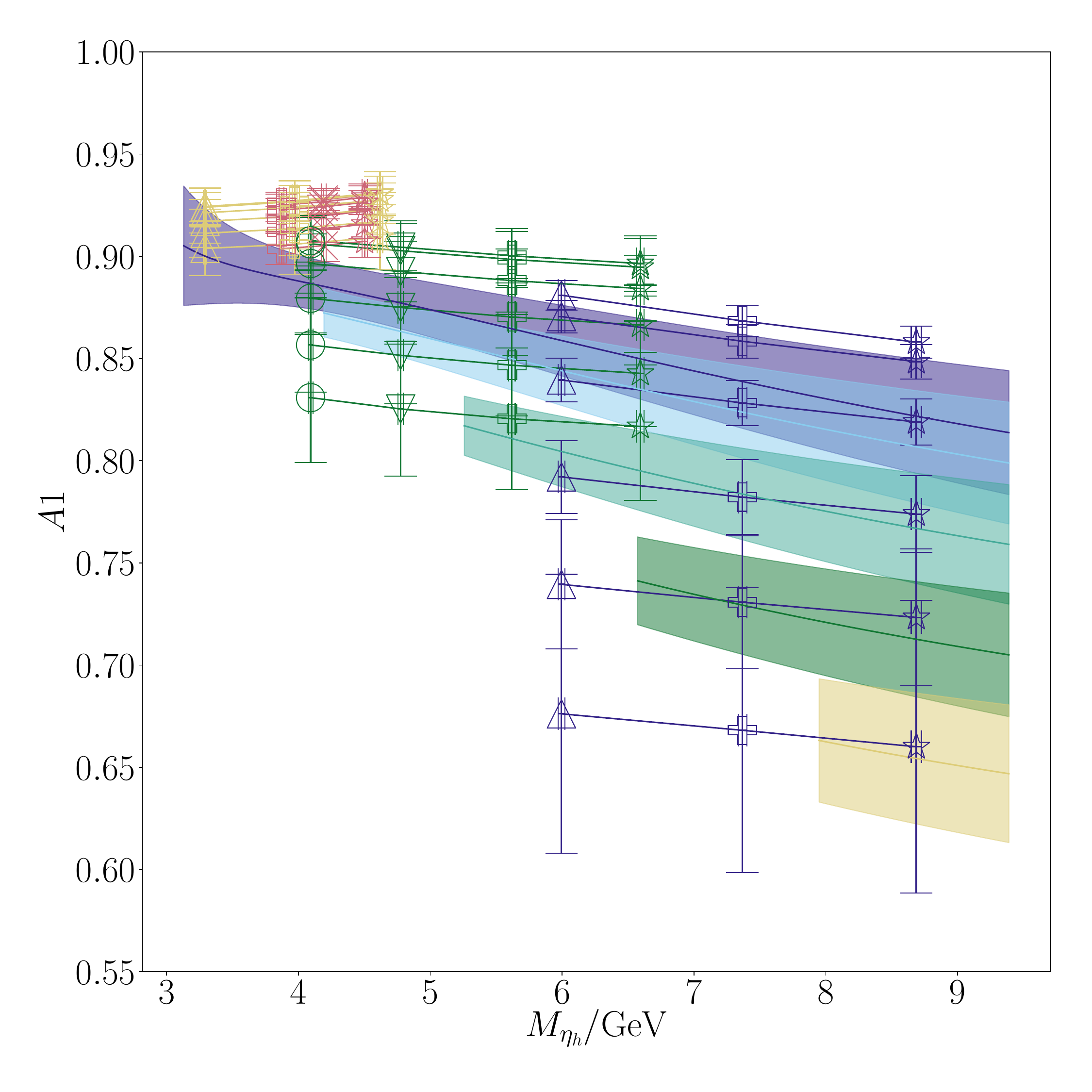}
\includegraphics[scale=0.225]{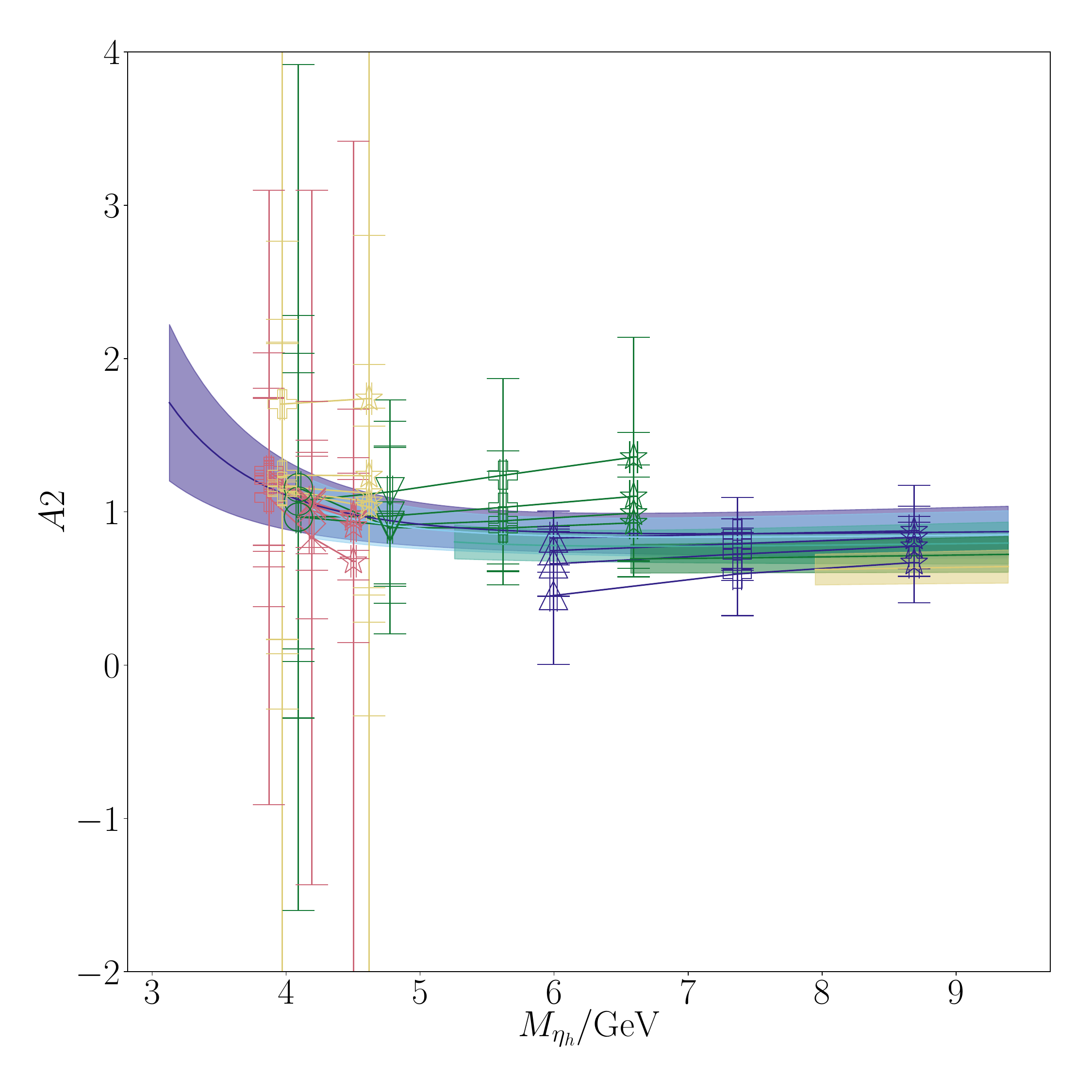}
\includegraphics[scale=0.225]{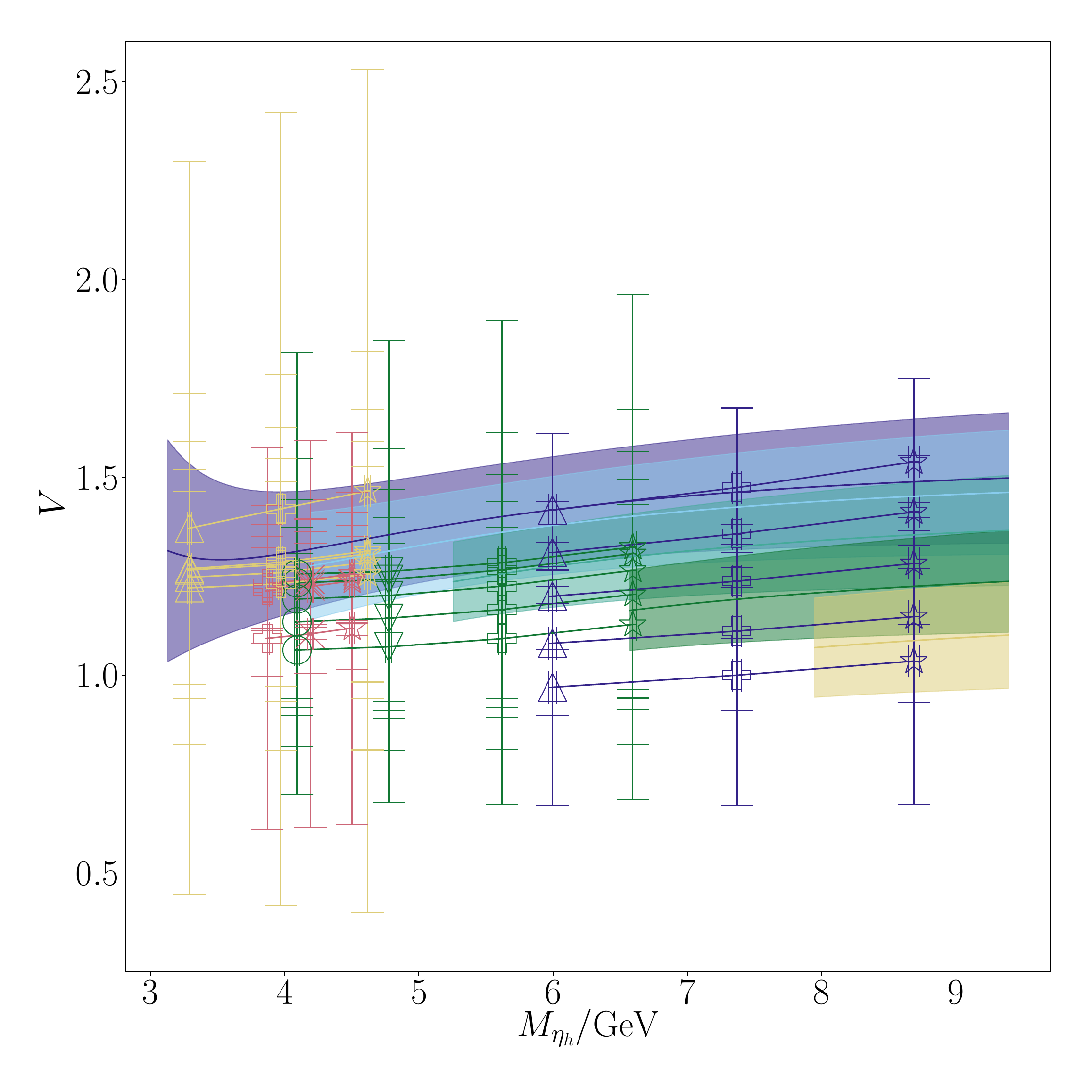}
\caption{\label{etahdepplot} The points show our lattice QCD results for each 
form factor as given in Tables~\ref{set1},~\ref{set2},~\ref{set3} 
and~\ref{set4} as a function of the $\eta_h$ mass $M_{\eta_h}$, with data points corresponding 
to the same $D_s^*$  spatial momentum (given in \cref{set1,set2,set3,set4}) connected. We also use Eq.~(\ref{mhcparam}) 
to plot our continuum result (solid coloured curves) at multiple, evenly spaced, fixed values of $D_s^*$ momentum within the semileptonic region $0\le q^2 \le q^2_\mathrm{max}$.
The legend gives the mapping between symbol colour and shape and the 
set of gluon field configurations used, as given by the lattice 
spacing, and the heavy quark in lattice units (see Tables~\ref{lattdets} and~\ref{masses}). Note that for the form factor $A_2$ we 
exclude from the plot the inaccurate lattice data for $am_h=0.5$ on set 4, as well as $ak=0.059$ and $ak=0.052$ on sets 2 and 3 respectively.
}
\end{figure*}

Our fit results allow us to examine the physical dependence of the form factors on the heavy quark mass. 
This allows us to check that the dependence is relatively benign and that our fit form Eq.~(\ref{fitfunctionequation}) effectively captures this dependence. 

Our fit function includes heavy mass dependence in several places. There are $\Lambda_\mathrm{QCD}/M_{\eta_h}$ terms in the $z$-expansion coefficients; and the $H_s$ mass enters through our choice of 
$t_0=q^2_\mathrm{max}=(M_{H_s}-M_{D_s^*})^2$ in the $q^2$ to $z$ mapping and the pole masses entering $P(q^2)$ depend on $M_{H_s}$. We therefore need to know how $M_{H_s}$ varies as a function of $M_{\eta_h}$. 
We fit our lattice data for the $\gamma_5 \otimes \gamma_5 $ $H_s$ 
against a simple function of the $\eta_h$ mass. We use the function
\begin{align}\label{mhsfitfunc}
M_{H_s}=&(M_{\eta_h}-M_{\eta_c}^\mathrm{phys})/2 +M_{D_s}^\mathrm{phys}  +\sum_{i=1}^4X_i\left(\frac{am_h}{\pi}\right)^{2i}\nonumber\\
&+ \sum_{i=1}^4Y_i\left(\frac{am_c}{\pi}\right)^{2i}+\sum_{i=1}^4Z_i\Delta^{(i)}_{hc}+\mathcal{N'}
\end{align}
where 
\begin{equation}
\Delta^{(i)}_{hc}=\left(\frac{2\Lambda_\mathrm{QCD}}{M^\mathrm{phys}_{\eta_c}}\right)^i-\left(\frac{2\Lambda_\mathrm{QCD}}{M_{\eta_h}}\right)^i
\end{equation}
and 
\begin{equation}
\mathcal{N}'_n = 1 + A'_n \delta_{m_c}^\mathrm{val}+ B'_n \delta_{m_c}^\mathrm{sea}+ C'_n \delta_{m_s}^\mathrm{val}+ D'_n \delta_{m_s}^\mathrm{sea} + E'_n \delta_{m_l}^\mathrm{sea}
\end{equation}
This form ensures the correct value of $M_{H_s}$ as $m_h\rightarrow m_c$. We take $M^\mathrm{phys}_{\eta_c}=2.9839\mathrm{GeV}$ from \cite{pdg20}, neglecting its very small uncertainty, 
and we also include the physical values of $M_{\eta_b}$ and $M_{B_s}$~from \cite{pdg20} in the fit as data points.
We take prior widths of $0(1)$ for $A'$, $B'$, $C'$, $D'$, $E'$, $X_i$, $Y_i$ and $Z_i$.
This gives a sensible fit with $\chi^2/\mathrm{dof}=1.3$ and $Q=0.2$.
We then use our fitted parameters $Z_i$ to estimate the continuum value of $M_{H_s}$ at a given $M_{\eta_h}$. Setting $\mathcal{N}'$, $X_i$, and $Y_i$ to zero in Eq.~(\ref{mhsfitfunc}) gives
\begin{equation}\label{mhcparam}
M_{H_s}=(M_{\eta_h}-M^\mathrm{phys}_{\eta_c})/2+M_{D_s}^\mathrm{phys}+\sum_{i=1}^4Z_i\Delta^{(i)}_{hc}.
\end{equation}
Note that this parameterisation of the $H_s$ mass is only used to demonstrate the heavy mass 
dependence of the form factors and will not have any impact on the physical $B_s\to D_s^*$ form factors.

In Figure~\ref{etahdepplot} we plot the form factors at fixed values of the $D_s^*$ momentum against $M_{\eta_h}$. We choose values of the $D_s^*$ momentum which evenly span the 
semileptonic range at the physical $b$ quark mass and only plot the mass region for which the 
resulting value of $q^2$ is between 0 and $q^2_\mathrm{max}$. We include in these plots our lattice 
data, connecting points on a given set which are at the same $D_s^*$ spatial momentum.
As for $B_c\to J/\psi$~\cite{Harrison:2020gvo}, we see that the continuum form factors have only mild 
heavy mass dependence across the range of masses we use here, and that our extrapolation to the $b$ mass using these points is reliable.
This is consistent with what is seen for other $b\to c$ form factors, e.g.~\cite{EuanBsDsstar,EuanBsDs,Cooper:2020wnj,Harrison:2020gvo}.

\subsection{Tests of the Stability of the Analysis}
\label{stabsec}

\begin{figure}
\centering
\includegraphics[scale=0.4]{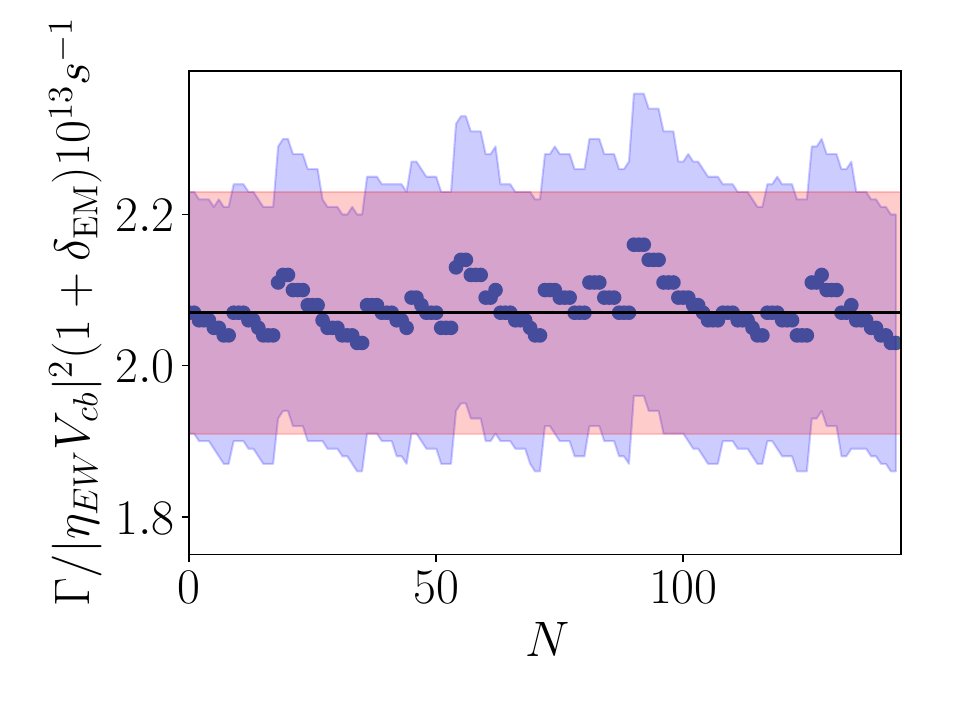}
\caption{\label{gammastabplot} Plot showing the stability of the total rate 
for ${B}_s^0\rightarrow D_s^{*-}\mu^+{\nu}_\mu$ under variations of the 
correlator fits. The $x$ axis value corresponds to 
$N = \delta_3 + 4\delta_2+16\delta_1+64\delta_4$ where $\delta_n$ is the 
value of $\delta$ corresponding to the fit given in Table~\ref{fitparams} 
for set $n$. The black horizontal line and red error band correspond to our 
final result and the blue points and blue error band correspond to the 
combination of fit variations associated to $N$. Our result for the total rate does not change significantly for these variations in the fits. Note that here we do not include the contribution of $\delta_\mathrm{EM}$ to the uncertainty. }
\end{figure}

\begin{figure}
\centering
\includegraphics[scale=0.4]{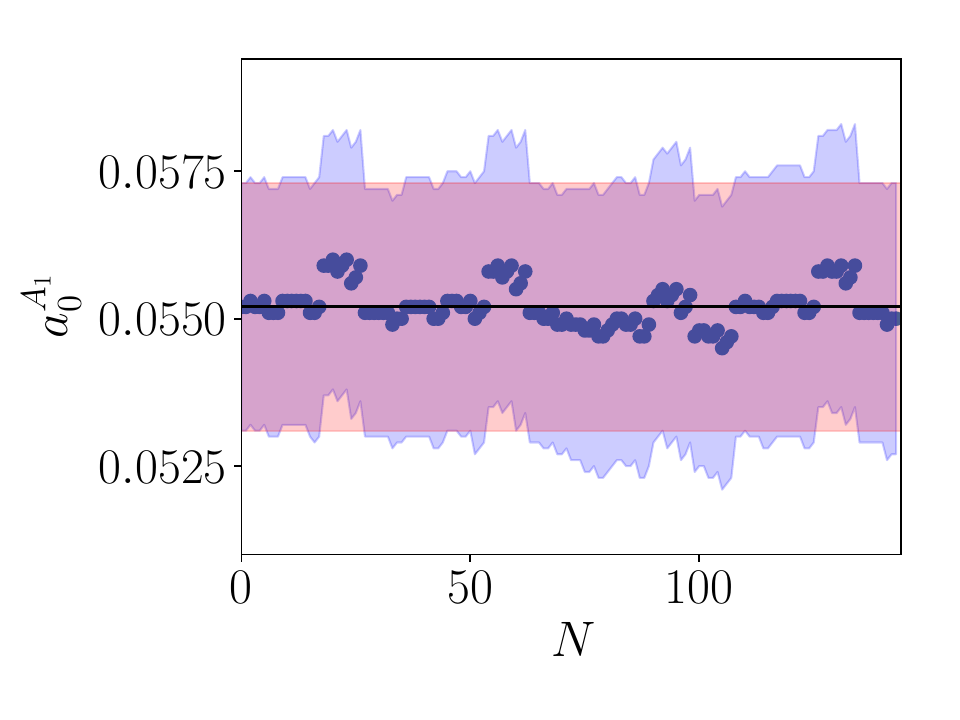}
\includegraphics[scale=0.4]{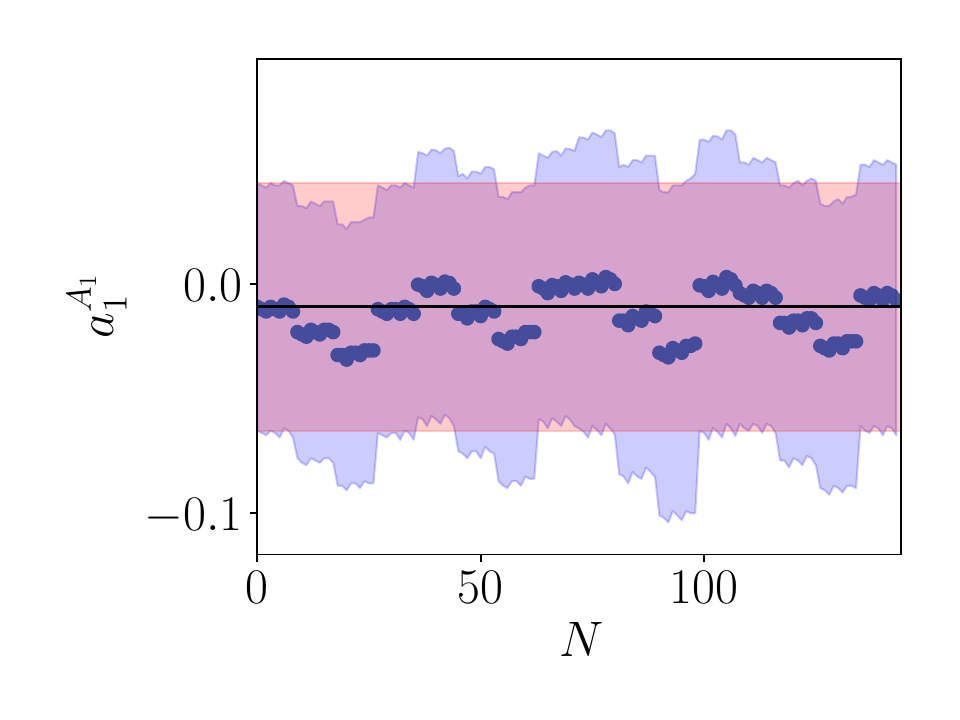}
\includegraphics[scale=0.4]{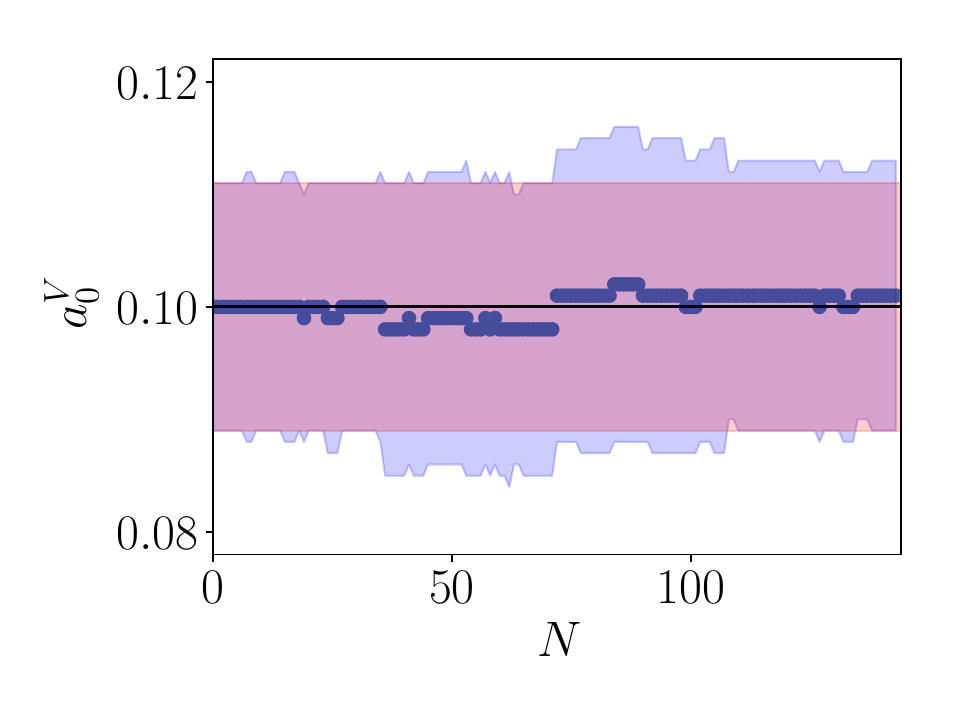}
\caption{\label{coeffstabplot} 
As for Figure~\ref{gammastabplot} showing the stability of the 
coefficients of the $z$-expansion for the form factors 
under variations of the correlator 
fits.  We include a subset of coefficients here; other plots look 
very similar. 
}
\end{figure}

\begin{figure}
\centering
\includegraphics[scale=0.4]{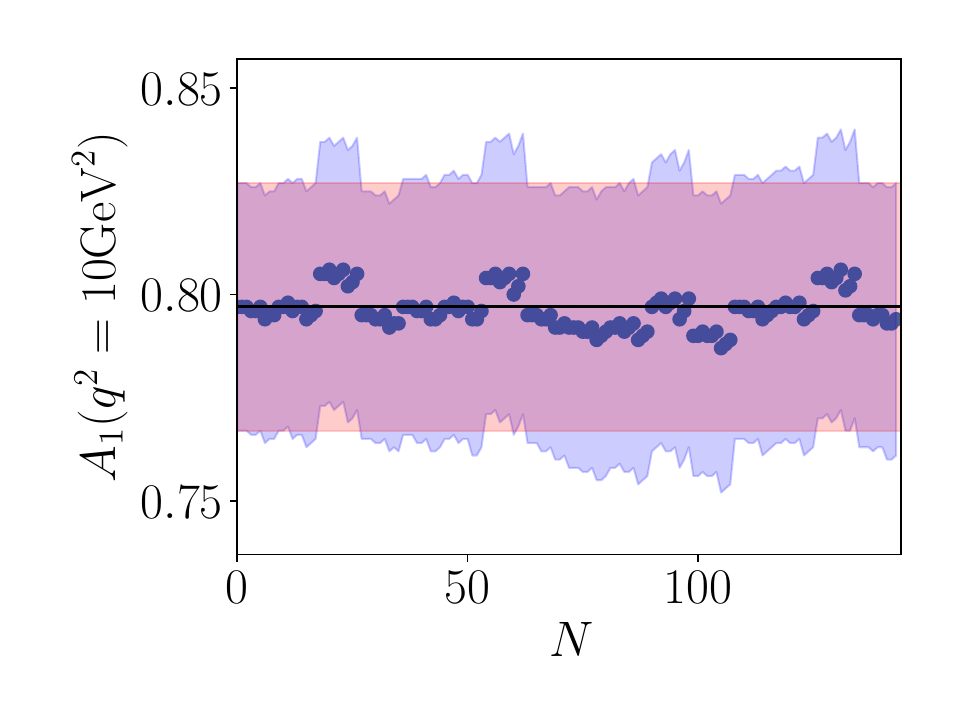}
\includegraphics[scale=0.4]{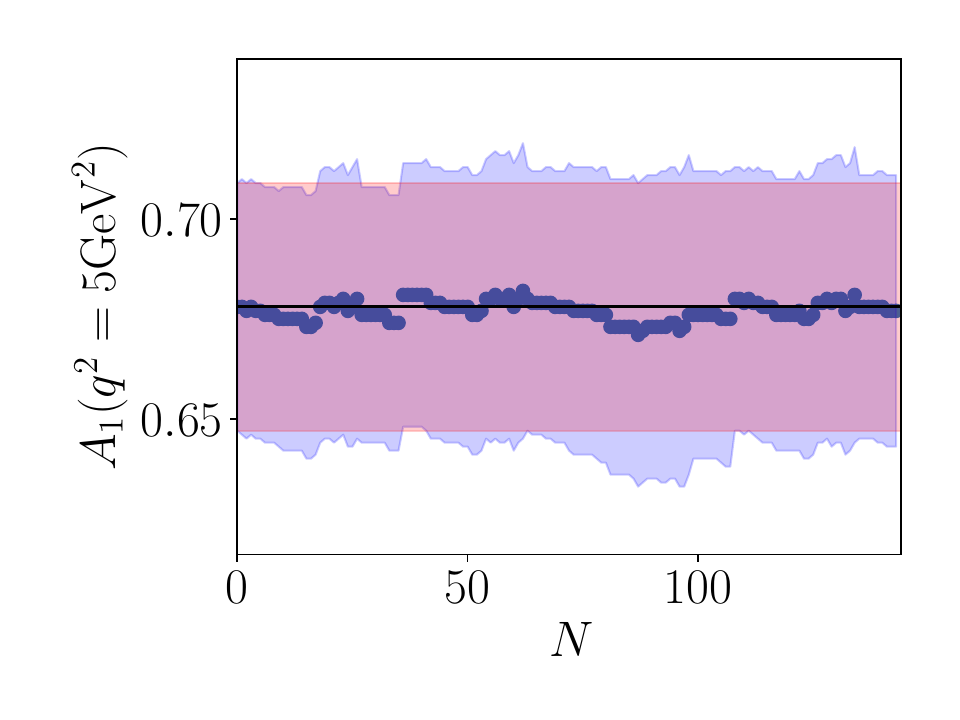}
\includegraphics[scale=0.4]{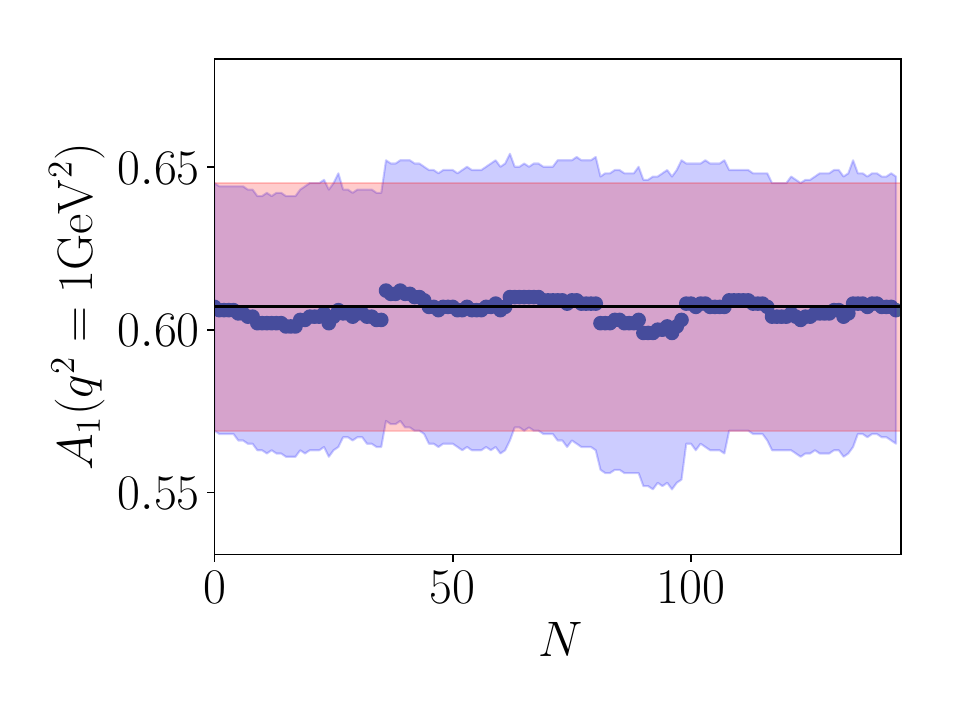}
\caption{\label{A1qsqpoints} 
As for Figure~\ref{gammastabplot} showing the stability of the 
form factor $A_1$ evaluated at $q^2=1~\mathrm{GeV}^2,5~\mathrm{GeV}^2,10~\mathrm{GeV}^2$.}
\end{figure}

\begin{figure}
\centering
\includegraphics[scale=0.4]{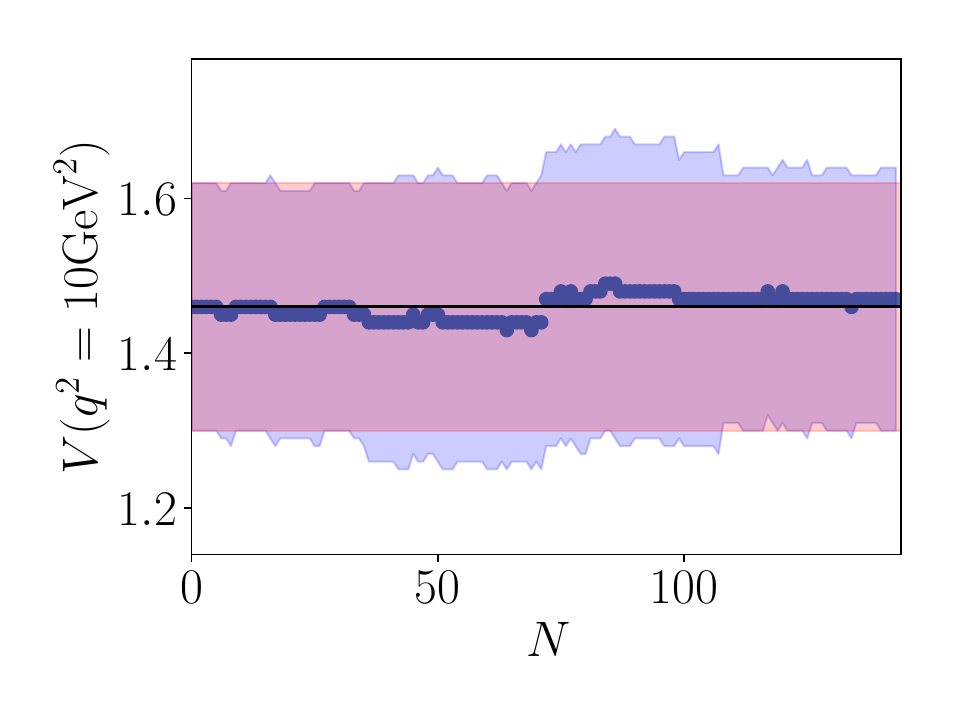}
\includegraphics[scale=0.4]{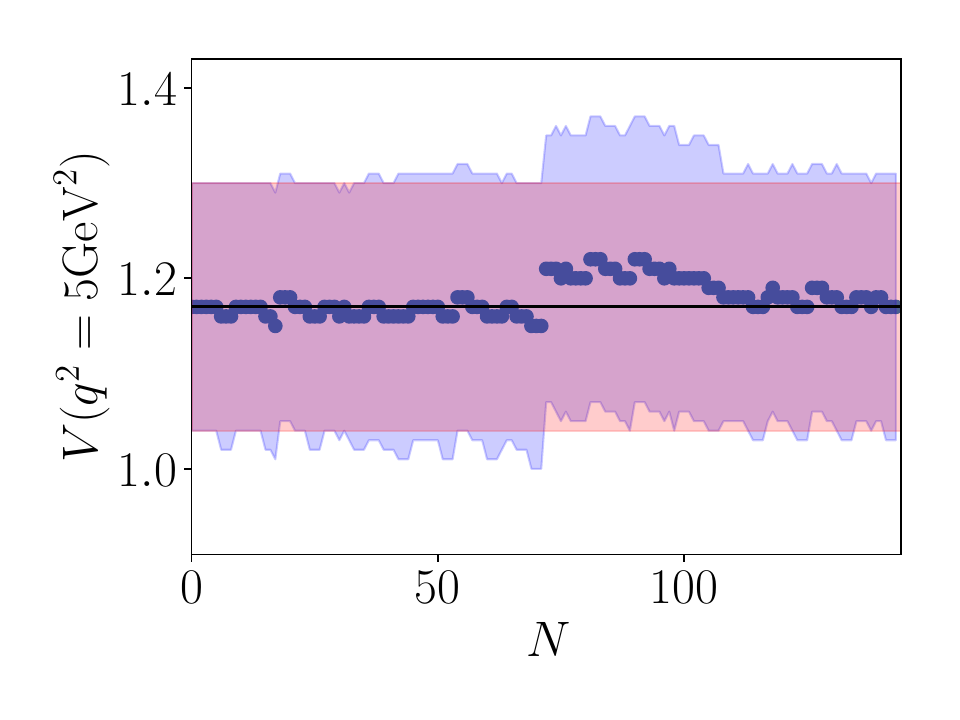}
\includegraphics[scale=0.4]{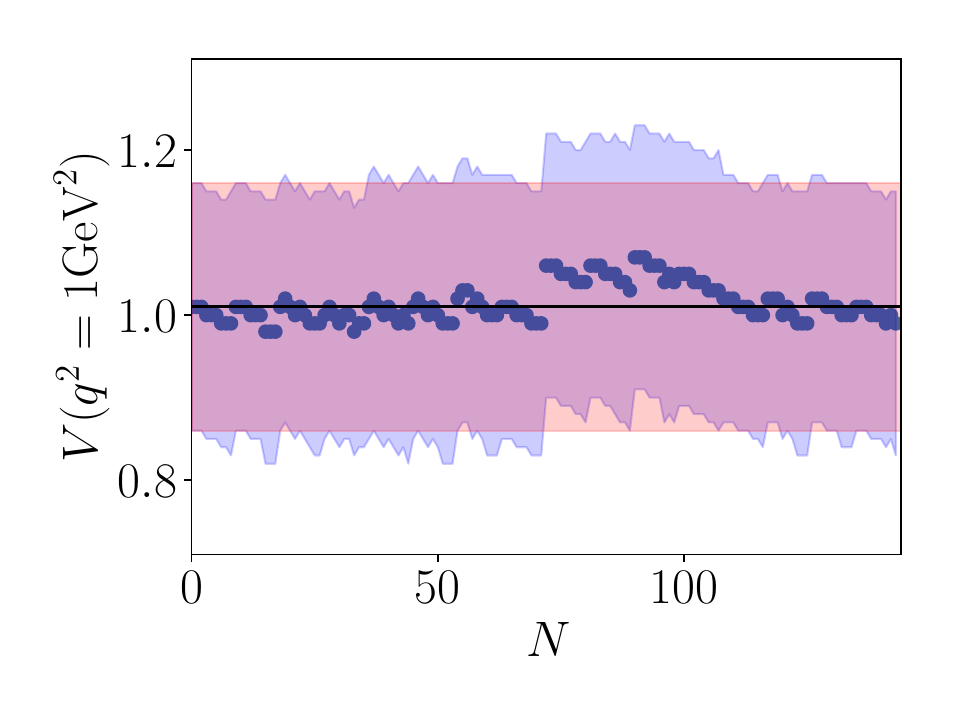}
\caption{\label{Vqsqpoints} 
As for Figure~\ref{gammastabplot} showing the stability of the 
form factor $V$ evaluated at $q^2=1~\mathrm{GeV}^2,5~\mathrm{GeV}^2,10~\mathrm{GeV}^2$.}
\end{figure}

Here we demonstrate the stability of our analysis to the different choices of correlator fit inputs given in Table~\ref{fitparams}. 
We show that under these variations the total rate of 
${B}_s^0\rightarrow D_s^{*-}\ell^+{\nu}_\ell$ 
decay, i.e. \
$\Gamma\left({B}_s^0\rightarrow D_s^{*-}\ell^+{\nu}_\ell\right)/|\eta_{\mathrm{EW}}V_{cb}|^2(1+\delta_\mathrm{EM})$, is stable.
This quantity is obtained by first determining the helicity amplitudes from our form 
factors and then integrating in $q^2$ over the differential rate 
they give (see Eqs~(\ref{helicityamplitudes}) and~(\ref{dgammadq2})). 
The results for the differential rates and total rate will be discussed 
in more detail in Section~\ref{sec:discussion}; 
here we focus on the stability of the final result under variations of fit 
choice. 

We first look at the choices of the correlator fit parameters:  
$\Delta T_\mathrm{3pt}$, $\Delta T_\mathrm{2pt}^{D_s^*}$, 
$\Delta T^{H_s}_\mathrm{2pt}$, the value of SVD cut and the number of 
exponentials used in the fit. In order to verify that our results 
are independent of such choices we repeat the full analysis using 
all combinations of the variations listed in Table~\ref{fitparams}. 
The total rate computed using each of these fit variations is 
plotted in Figure~\ref{gammastabplot}, where we see that our 
final result is not sensitive to such variations. 

We also look at the dependence of the physical continuum $z$-expansion coefficients, $a_n=b_n^{000}$, on these variations (see Eq.~(\ref{fitfunctionequation})). In Figure~\ref{coeffstabplot} we show the variation of the fitted $a_0$ coefficient for the form factors $V$ and $A_1$, as well as the $a_1$ term for $A_1$. In these plots we see that these coefficients are very stable to variations of correlator fit inputs.

 We also show that the form factors themselves are stable under these variations. In Figure~\ref{A1qsqpoints} and Figure~\ref{Vqsqpoints} we show the variation of the form factors $V$ and $A_1$, evaluated at $q^2=1~\mathrm{GeV}^2,5~\mathrm{GeV}^2,10~\mathrm{GeV}^2$, where we see that $A_1$ and $V$ are very stable to variations of correlator fit inputs. Similar plots for $A_0$ and $A_2$ are given in Appendix~\ref{qsqstabappendix}.

We also convert our results to the Boyd, Grinstein and Lebed (BGL) scheme~\cite{Boyd:1997kz} and check the unitarity constraints. This analysis is given in Appendix~\ref{fullBGL} where we see that these constraints are far from saturation. In~\cite{Harrison:2020gvo} we also studied the effect of including fewer resonances in the pole term, Eq.~(\ref{poleformeq}). Here, in addition to this analysis, we also investigate using alternative parameterisations when performing the heavy-HISQ fit. We show the results of these fits in appendix~\ref{variations}, where we see no significant variation in the physical form factors or differential decay rate.

\section{Discussion of Lattice Results}
\label{sec:discussion}
\subsection{Differential and Total Rates for Each Lepton Flavor}
\begin{figure}
\centering
\includegraphics[scale=0.225]{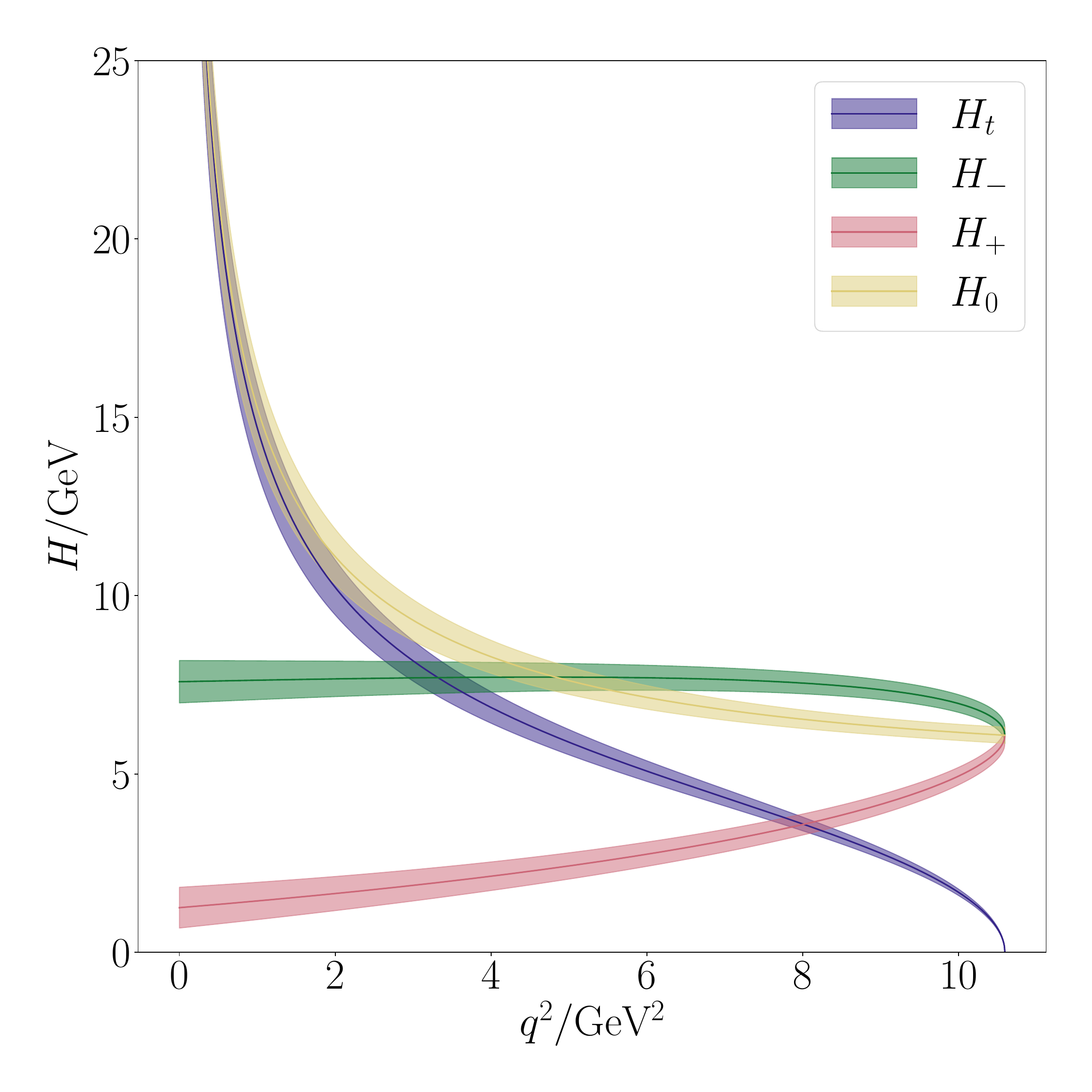}
\caption{\label{helicityplot} 
Helicity amplitudes for $B_s\to D_s^{*}$ plotted as a function of $q^2$.}
\end{figure}

\begin{figure}
\centering
\includegraphics[scale=0.225]{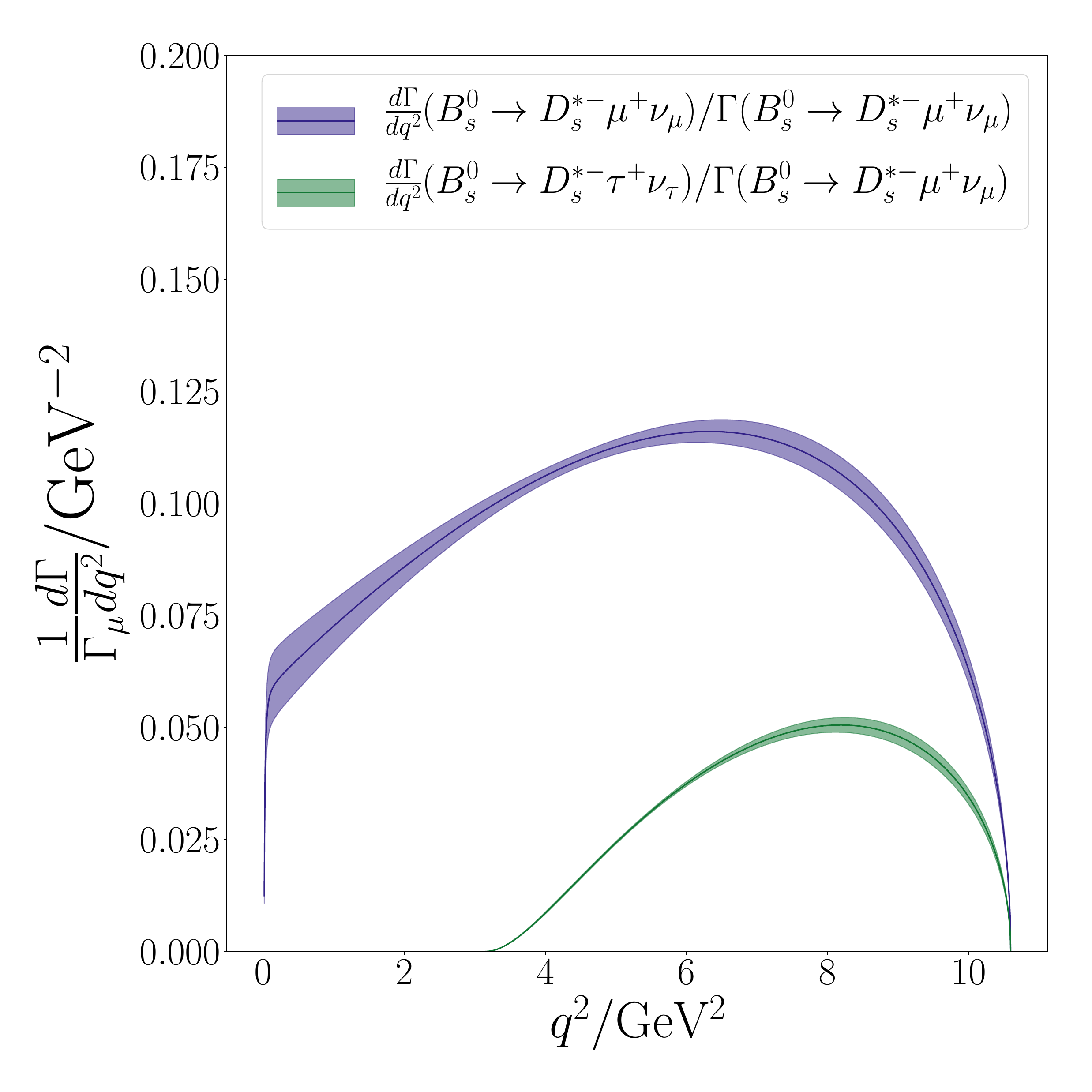}
\caption{\label{dgammadq2plot} 
The differential rate $d\Gamma/dq^2$ for $B_s^0\to D_s^{*-}\ell^+{\nu}_\ell$ for $\ell=\mu$ and $\ell=\tau$ as a function of $q^2$, normalised by the total decay rate for the $\ell=\mu$ 
case. Note that here, for the $\ell=\tau$ curve, the error bands do not include the contribution from $\delta_\mathrm{EM}$.}
\end{figure}
\begin{figure}
\centering
\includegraphics[scale=0.18]{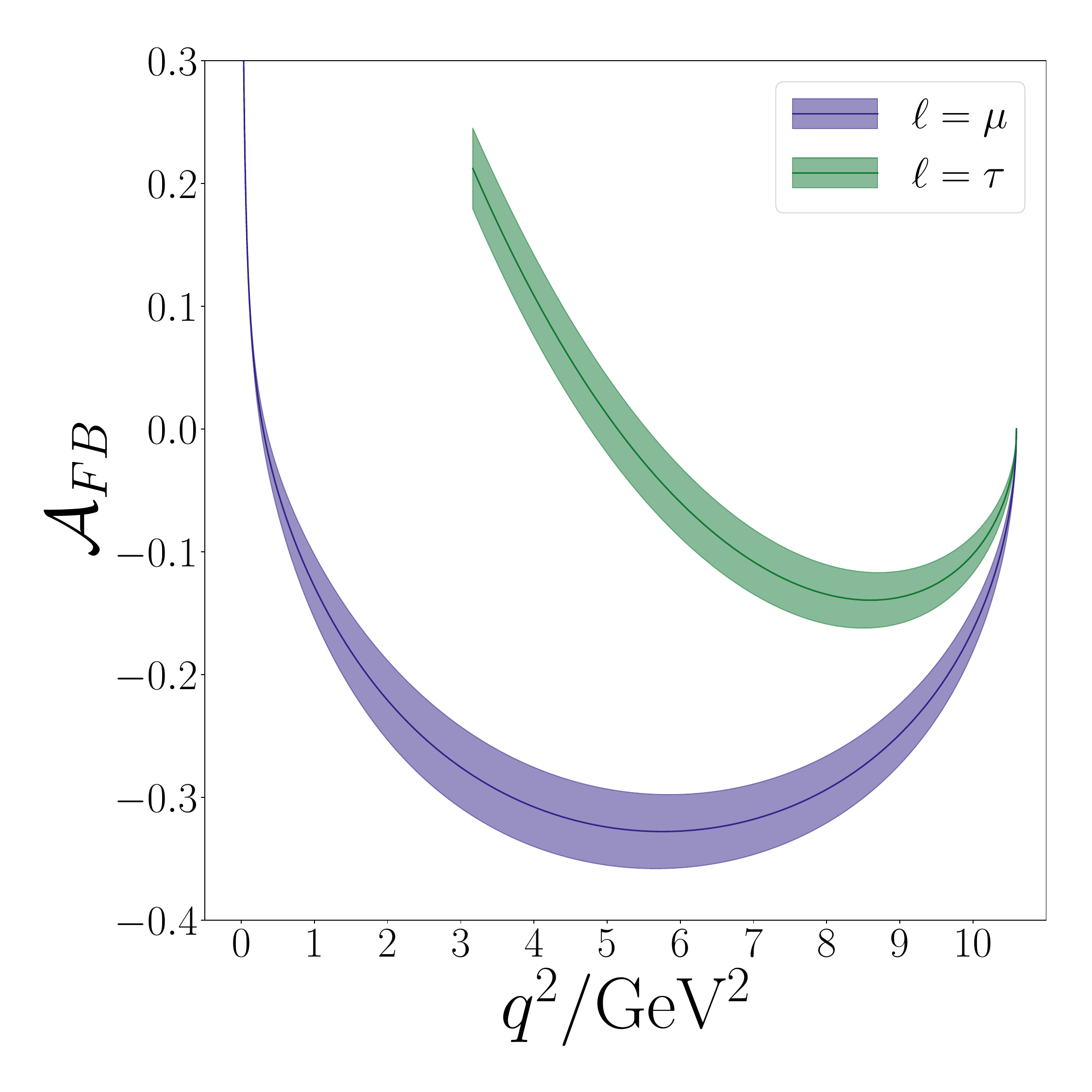}
\includegraphics[scale=0.18]{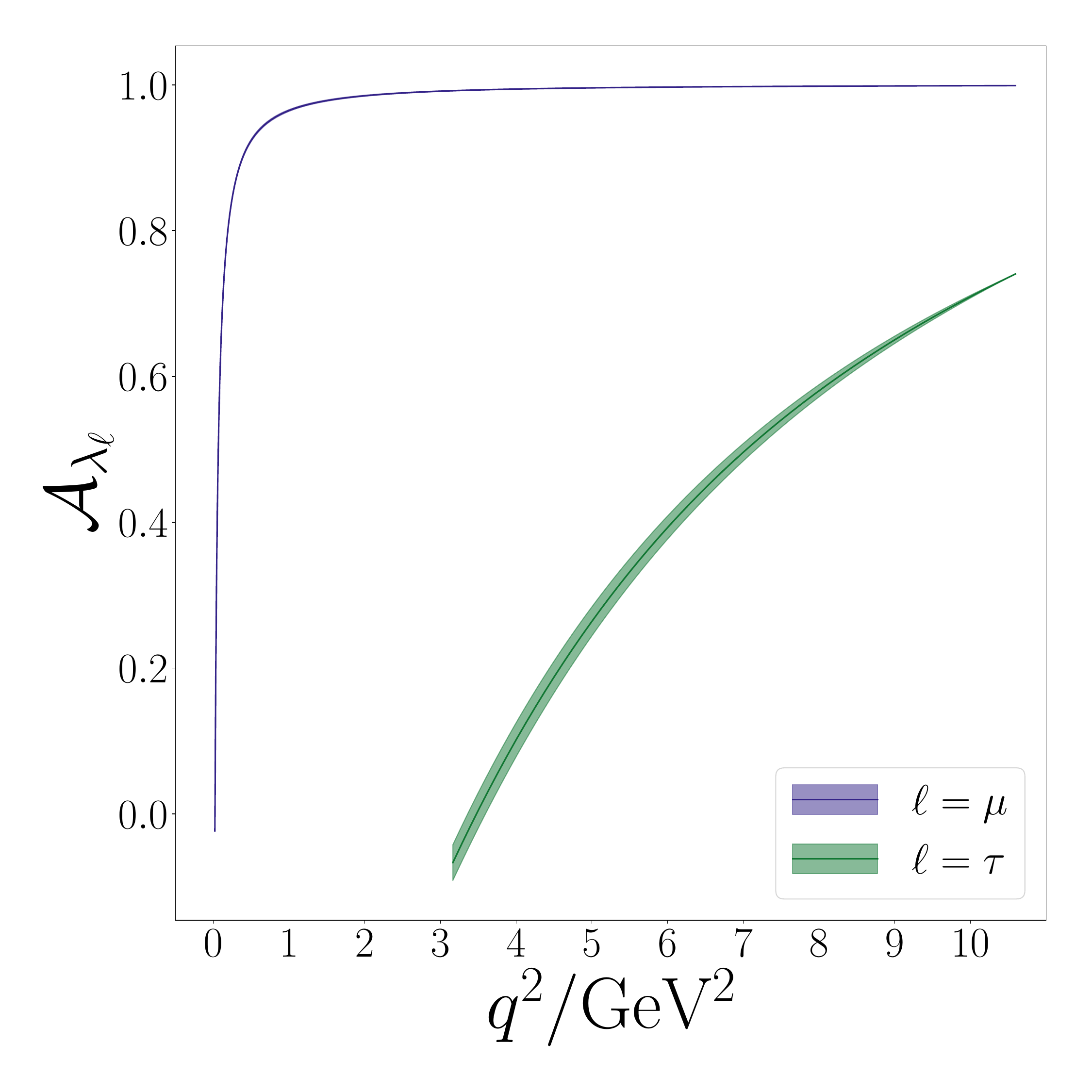}
\includegraphics[scale=0.18]{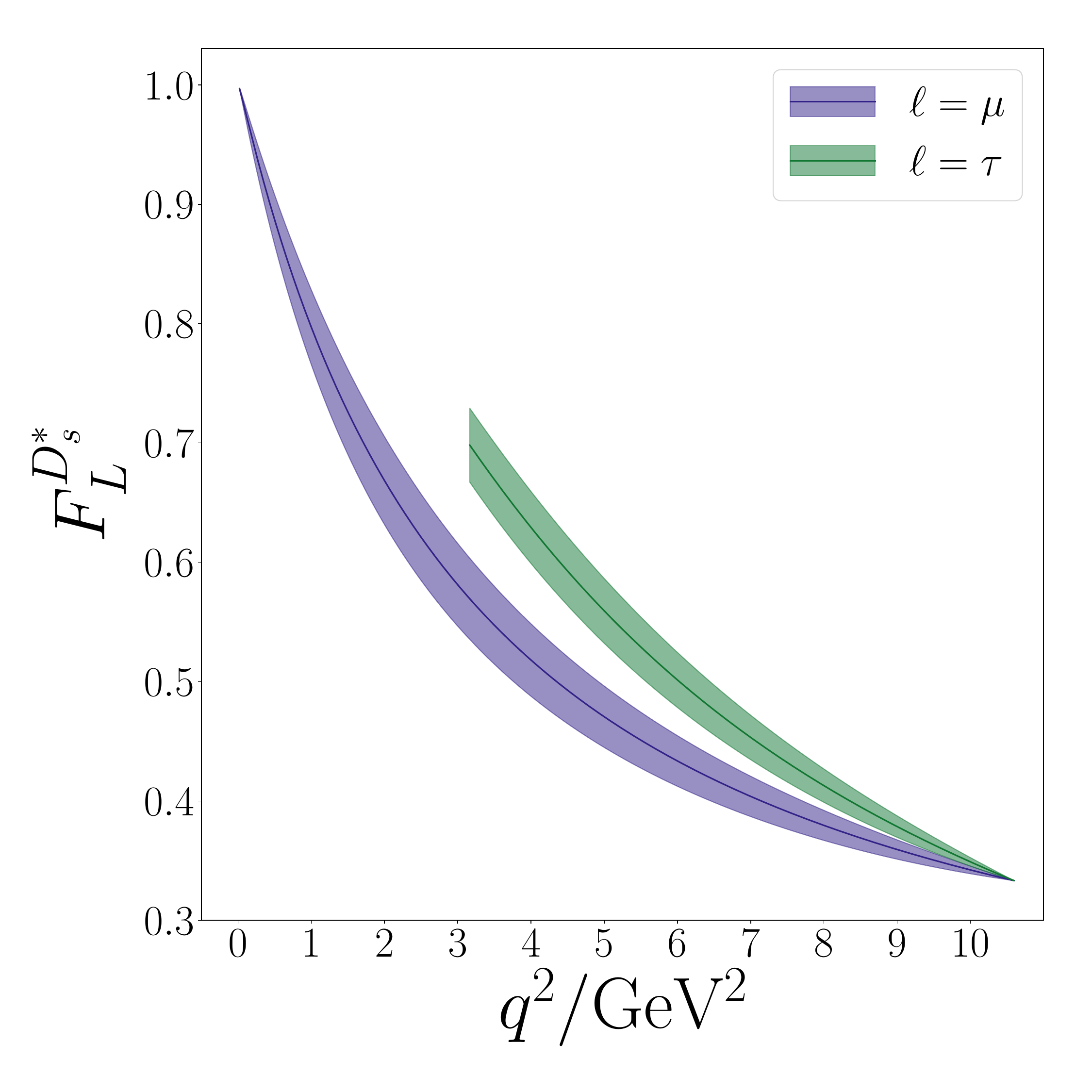}
\caption{\label{angasym} Angular asymmetry variables for $\bar{B}_s^0\to D_s^{*+}\ell^-\bar{\nu}_\ell$ decay defined in Eq.~(\ref{angasymeq}) for the cases $\ell=\mu$ and $\ell=\tau$.}
\end{figure}

In this section we first use our form factors to construct the helicity amplitudes defined in Eq.~(\ref{helicityamplitudes}). These are plotted in Figure~\ref{helicityplot} where we see that $H_0(q^2_\mathrm{max})=H_\pm(q^2_\mathrm{max})$ and that $H_0$ and $H_t$ are singular at $q^2=0$ as we would expect from the factors of $1/\sqrt{q^2}$ appearing in their definitions. This singular behaviour is cancelled in the physical differential decay rate by the factor of $(q^2-m_\ell^2)^2$ appearing in Eq.~(\ref{normfacdiff}).

From these helicity amplitudes we compute the differential rate with respect to $q^2$, given by Eq.~(\ref{dgammadq2}). This is plotted in Figure~\ref{dgammadq2plot} for 
the $\ell=\mu$ and $\ell=\tau$ cases, where in the plot we normalise both curves by the total rate $\Gamma$ for the $\ell=\mu$ case. We integrate these, as well as the rate for $\ell=e$, to find the total rate to each lepton flavor. When integrating our results we use a simple (but accurate) trapezoidal approximation in order to ensure we carry through correlations correctly. We find

\begin{align}\label{gammaelectoteq}
\frac{\Gamma(B_s^0\to D_s^{*-}e^+{\nu}_e)}{|\eta_\mathrm{EW}V_{cb}|^2}=&~\GAMMAeTOTAL\nonumber\\
=&~\GAMMAeGEV
\end{align}
and
\begin{align}\label{gammamutoteq}
\frac{\Gamma(B_s^0\to D_s^{*-}\mu^+{\nu}_\mu)}{|\eta_\mathrm{EW}V_{cb}|^2}=&~\GAMMAmuTOTAL\nonumber\\
=&~\GAMMAmuGEV
\end{align}
with the ratio $\Gamma_{\ell=e}/\Gamma_{\ell=\mu}=1.00443(16)$, amounting to an effect of 0.4\% in the total rate from the muon mass. Note that we are ignoring differences in $\delta_\mathrm{EM}$ between the two cases in this ratio. For the $\ell=\tau$ case the effect of including the mass is much more substantial, we find
\begin{align}\label{gammatautoteq}
\frac{\Gamma(B_s^0\to D_s^{*-}\tau^+{\nu}_\tau)}{|\eta_\mathrm{EW}V_{cb}|^2}=&~\GAMMAtauTOTAL\nonumber\\
=&~\GAMMAtauGEV.
\end{align}

We can also readily construct $R(D_s^*)$, the ratio of the total rates for the $\ell=\tau$ and $\ell=\mu$ cases, where many uncertainties which are correlated between the two cancel. We find
\begin{equation}\label{rdsstar}
R(D_s^*) = \Gamma_{\ell=\tau}/\Gamma_{\ell=\mu} = \RDsstar.
\end{equation}
This value is $\approx 1.6\sigma$ below both the value of $R(J/\psi)$ computed in~\cite{Harrison:2020nrv} as well as the HFLAV average SM value of $R(D^*)$~\cite{Amhis:2019ckw}. Note that our value is consistent with the value computed in~\cite{Bordone:2019guc} using the Heavy-Quark expansion of $R(D_s^*)=0.2472(77)$. Note also that unlike the total rate $\Gamma$, for which the contribution of $\delta_\mathrm{EM}$ to the uncertainty is relatively small, the lattice uncertainty in $R(D_s^*)$ is the same order of magnitude as the uncertaintly resulting from long-range QED effects, at least for charged final-state mesons. These QED effects are often ignored but must be addressed in future calculations in order to produce reliable SM results with sub-percent level uncertainties.

We also construct the improved ratio~\cite{Isidori:2020eyd}
\begin{equation}
R^\mathrm{imp}(D_s^*)=\frac{\int_{m_\tau^2}^{q^2_\mathrm{max}}dq^2 \frac{d\Gamma}{dq^2}({B}_s^0\rightarrow D_s^{*-}\tau^+{\nu}_\tau)}{\int_{m_\tau^2}^{q^2_\mathrm{max}}dq^2 \frac{d\Gamma}{dq^2}({B}_s^0\rightarrow D_s^{*-}\mu^+{\nu}_\mu)}.
\end{equation}
We find
\begin{equation}
R^\mathrm{imp}(D_s^*)=\RDsstarIMP
\end{equation}
where now the uncertainty resulting from electromagnetic effects is dominant due to the improved cancellation of correlated lattice uncertainties.

We may use our value of $\Gamma/|\eta_\mathrm{EW}V_{cb}|^2$ in Eq.~(\ref{gammamutoteq}), together with values of $V_{cb}$ and $\eta_\mathrm{EW}$, to derive a result for the total width of the decay. We take $\eta_\mathrm{EW}=1.0066$ following~\cite{Aaij:2020xjy} and $|V_{cb}|=41.0(1.4)\times 10^{-3}$ using an average of inclusive and exclusive determinations with the error scaled by 2.4 to allow for their inconsistency~\cite{pdg20}. Note that here we neglect the small uncertainty in $\eta_\mathrm{EW}$. This gives
\begin{equation}
\Gamma(B_s^0\to D_s^{*-}\mu^+{\nu}_\mu) = 3.53(27)_\mathrm{latt}(24)_\mathrm{V_{cb}}(4)_\mathrm{EM} \times 10^{10}s^{-1}
\end{equation}
where the first uncertainty is from our lattice QCD calculation, the second is from the uncertainty in $|V_{cb}|$ and the final error is from $\delta_\mathrm{EM}$. We may combine this with the experimental average of the $B_s^0$ mean life time~\cite{pdg20,Amhis:2019ckw} $\tau(B_s^0) = 1.515(4)\times 10^{-12}s$ to find the branching fraction
\begin{equation}
\mathrm{Br}({B}_s^0\rightarrow D_s^{*-}\mu^+{\nu}_\mu)=0.0534(42)_\mathrm{latt}(36)_{V_{cb}}(1)_{\tau}(5)_\mathrm{EM}
\end{equation}
where the uncertainties are from our lattice calculation, from the uncertainty in $V_{cb}$, from the uncertainty in the $B_s^0$ lifetime and from $\delta_\mathrm{EM}$ respectively. This is in good agreement with, but much more accurate than, the value of the more inclusive branching fraction measured by Belle, $\mathrm{Br}({B}_s^0\rightarrow D_s^{*-} X \ell^+{\nu})=0.054(11)$~\cite{Oswald:2015dma}.

\subsection{Angular and Polarisation Asymmetries and Ratios}
We can also construct the lepton polarisation asymmetry, as well as the longitudinal polarisation fraction and the forward-backward asymmetry. 
Note that as for $B\to D^*$~\cite{Becirevic:2019tpx} and $B_c\to J/\psi$~\cite{Harrison:2020nrv} these are conventionally defined for the charge conjugate mode, which here is $\bar{B}_s^0\to D_s^{*+}\ell^-\bar{\nu}_\ell$. They are given by
\begin{align}
\label{angasymeq}
\mathcal{A}_{\lambda_\ell}(q^2) =& \frac{d\Gamma^{\lambda_\ell=-1/2}/dq^2-d\Gamma^{\lambda_\ell=+1/2}/dq^2}{d\Gamma/dq^2},\nonumber\\
F_{L}^{D_s^*}(q^2) =& \frac{d\Gamma^{\lambda_{D_s^*}=0}/dq^2}{d\Gamma/dq^2},\nonumber\\
\mathcal{A}_{FB}(q^2) =& \frac{1}{d\Gamma/dq^2} \frac{2}{\pi}\int_0^\pi\frac{d\Gamma}{dq^2d\cos(\theta_W)}\cos(\pi-\theta_W)d\theta_W
\end{align}
respectively, where we have chosen the forward direction for the purpose of $\mathcal{A}_{FB}$ as being in the direction of the $D_s^*$ momentum in the $B_s$ rest frame. 
We plot these for the $\ell=\mu$ and $\ell=\tau$ cases in Figure~\ref{angasym}. Following the notation used in~\cite{Harrison:2020nrv} and~\cite{Becirevic:2019tpx} for the integrated observables and lepton flavor universality violating ratios we find for the $\ell=\tau$ case
\begin{align}\label{lfuvexpectations}
\langle \mathcal{A}_{\lambda_\tau} \rangle &= 0.520(12),\nonumber\\
\langle F_{L}^{D_s^*}              \rangle &= 0.440(16),\nonumber\\
\langle \mathcal{A}_{FB}           \rangle &= -0.092(24).
\end{align}
Note that these are consistent with the values given in~\cite{Bordone:2019guc} of $\langle F_{L}^{D_s^*}\rangle=0.471(16)$ and $\langle \mathcal{A}_{\lambda_\tau} \rangle =0.486(23)$. For the ratios of $\ell=\tau$ to $\ell=\mu$ cases we find
\begin{align}\label{lfuvratios}
R( \mathcal{A}_{\lambda_\tau} ) &= 0.524(12),\nonumber\\
R( F_{L}^{D_s^*}              ) &= 0.880(18),\nonumber\\
R( \mathcal{A}_{FB}           ) &= 0.345(56).
\end{align}

\subsection{Ratio of $B_s \to D_s$ and $B_s \to D_s^*$ Rates}
We may also use our results in combination with the results of~\cite{EuanBsDs} to compute the ratio $\Gamma(B^0_s\to D_s^-\mu^+\nu_\mu)/\Gamma(B_s^0\to D_s^{*-}\mu^+\nu_\mu)$. In doing so, we neglect correlations between the two calculations. Note that here, as for $R(D_s^*)$, the $|\eta_{EW}V_{cb}|^2$ factors cancel in the ratio. The uncertainty in $\Gamma(B_s\to D_s)$ computed using the results in~\cite{EuanBsDs}, $\Gamma(B_s\to D_s)/|\eta_{EW}V_{cb}|^2=6.02(24)\times 10^{-12}\mathrm{GeV}$, is a factor of $\approx 2$ smaller than the uncertainty in $\Gamma(B_s\to D_s^*)$, and the correlations between the two results are expected to be relatively small. Neglecting these correlations is therefore not expected to significantly affect the uncertainty in the ratio. We find
\begin{equation}\label{bsdsbsdss}
\frac{\Gamma(B^0_s\to D_s^-\mu^+\nu_\mu)}{\Gamma(B_s^0\to D_s^{*-}\mu^+\nu_\mu)}=0.443(40)_\mathrm{latt}(4)_\mathrm{EM}.
\end{equation}
This is in good agreement with the experimental value of $\Gamma(B^0_s\to D_s^-\mu^+\nu_\mu)/\Gamma(B_s^0\to D_s^{*-}\mu^+\nu_\mu)=0.464(45)$ recently measured by LHCb~\cite{Aaij:2020hsi} and has a comparable uncertainty.

\subsection{Error Budget}
\label{sec:errorbudget}

We use the \textbf{lsqfit}~\cite{lsqfit} inbuilt error budget function, which computes the partial variance of our result with respect to priors and data, 
to estimate the contributions of systematic uncertainties (see e.g.~\cite{Bouchard:2014ypa} Appendix A). 
The error budget for the total rate $\Gamma$, for both the $\ell=\mu$ and $\ell=\tau$ cases, excluding the contribution from $\delta_\mathrm{EM}$, is given in Table~\ref{errorbudget} together with the budget for $R(D_s^*)$. We see here, 
as we would expect from~\cite{Harrison:2020nrv}, that the largest uncertainties originate from the statistics on Set 3, from taking $am_h\to 0$ and from taking $M_{\eta_h}\to M_{\eta_b}$.

Note that these uncertainties may also be straightforwardly and systematically 
improved, as discussed in~\cite{Harrison:2020nrv}. 
The key improvements would be to include more gluon field configurations 
to reduce the statistical errors on the current finest Set 3. Adding in 
results from `exafine' lattices with $a\approx 0.03~\mathrm{fm}$ is also feasible. 
These  would allow for calculations directly at the physical $b$ quark mass 
with $am_h\approx 0.6$~\cite{Hatton:2021dvg}, reducing the uncertainties associated with taking $am_h\to 0$ and $M_{\eta_h}\to M_{\eta_b}$ significantly. 

\begin{table}
\centering
\caption{\label{errorbudget} Error budget for the total rate $\Gamma$ for the cases $\ell=\mu$ and $\ell=\tau$, given in Eq.~(\ref{gammamutoteq}) and Eq.~(\ref{gammatautoteq}) respectively, as well as for $R(D_s^*)$ given in Eq.~(\ref{rdsstar}), excluding the contribution from $\delta_\mathrm{EM}$. Errors are given as a percentage of the final answer. The top half gives the contributions of systematic uncertainties originating from the dependence of the form factors on $M_{\eta_h}$, from discretisation effects going as $am_h$ and $am_c$, from sea and valence quark mass mistunings and from uncertainties in the determination of the lattice spacing. The second half of the table gives the contributions of the statistical uncertainty in our lattice correlator data, broken down by set. Finally, `Other Priors' includes all of the remaining sources of uncertainty, such as $\Delta_\mathrm{kin}$ and the current renormalisation factors. `Other Priors' also includes the uncertainty of mixed terms in the fit which cannot be attributed uniquely to any of the categories in the first half of the table (e.g. from the prior uncertainty of $b_n^{011}$, the coefficient in Eq.~(\ref{fitfunctionequation}) which mixes $am_h$ and $am_c$ dependence).}
\begin{tabular}{ c | c c | c }
\hline
& $\Gamma/|\eta_\mathrm{EW}V_{cb}|^2(1+\delta_\mathrm{EM})$ & &  \\\hline
 Source  & ${\ell=\mu}$&${\ell=\tau}$ & $R(D_s^*)$\\\hline
$M_{\eta_h}\rightarrow M_{\eta_b}$ & 2.62 & 2.26 & 1.12 \\ 
$am_c\rightarrow 0$ & 2.0 & 1.8 & 0.3 \\ 
$am_h\rightarrow 0$ & 3.7 & 3.6 & 0.52 \\ 
$\delta_{m_c}^\mathrm{val}$ & 0.20 & 0.18 & 0.058 \\ 
$\delta_{m_c}^\mathrm{sea}$ & 2.2 & 2.3 & 0.087 \\ 
$\delta_{m_s}^\mathrm{val}$ & 0.02 & 0.02 & 0.01 \\ 
$\delta_{m_s}^\mathrm{sea}$ & 1.1 & 1.1 & 0.03 \\ 
$\delta_{m_l}^\mathrm{sea}$ & 0.75 & 0.71 & 0.10 \\ 
$w_0/a$, $w_0$ & 0.46 & 0.54 & 0.17 \\ 
 \hline
Statistics & \\\hline
Set 1 & 1.3 & 1.0 & 0.51 \\ 
Set 2 & 2.5 & 2.2 & 0.76 \\ 
Set 3 & 4.3 & 3.6 & 1.6 \\ 
Set 4 & 0.69 & 0.51 & 0.26 \\ 
 \hline
Other Priors & 2.3 & 2.1 & 0.88\\ 
 \hline
Total & 8.0 & 7.2 & 2.4 \\ 
 \hline
\end{tabular}
\end{table}

\section{Comparison to LHCb Results for the Differential Decay Rate}
\label{sec:lhcbcomp}

\subsection{The Shape of the differential Decay rate}
\begin{figure}
\centering
\includegraphics[scale=0.215]{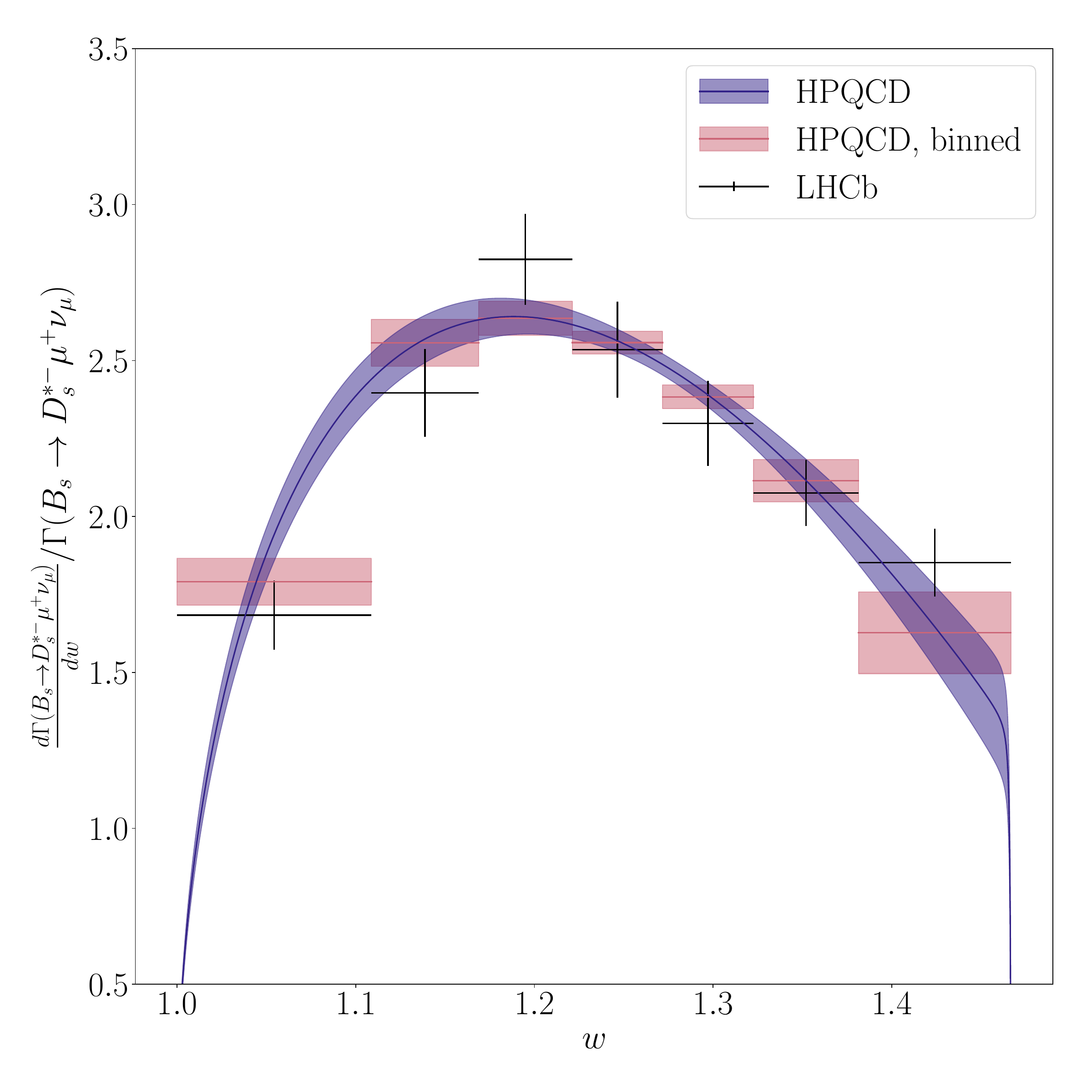}
\caption{\label{lhcbshapeplot} 
The differential rate $d\Gamma/dw$ for $B_s^0\to D_s^{*-}\mu^+{\nu}_\mu$ as a function of the recoil $w=v_{B_s}\cdot v_{D_s^*}$ and normalised by the total decay rate calculated from our form factors is given by the purple band. We also show our rate integrated across bins and measurements by LHCb~\cite{Aaij:2020xjy}.}
\end{figure}

The predicted shape of the differential rate $d\Gamma/dq^2$ from our form factors, plotted in Figure~\ref{dgammadq2plot}, may be compared directly to recent experimental measurements by LHCb. 
The results of these measurements are given in~\cite{Aaij:2020xjy}, where unfolded normalised data is binned according to the recoil parameter, $w=v_{B_s}\cdot v_{D_s^*}$, and includes correlations. Here $v_{D_s^*}$ and $v_{B_s}$ are the four-velocities of the $D_s^*$ and $B_s$ respectively. In the $B_s$ rest frame this gives the simple form $w=E_{D_s^*}(p')/M_{D_s^*}=(M_{B_s}^2 + M_{D_s^*}^2 -q^2)/(2M_{B_s}M_{D_s^*})$.
Here we integrate our computed differential rate normalised by the total rate 
over the bins used in~\cite{Aaij:2020xjy}. The $w$ limits of these bins, together with our integrated normalised rates for each bin, are given in Table~\ref{bintable} 
together with the measured values from LHCb. 

Our results and those of LHCb are plotted together in Figure~\ref{lhcbshapeplot}. We see that our results largely agree with the LHCb measurement.
We compute the value of $\chi^2/\mathrm{dof}$ for these measured values compared to our predicted values in the usual way using $\chi^2=\delta g\sigma^{-1}\delta g$ where the vector
$\delta g$ is made up of the differences between our values and the measured values and $\sigma^{-1}$ is the inverse of the covariance matrix for $\delta g$ including correlations from this calculation and those from experiment. We find $\chi^2/\mathrm{dof}=1.8$ 
with a $Q$-value of 0.1. In Figure~\ref{lhcbshapeplot} we see that the third bin with $1.1688 < w < 1.2212$ seems to be furthest from our predicted rate. 
Excluding this bin from the computation of $\chi^2/\mathrm{dof}$ results in a $\chi^2/\mathrm{dof}$ of 0.62, with a $Q$-value of 0.78.

\begin{table*}
\centering
\caption{\label{bintable} $w$ bins together with normalised integrated rates for $B_s^0\to D_s^{*-}\mu^+{\nu}_\mu$ for each bin. The row labelled this work are those values computed using our form factors, discussed in the text, and the LHCb values are those given in~\cite{Aaij:2020xjy}. These are in reasonable agreement, with $\chi^2/\mathrm{dof}=1.8$ and a $Q$-value of 0.1. Note that the majority of the tension with our results originates from the LHCb value for the bin $1.1688 < w < 1.2212$.}
\begin{tabular}{ c | c | c c c c c c c }\hline
$w$ bin & & 1.0-1.1087 &1.1087-1.1688 &1.1688-1.2212 & 1.2212-1.2717 & 1.2717-1.3226 & 1.3226-1.3814 & 1.3814-1.4667 \\ \hline
$\Gamma_\mathrm{bin}/\Gamma_\mathrm{tot}$ & This work &0.1946(82)& 0.1537(45)& 0.1381(29)& 0.1291(18)& 0.1213(19)& 0.1243(40)& 0.139(11) \\
 & LHCb ~\cite{Aaij:2020xjy}& 0.183(12)& 0.1440(84)& 0.1480(76) &0.1280(77) &0.1170(69) & 0.1220(62) & 0.1580(93) \\ \hline
\end{tabular}
\end{table*}

For comparison to others it is useful to give our results in the Caprini, Lellouch and Neubert (CLN) form factor parameterisation~\cite{Caprini:1997mu}. In this scheme the form factors are rewritten in terms of a single leading form factor $h_{A_1}$ together with three ratios. These are related to our form factors by
\begin{align}
h_{A_1}(w)=&A_1(w)\frac{2}{w+1}\frac{1}{R_{D_s^*}}\nonumber\\
R_0(w)=&A_0(w)\frac{R_{D_s^*}}{h_{A_1}(w)}\nonumber\\
R_1(w)=&V(w)\frac{R_{D_s^*}}{h_{A_1}(w)}\nonumber\\
R_2(w)=&A_2(w)\frac{R_{D_s^*}}{h_{A_1}(w)}
\end{align}
where $R_{D_s^*}=2\sqrt{r}/(1+r)$, with $r=M_{D_s^*}/M_{B_s}$. These are then parameterised as
\begin{align}
h_{A_1}(w)=&h_{A_1}(1)[1-8\rho^2 z(w) + (53\rho^2-15)z^2(w)\nonumber\\
&-(231\rho^2-91)z^3(w)]\nonumber\\
R_0(w)=&R_0(1)-0.11(w-1)+0.01(w-1)^2\nonumber\\
R_1(w)=&R_1(1)-0.12(w-1)+0.05(w-1)^2\nonumber\\
R_2(w)=&R_2(1)+0.11(w-1)-0.06(w-1)^2
\end{align}
Where the expressions for $h_{A_1}(w)$, $R_1(w)$ and $R_2(w)$ may be found in~\cite{Caprini:1997mu} and the expression for $R_0(w)$ is derived from the results of~\cite{Caprini:1997mu} in~\cite{Fajfer:2012vx}. In this case
\begin{equation}
z(w)=\frac{\sqrt{w+1}-\sqrt{2}}{\sqrt{w+1}+\sqrt{2}}
\end{equation}
and $h_{A_1}(1)$, $R_0(1)$, $R_1(1)$, $R_2(1)$ and $\rho^2$ are the free parameters. Converting our continuum form factor results to this scheme and then fitting the CLN parameters gives
\begin{align}\label{eq:clnparams}
h_{A_1}(1)=&0.902 (36)\nonumber\\
R_0(1)=&1.057 (58)\nonumber\\
R_1(1)=&1.52 (16)\nonumber\\
R_2(1)=&0.93 (11)\nonumber\\
\rho^2=&1.23 (12).
\end{align}
The value of $\rho^2$ may be compared to the LHCb experimental result given in~\cite{Aaij:2020xjy}, $\rho^2_\mathrm{exp}=1.16(9)$, and we see that our result is in good agreement.
In~\cite{Aaij:2020xjy} the term $h_{A_1}(1)$ is absorbed into the normalisation, and values for $R_1(1)$ and $R_2(1)$ are taken from the HFLAV average of the corresponding parameters determined from experimental measurements of $B \to D^*$ decay~\cite{Amhis:2019ckw}. These are given by
\begin{align}
R_1^{B\to D^*}(1)=&1.270(26),\nonumber\\
R_2^{B\to D^*}(1)=&0.852(18).
\end{align}
\cite{Amhis:2019ckw} also gives $\rho^2_{B\to D^*}=1.122 (24)$. Our results agree with these values within uncertainties and so we see no significant $SU(3)_\mathrm{flavor}$ symmetry breaking between the measured shape of $B\to D^*$ decay and our results for $B_s\to D_s^*$ using the CLN parameterisation scheme. We may also compare our values to those measured by LHCb in~\cite{Aaij:2020hsi}: $\rho^{2,\mathrm{LHCb}}=1.23(17)$, $R_1^\mathrm{LHCb}(1)=1.34(25)$ and $R^\mathrm{LHCb}_2(1)=0.83(16)$, where again we see good agreement. 

\begin{figure}
\includegraphics[scale=0.19]{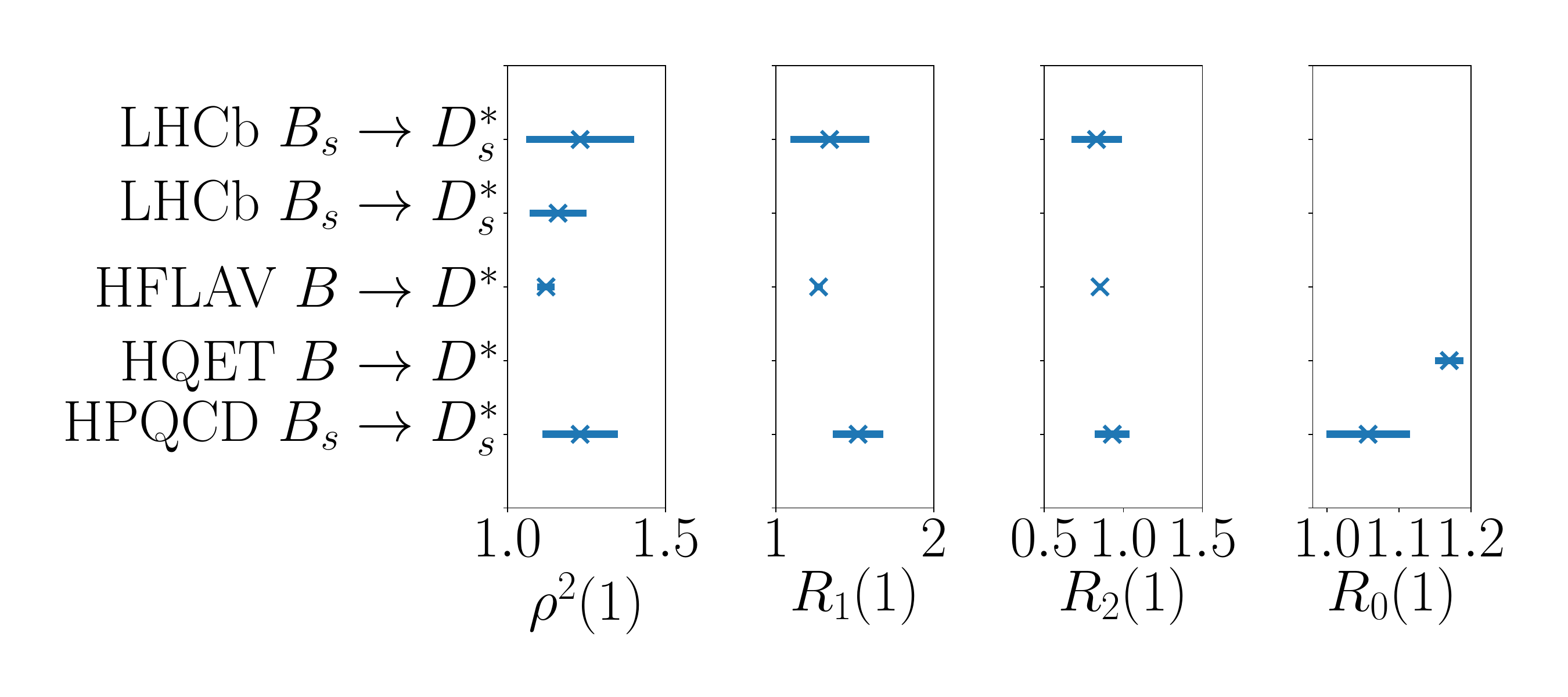}
\caption{\label{shapeparamsplot} Plot comparing our results for the CLN shape parameters $\rho^2$, $R_1(1)$, $R_2(1)$ and $R_0(1)$ to those determined by LHCb in~\cite{Aaij:2020xjy,Aaij:2020hsi}. We also include in this figure the HFLAV values for the $B\to D^*$ shape parameters~\cite{Amhis:2019ckw} as well as the HQET result for $R_0(1)$ for $B\to D^*$~\cite{Bernlochner:2017jka}. We see good agreement, except for our value of $R_0(1)$, which is in slight tension with the HQET result.} 
\end{figure}

We may also compare our result for $R_0(1)$ to the value for $B\to D^*$ decays, which is suppressed by a factor of $m_\ell^2/q^2$ in experimentally measured rates and so is instead determined from HQET~\cite{Bernlochner:2017jka}. We find this value, $R_0^{B\to D^*,\mathrm{HQET}}(1)=1.17(2)$, is in slight tension with our result, with the central value higher by $\approx 1.5\sigma$. These comparisons to experiment and HQET results are summarised in Figure~\ref{shapeparamsplot}.

\subsection{Determination of $|V_{cb}|$}
\label{sec:vcb}

\begin{figure}
\includegraphics[scale=0.195]{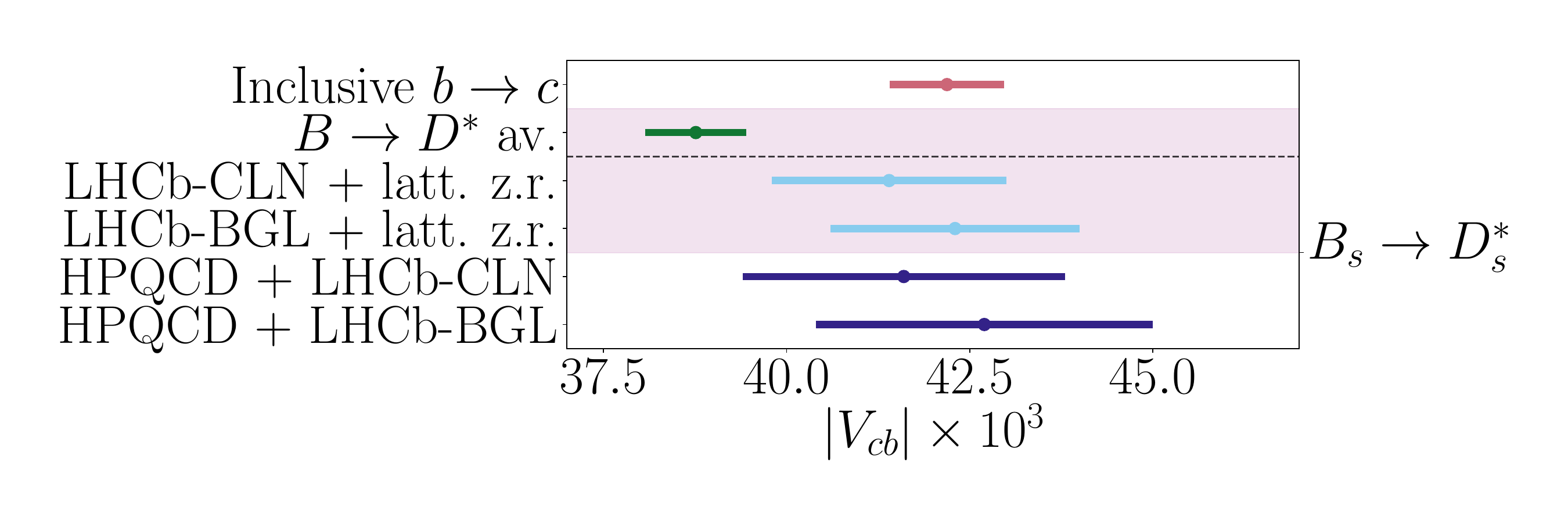}
\caption{\label{vcbplot} Comparison of results for $|V_{cb}|$. 
The values obtained using our lattice results across the full physical $q^2$ range
and LHCb results parameterised using the BGL and CLN schemes given in~\cite{Aaij:2020hsi},
are shown in dark blue. We plot the values determined by LHCb~\cite{Aaij:2020hsi}, using BGL and CLN parameterisations and lattice input only at zero recoil, in light blue. 
We also show the average value determined using $B\to D^{(*)}$ decay, again with 
lattice results at zero recoil only, in green and the value determined 
from inclusive $b \rightarrow c$ measurements in red. 
Both of these latter values are taken from~\cite{Amhis:2019ckw}. 
The pink shaded band indicates which results use lattice input only at zero recoil.}
\end{figure}

In this section we use the results of~\cite{Aaij:2020hsi} to reconstruct LHCb's measured differential decay rate, using the CLN parameterisation scheme, and then compare this to our results here in order to extract a value of $|V_{cb}|$. 
%In~\cite{Aaij:2020hsi} values are given for $h_{A_1}(1)$ and $\eta_\mathrm{EW}$, which are constrained using external inputs from the previous HPQCD lattice computation of $h_{A_1}(1)$~\cite{EuanBsDsstar} and from~\cite{Sirlin:1981ie} respectively, as well as values for $\rho^2$, $R_1(1)$, $R_2(1)$ and $|V_{cb}|$ extracted from fits to experimental data. 
We use the values of $\rho^2$, $R_1(1)$, $R_2(1)$, $\eta_\mathrm{EW}$, $h_{A_1}(1)$ and $|V_{cb}|$ given in~\cite{Aaij:2020hsi}, including their correlations, to reconstruct the measured differential rate, $d\Gamma^\mathrm{exp}/dq^2$, parameterised in the CLN scheme. We then fit this using our computed $({1}/{|V_{cb}\eta_\mathrm{EW}|^2})d\Gamma/dq^2$ in order to extract a value of $|V_{cb}|$. We find
\begin{equation}\label{vcbcln}
|V_{cb}|^\mathrm{CLN}=41.6 (1.5)_\mathrm{latt}(1.6)_\mathrm{exp}(0.4)_\mathrm{EM} \times 10^{-3},
\end{equation}
where the first uncertainty is from our form factor calculation, the second is from the uncertainty in $d\Gamma^\mathrm{exp}/dq^2$ and the final uncertainty is from $\delta_\mathrm{EM}$. 

Repeating this analysis using the BGL parameters given in~\cite{Aaij:2020hsi} to reconstruct $d\Gamma^\mathrm{exp}/dq^2$ yields a value of 
\begin{equation}\label{vcbbgl}
|V_{cb}|^\mathrm{BGL}=42.7 (1.5)_\mathrm{latt}(1.7)_\mathrm{exp}(0.4)_\mathrm{EM} \times 10^{-3}, 
\end{equation}
where again the first uncertainty is from our form factors and the second is from the uncertainty in $d\Gamma^\mathrm{exp}/dq^2$. Note that the difference between $|V_{cb}|^\mathrm{BGL}$ and $|V_{cb}|^\mathrm{CLN}$ is compatible with the difference between $|V_{cb}|$ determined using the two schemes observed in~\cite{Aaij:2020hsi}, using just the zero recoil lattice result, $h_{A_1}(1)$, from~\cite{Harrison:2017fmw, EuanBsDsstar}. 

Since the fits to both CLN and BGL schemes in~\cite{Aaij:2020hsi} have similar $\chi^2/\mathrm{dof}$ we take the average central value of $|V_{cb}|^\mathrm{BGL}$ and $|V_{cb}|^\mathrm{CLN}$, together with the larger of the two uncertainties from Eq.~(\ref{vcbbgl}), to give
\begin{equation}\label{vcbfinal}
|V_{cb}|=42.2 (1.5)_\mathrm{latt}(1.7)_\mathrm{exp}(0.4)_\mathrm{EM} \times 10^{-3}.
\end{equation}
The uncertainty is split approximately equally between lattice QCD and experiment.  

We show a comparison of our results to those of~\cite{Aaij:2020hsi}, together 
with the average values computed using $B\to D^{(*)}$ decay and from 
$b \rightarrow c$ inclusive measurements, both taken from~\cite{Amhis:2019ckw}, 
in Figure~\ref{vcbplot}. 
Our result from this first calculation 
is not sufficiently accurate to resolve the long-standing tension between the 
inclusive and exclusive $V_{cb}$ results. 

We see from Figure~\ref{vcbplot} 
that a model-independent determination of $|V_{cb}|$ using $B_s\to D_s^*$ 
will require a reduction in uncertainty by a factor of $\approx 3$ to reach the 
same precision as that quoted for the exclusive determination using $B\to D^*$ 
at zero-recoil. This reduction is feasible with a direct comparison of improved 
lattice and experimental results that would enable a joint fit. 
Here we have used experimental results indirectly, 
through the fitted BGL and CLN parameters provided by LHCb in~\cite{Aaij:2020hsi}. 
Fitting our results directly to binned experimental data for $d\Gamma/dq^2$ would 
reduce or remove dependence on the parameterisation scheme used by the experiment 
and would certainly be preferable if this data was available. 
A similar comparison of future lattice results for the $B\to D^*$ differential 
rate against binned experimental values will be important for determining 
$V_{cb}$ with reduced dependence on the parameterisation scheme used.

\subsection{Determining $d\mathcal{F}(w)/dw|_{w=1}$}
\label{sec:slope}

\begin{figure}
\includegraphics[scale=0.19]{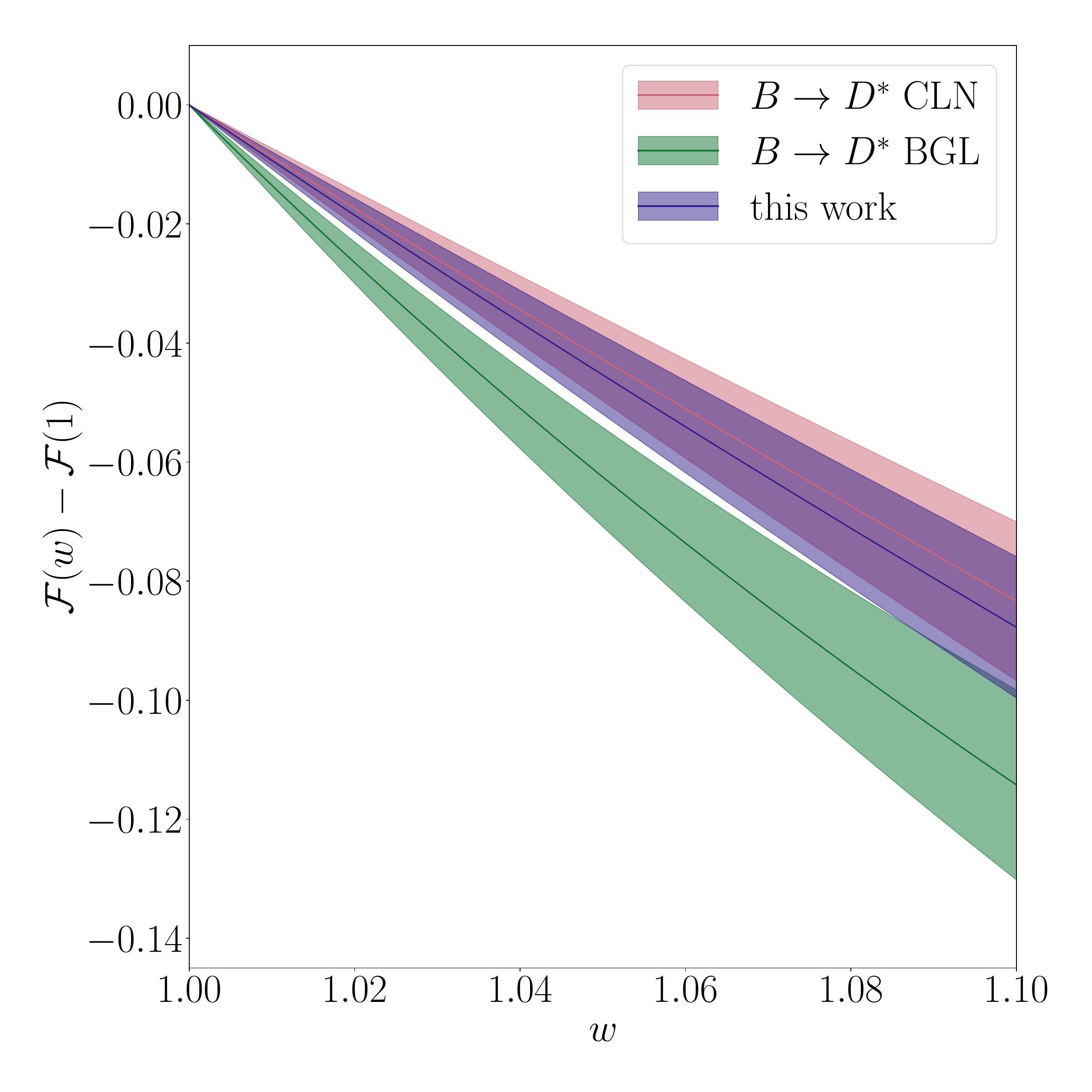}
\caption{\label{slopeplot}Plot showing $\mathcal{F}(w)-\mathcal{F}(1)$, defined in Eq.~(\ref{Fdefeq}), against $w$ for our $B_s \to D_s^*$ form factors computed here, together with the corresponding values from fits to experimental results for $B\to D^*$ including light cone sum rule constraints from~\cite{Bigi:2017njr}. Here we see that our results for the slope of $\mathcal{F}(w)$ in the $B_s \to D_s^*$ case, which we expect to be close to the slope for $B\to D^*$, are consistent with the CLN fit but in tension with the BGL fit at the level of $\approx 2.5\sigma$.}
\end{figure}

After integrating over the angular variables, the differential rate with respect to the recoil $w$ for $B_{(s)}\to D_{(s)}^*$ may be written~\cite{Caprini:1997mu}:
\begin{align}\label{Fdefeq}
&\frac{d\Gamma(B_{(s)}\to D_{(s)}^*\ell\bar{\nu}_\ell)}{dw} =\nonumber\\
& \frac{G_F^2|\eta_\mathrm{EW}V_{cb}|^2}{48\pi^3}(M_{B_{(s)}}-M_{D_{(s)}^*})^2 M_{D_{(s)}^*}^3\sqrt{w^2-1}(w+1)^2\nonumber\\
&\times \left[1+\frac{4w}{w+1}\frac{M_{B_{(s)}}^2-2wM_{B_{(s)}}M_{D_{(s)}^*}+M_{D_{(s)}^*}^2}{(M_{B_{(s)}}-M_{D_{(s)}^*})^2} \right]\nonumber\\
&\times|\mathcal{F}^{B_{(s)}\to D_{(s)}^*}(w)|^2.
\end{align}
In~\cite{Bigi:2017njr} it was emphasised that, for determining $V_{cb}$ from $B\to D^*\ell\bar{\nu}_\ell$ decay, information about the slope of $\mathcal{F^{B\to D^*}}(w)$ at zero recoil, $d\mathcal{F}^{B\to D^*}(w)/dw|_{w=1}$, could significantly reduce the uncertainty in $|V_{cb}|$. They consider a hypothetical lattice determination of $d\mathcal{F}^{B\to D^*}(w)/dw|_{w=1}=-1.44\pm 0.07$, which results in a $\approx 25\%$ reduction in the uncertainty of $V_{cb}$ determined using the BGL parameterisation and also moves the values of $V_{cb}$ determined using both CLN and BGL schemes to within $\approx 0.2\sigma$ of one another.

Here we find a value for the slope of 
\begin{equation}
\frac{d\mathcal{F}^{B_s\to D_s^*}(w)}{dw}\Big{|}_{w=1}=-0.94\pm 0.15,
\end{equation} 
which we determine using a simple finite difference method. This is consistent with the slope for ${B\to D^*}$ determined from the CLN parameters extracted from experimental data including light cone sum rule constraints in~\cite{Bigi:2017njr} of $d\mathcal{F}^{B\to D^*}_\mathrm{CLN}(w)/dw|_{w=1}=-0.84(16)$, where we have estimated the uncertainty from the uncertainties of the CLN parameters in~\cite{Bigi:2017njr} excluding correlations. 

In Figure~\ref{slopeplot} we have plotted $\mathcal{F}(w)-\mathcal{F}(1)$ against $w$ where we see the difference between our results for the slope of $\mathcal{F}(w)$ near zero recoil for $B_s \to D_s^*$ compared to the slope computed from CLN and BGL fits to $B\to D^*$ experimental data in~\cite{Bigi:2017njr}. Note that our result for the slope is $\approx 2.5\sigma$ from the slope computed from the BGL fit, which we would expect to be similar for both $B\to D^*$ and $B_s\to D_s^*$ based on the small expected size of $SU(3)_\mathrm{flavor}$ breaking effects~\cite{Bordone:2019guc}.

\section{LFUV Observables and Impact of New Physics Couplings}
\label{sec:lfuv}
\begin{figure*}
\includegraphics[scale=0.15]{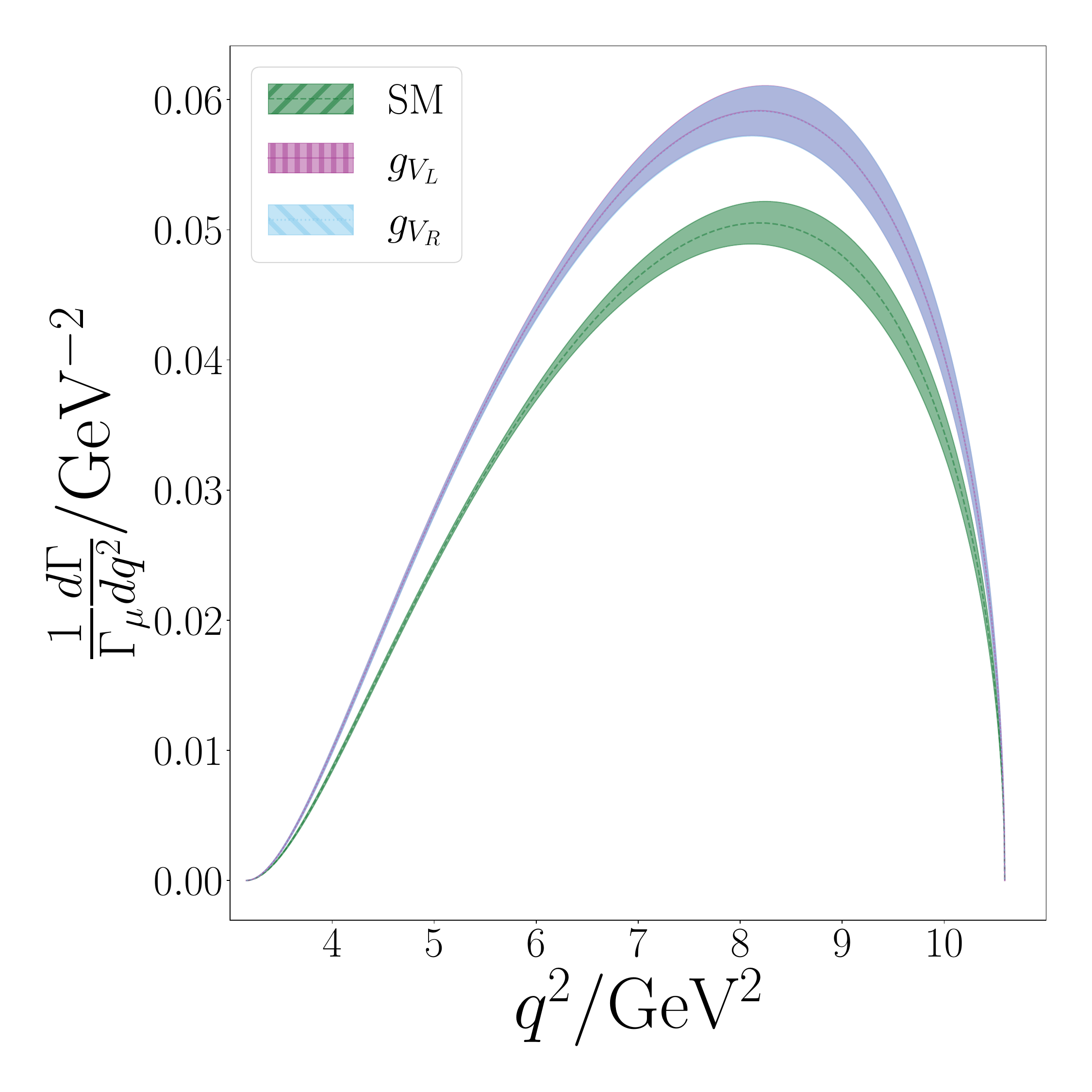}
\includegraphics[scale=0.15]{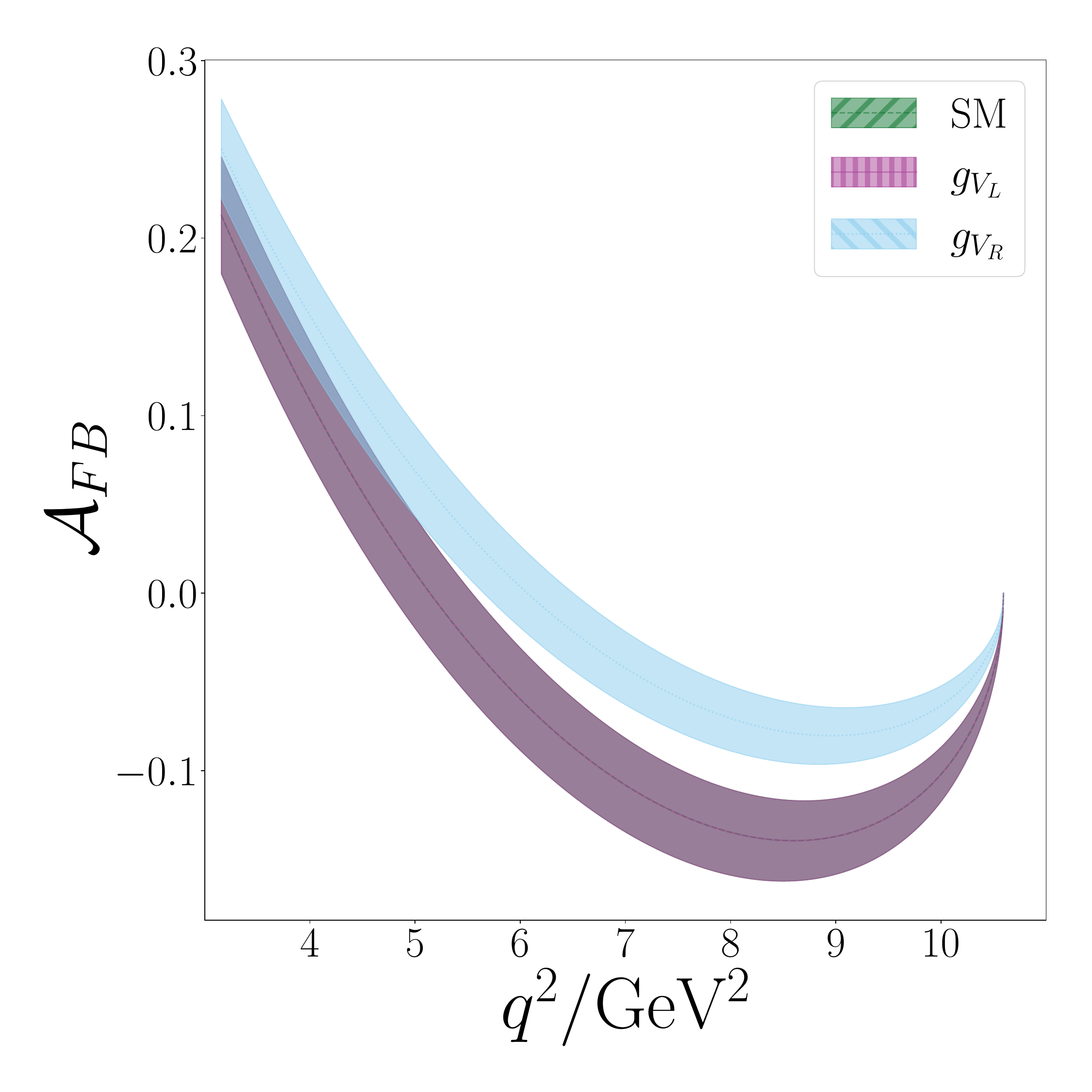}
\includegraphics[scale=0.15]{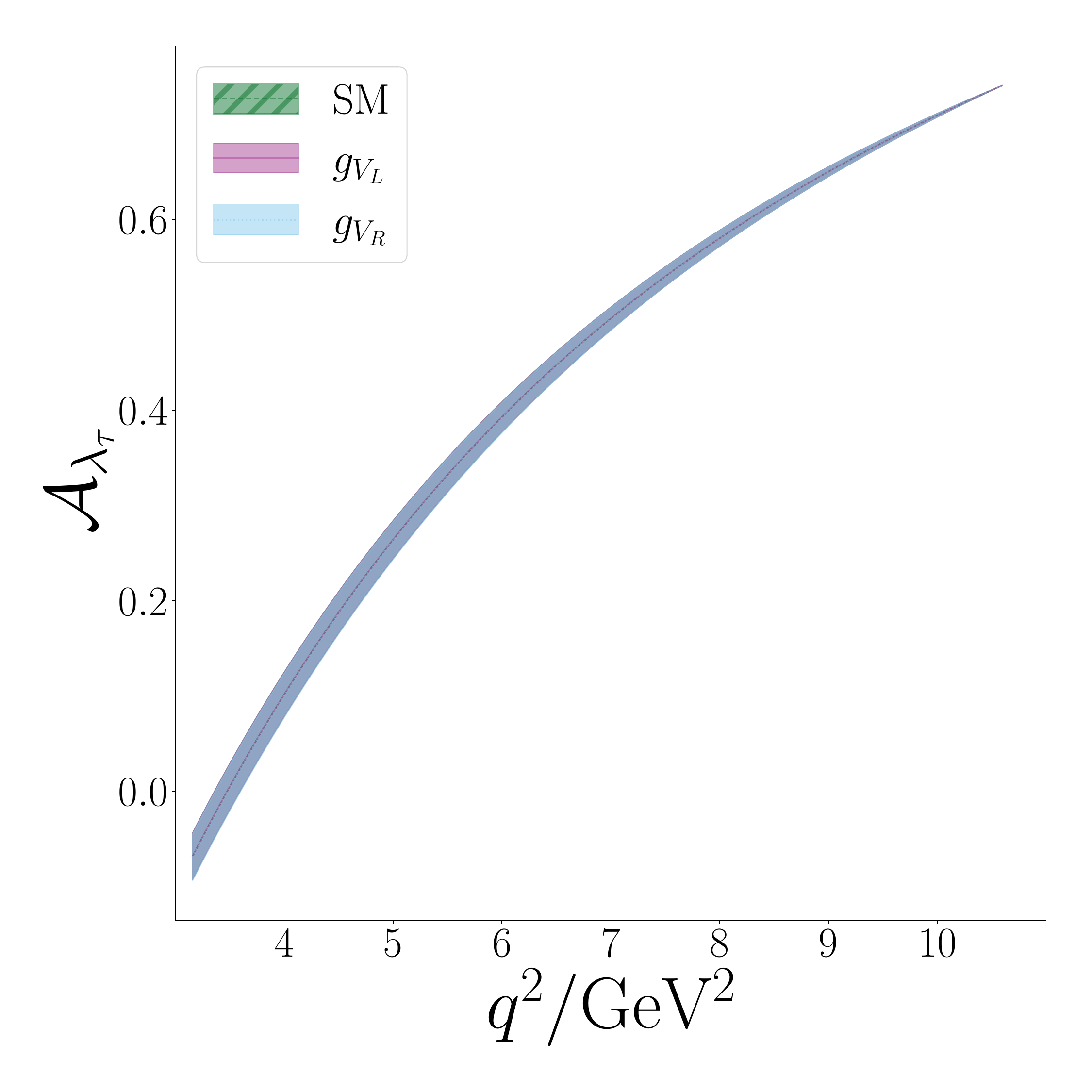}
\caption{\label{NPplots} $d\Gamma/dq^2$, $\mathcal{A}_{FB}$ and $\mathcal{A}_{\lambda_\tau}$ 
for $\bar{B}_s^0 \rightarrow D_s^{*+} \tau^- \bar{\nu}_{\tau}$ in the   
SM and for the values of $g_{V_R}$ and $g_{V_L}$ given in Eq.~(\ref{NPcouplings}) from~\cite{Becirevic:2019tpx}. $d\Gamma/dq^2$ is normalised to the total rate in the $\ell=\mu$ case, $\Gamma_{\mu}$, and the $g_{V_L}$ and $g_{V_R}$ curves overlap. For $A_{F,B}$ the SM and $g_{V_L}$ curves overlap and for $A_{\lambda_{\tau}}$ all three curves overlap. Note that here we do not include the contribution of $\delta_\mathrm{EM}$ to the uncertainty.}
\end{figure*}
In this section we study the impact of new physics (NP) couplings on observables in $B_s\to D_s^*$ decay. We do this by extending the analysis performed in~\cite{Harrison:2020nrv}, where the NP couplings extracted from fits to $R(D)$ and $R(D^*)$ in~\cite{Becirevic:2019tpx} were used to predict the variations of the relevant observables defined in~\cite{Becirevic:2019tpx} away from their Standard Model (SM) values for the case of $B_c\to J/\psi\ell\bar{\nu}_\ell$. 
For $B_s \to D_s^*$ the relevant observables are the same, specifically those defined in Eq.~(\ref{angasymeq}) as well as their integrated values and ratios given in Eq.~(\ref{lfuvexpectations}) and Eq.~(\ref{lfuvratios}) respectively for the SM couplings. We also compute here the value of the tauonic/muonic ratio, $R(D_s^{*})$, for different values of new physics couplings. Following~\cite{Harrison:2020nrv}, we consider nonzero values for $g_{V_R}$ and $g_{V_L}$, the complex-valued NP couplings multiplying left and right handed NP vector currents additional to those present in the SM. Here, as in~\cite{Becirevic:2019tpx} and~\cite{Harrison:2020nrv}, we only take these couplings to modify the tauonic decay. 

In the left hand plot of Figure~\ref{NPplots} we plot the tauonic differential rate, normalised by the muonic rate, where we see that both the left and right handed NP vector currents increase the tauonic differential rate markedly. The corresponding values of $R^{g_i}(D_s^*)$ for each NP coupling, together with the numerical values of $g_{V_R}$ and $g_{V_L}$, are;
\begin{align}\label{NPcouplings}
g_{V_R} =& -0.01 - i\,0.39;\, \nonumber\\
&R^{g_{V_R}}(D_s^*)=0.2912(71)_\mathrm{latt}(40)_\mathrm{EM},\nonumber\\
g_{V_L} =& \,0.07 - i\,0.16;\, \nonumber\\
&R^{g_{V_L}}(D_s^*)=0.2915(71)_\mathrm{latt}(40)_\mathrm{EM}.
\end{align}
These values are both larger than our SM value given in Eq.~(\ref{rdsstar}) and are both consistent with the HFLAV average experimental value for $R(D^*)=0.295(14)$~\cite{Amhis:2019ckw}.

The middle and right hand plots in Figure~\ref{NPplots} show $\mathcal{A}_{FB}$ and $\mathcal{A}_{\lambda_\tau}$ respectively. Here we see that only $\mathcal{A}_{FB}$, the forward-backward asymmetry of the final state tau lepton, is sensitive to changes in $g_{V_R}$, only. As expected modifications to the left handed current do not result in a change away from the SM. For $\mathcal{A}_{\lambda_\tau}$ neither $g_{V_R}$ or $g_{V_L}$ produce any change. The integrated observables and ratios for these quantities are given in Table~\ref{NPlfuvintegrated} and Table~\ref{NPlfuvratios} respectively, where we repeat the SM results given in Table~\ref{lfuvexpectations} and Table~\ref{lfuvratios} for reference. Note that these are all within $1\sigma$ of the equivalent quantities computed for $B_c\to J/\psi\ell\bar{\nu}_\ell$ in~\cite{Harrison:2020nrv}.

\begin{table}
\centering
\caption{\label{NPlfuvintegrated} Integrated angular variables defined in~\cite{Becirevic:2019tpx} for the NP couplings given in Eq.~(\ref{NPcouplings}).}
\begin{tabular}{ c | c c c }
\hline
 & SM & $g_{V_R}$&$g_{V_L}$ \\ \hline
$\langle \mathcal{A}_{\lambda_\tau} \rangle$ & 0.520(12) & 0.520(12) & 0.520(12) \\
$\langle F^{D_s^*}_L$ &  0.440(16) & 0.441(16) & 0.440(16) \\
$\langle \mathcal{A}_{FB}\rangle$ & -0.092(24) & -0.033(18) & -0.092(24)\\\hline
\end{tabular}
\end{table}

\begin{table}
\centering
\caption{\label{NPlfuvratios} LFUV ratios defined in~\cite{Becirevic:2019tpx} for the NP couplings given in Eq.~(\ref{NPcouplings}). Note that as expected from the form of the SM current, a modification to the left handed vector current does not change any of these ratios away from their SM values.}
\begin{tabular}{ c | c c c }
\hline
 & SM & $g_{V_R}$&$g_{V_L}$ \\ \hline
$R(\mathcal{A}_{\lambda_\tau})$ & 0.524(12) & 0.524(12) & 0.524(12)\\
$R(F_{L}^{J/\psi})$ & 0.878(18) & 0.880(18) & 0.878(18)  \\
$R(\mathcal{A}_{FB})$ & 0.345(56) & 0.126(57) & 0.345(56)  \\\hline
\end{tabular}
\end{table}

\section{Conclusions}
\label{sec:conclusions}
We have extended the heavy-HISQ methods of~\cite{Harrison:2020gvo} to compute the four form factors, $A_0(q^2)$, $A_1(q^2)$, $A_2(q^2)$ and $V(q^2)$, for $B_s\to D_s^*\ell\bar{\nu}_\ell$ across the full kinematic range of the decay for the first time using lattice QCD. As in~\cite{Harrison:2020gvo} our calculation uses the HISQ action for all quarks, allowing us to normalise the lattice weak current operators that couple the $b$ and $c$ quarks fully nonperturbatively. We have included with this work the complete set of parameters and their correlations needed to reconstruct our form factors, which are expected to be useful in upcoming improved analyses by the LHCb experiment.

Using these form factors we have presented the first computation in lattice QCD of the total decay rate to each of the three different final state leptons, as well as the tauonic/muonic ratio.  We find (repeating \cref{gammaelectoteq,gammamutoteq,gammatautoteq,rdsstar})
\begin{align}
\frac{\Gamma(B_s^0\to D_s^{*-}e^+{\nu}_e)}{|\eta_\mathrm{EW}V_{cb}|^2}=&\GAMMAeTOTAL\nonumber\\
\frac{\Gamma(B_s^0\to D_s^{*-}\mu^+{\nu}_\mu)}{|\eta_\mathrm{EW}V_{cb}|^2}=&\GAMMAmuTOTAL\nonumber\\
\frac{\Gamma(B_s^0\to D_s^{*-}\tau^+{\nu}_\tau)}{|\eta_\mathrm{EW}V_{cb}|^2}=&\GAMMAtauTOTAL\nonumber\\
R(D_s^*) = & \RDsstar.
\end{align}
where the full error budget for these quantities is given in Table~\ref{errorbudget}. 
Note that we have included an uncertainty in $R(D_s^*)$ to allow for long-distance QED 
corrections that could differ between the $\tau$ and $\mu$ cases. These QED effects 
must be addressed in future calculations in order to produce results with reliable 
percent-level uncertainties.

Since the current experimental average for $R(D^*)$ is causing some tension with 
the SM, we have considered the impact of NP scenarios 
using our form factors (Section~\ref{sec:lfuv}), 
and illustrated how a modified left or right handed vector current consistent with measurements of $B\to D^{(*)}$ might show up in $B_s\to D_s^*$ decay. We have shown that these NP currents result in a modified value of $R(D_s^*)$ which is larger than in the SM and is within $\approx 0.6\sigma$ of the current experimental average value of $R(D^*)$.

We have also computed the ratio of total SM rates for $B_s\to D_s$ and $B_s\to D_s^*$. We find (repeating \cref{bsdsbsdss}) 
\begin{equation}
\frac{\Gamma(B^0_s\to D_s^-\mu^+\nu_\mu)}{\Gamma(B_s^0\to D_s^{*-}\mu^+\nu_\mu)}=0.443(40)_\mathrm{latt}(4)_\mathrm{EM}
\end{equation}
which is in good agreement with the experimental value from LHCb~\cite{Aaij:2020hsi}. 
We give the forward-backward asymmetry of the final state lepton, the lepton polarisation asymmetry and the longitudinal polarisation fraction in Eq.~(\ref{lfuvexpectations}).

We have compared the normalised differential decay rate, $(1/\Gamma) d\Gamma/dq^2$, computed using our form factors to the recent measurement of the shape of the decay by LHCb. 
We find that the measurement is broadly consistent with our computed shape, 
showing only very mild tension which can be seen to originate from a single $w$ bin. 
We have also used our results to determine the SM CLN parameters $\rho^2$, $R_0(1)$, $R_1(1)$, $R_2(1)$ and $h_{A_1}(1)$ for $B_s\to D_s^*$. The value of the slope of $h_{A_1}(w)$, $\rho^2$, measured by LHCb for $B_s\to D_s^*$ is found to be in good agreement with our computed value, while our values for the other parameters are seen to be consistent with values for the corresponding parameters for $B\to D^*$.

Finally, we have used our result together with the reconstructed experimentally measured differential decay rate, parameterised using both the CLN and BGL schemes, to compute a value of $|V_{cb}|$. We find the value computed in this way using lattice results across the full kinematic range is consistent with that computed using lattice input for only $h_{A_1}(1)$. Our values for $|V_{cb}|$ computed using experimental results parameterised in the CLN and BGL schemes respectively are given in \cref{vcbcln,vcbcln}. For $|V_{cb}|$ we take the average of their central values, with the larger of the two errors, to find (repeating Eq.~(\ref{vcbfinal})) 
\begin{equation}\label{vcbfinal2}
|V_{cb}|=42.2 (1.5)_\mathrm{latt}(1.7)_\mathrm{exp}(0.4)_\mathrm{EM} \times 10^{-3}.
\end{equation}
This value is not yet accurate enough to resolve the tension between inclusive and 
exclusive (using $B \rightarrow D^*$ at zero recoil) values for $V_{cb}$. 
Note that here, if binned experimental data for the differential rate was available, 
$|V_{cb}|$ could be determined by comparing our lattice results directly to 
experiment in a joint fit, without the need to use a parameterisation scheme. 

Such a model-independent determination of $|V_{cb}|$ using $B_s\to D_s^*$ would require a reduction in uncertainty by a factor of $\approx 3$ in order to be competitive. 
Such a reduction in the uncertainty of our form factor results is feasible and 
may be achieved by working on `exafine' lattices with lattice spacings $\approx 0.03\mathrm{fm}$, allowing us to work directly at the physical $b$ quark mass, as well as by including additonal configurations to reduce statistical uncertainties at other values of 
the lattice spacing, as discussed in Section~\ref{sec:errorbudget}.

Our work paves the way for the calculation of form factors for $B \rightarrow D^*$ decay 
across the full range of $q^2$. This would also allow a direct determination of 
$|V_{cb}|$ from unfolded experimental results binned in $q^2$, reducing the reliance 
on extrapolation to the zero recoil point. 

\subsection*{\bf{Acknowledgements}}

We are grateful to the MILC collaboration for the use of
their configurations and code. 
We thank C. Bouchard, B. Colquhoun, J. Koponen, P. Lepage, 
E. McLean and C. McNeile for useful discussions.
Computing was done on the Cambridge service for Data 
Driven Discovery (CSD3), part of which is operated by the 
University of Cambridge Research Computing on behalf of 
the DIRAC 
HPC Facility of the Science and Technology Facilities 
Council (STFC). The DIRAC component of CSD3 was funded by 
BEIS capital funding via STFC capital grants ST/P002307/1 and 
ST/R002452/1 and STFC operations grant ST/R00689X/1. 
DiRAC is part of the national e-infrastructure.  
We are grateful to the CSD3 support staff for assistance.
Funding for this work came from the UK
Science and Technology Facilities Council grants 
ST/L000466/1 and ST/P000746/1.

\begin{appendix}

\section{Lattice Results}
\label{app:lattice_results}

Here, in \cref{set1,set2,set3,set4}, we give the lattice results for the form factors, including renormalisation factors, which were extracted from the matrix elements in Eq.~(\ref{relnorm}) resulting from the fits discussed in Section~\ref{sec:lattcalc}.

\begin{table}
\centering
\caption{Lattice form factor results for set 1. $ak$ here is the value of the $x$ and $y$ components of the lattice momentum for the $D_s^*$. $ak$ is calculated from the corresponding twist in Table \ref{twists}.  \label{set1}}
\begin{tabular}{ c c c c c c }
\hline
$am_h$& $ak$ & $A_0$& $A_1$& $A_2$& $V$\\\hline
0.65&	0.0	&$-$	&0.9261(53)	&$-$	&$-$	\\
&	0.0358913	&0.90(13)	&0.9251(54)	&1.1(2.0)	&1.09(48)	\\
&	0.0717826	&0.898(68)	&0.9226(57)	&1.21(83)	&1.21(22)	\\
&	0.107674	&0.893(48)	&0.9183(61)	&1.22(58)	&1.23(15)	\\
&	0.143565	&0.882(38)	&0.9120(67)	&1.24(50)	&1.23(12)	\\
&	0.179456	&0.867(33)	&0.9037(76)	&1.26(48)	&1.22(10)	\\
\hline
0.725&	0.0	&$-$	&0.9277(54)	&$-$	&$-$	\\
&	0.0358913	&0.92(14)	&0.9267(55)	&0.8(2.3)	&1.10(49)	\\
&	0.0717826	&0.913(71)	&0.9242(58)	&1.01(71)	&1.22(22)	\\
&	0.107674	&0.907(50)	&0.9199(62)	&1.04(42)	&1.24(15)	\\
&	0.143565	&0.897(40)	&0.9136(68)	&1.06(33)	&1.24(12)	\\
&	0.179456	&0.882(35)	&0.9052(78)	&1.07(30)	&1.23(10)	\\
\hline
0.8&	0.0	&$-$	&0.9299(56)	&$-$	&$-$	\\
&	0.0358913	&0.93(14)	&0.9289(57)	&0.7(2.7)	&1.12(49)	\\
&	0.0717826	&0.930(75)	&0.9264(59)	&0.91(76)	&1.24(22)	\\
&	0.107674	&0.924(53)	&0.9220(63)	&0.95(40)	&1.26(16)	\\
&	0.143565	&0.913(43)	&0.9157(70)	&0.97(28)	&1.25(12)	\\
&	0.179456	&0.898(37)	&0.9073(80)	&0.98(23)	&1.24(11)	\\
\hline
\end{tabular}
\end{table}

\begin{table}
\centering
\caption{Lattice form factor results for set 2. $ak$ here is the value of the $x$ and $y$ components of the lattice momentum for the $D_s^*$. $ak$ is calculated from the corresponding twist in Table \ref{twists}.  \label{set2}}
\begin{tabular}{ c c c c c c }
\hline
$am_h$& $ak$ & $A_0$& $A_1$& $A_2$& $V$\\\hline
0.427&	0.0	&$-$	&0.908(12)	&$-$	&$-$	\\
&	0.0524832	&0.89(18)	&0.906(13)	&1.6(1.7)	&1.26(56)	\\
&	0.104966	&0.87(10)	&0.896(15)	&1.07(96)	&1.23(31)	\\
&	0.15745	&0.824(79)	&0.879(18)	&0.97(94)	&1.19(25)	\\
&	0.209933	&0.768(75)	&0.857(23)	&1.0(1.3)	&1.13(24)	\\
&	0.262416	&0.705(78)	&0.831(32)	&1.2(2.8)	&1.06(24)	\\
\hline
0.525&	0.0	&$-$	&0.904(13)	&$-$	&$-$	\\
&	0.0524832	&0.92(20)	&0.903(13)	&1.9(1.7)	&1.26(58)	\\
&	0.104966	&0.89(11)	&0.893(15)	&1.13(60)	&1.24(33)	\\
&	0.15745	&0.851(87)	&0.875(18)	&0.97(46)	&1.20(27)	\\
&	0.209933	&0.792(83)	&0.852(24)	&0.91(51)	&1.14(25)	\\
&	0.262416	&0.727(86)	&0.825(33)	&0.90(69)	&1.07(26)	\\
\hline
0.65&	0.0	&$-$	&0.900(13)	&$-$	&$-$	\\
&	0.0524832	&0.96(22)	&0.898(14)	&2.3(2.2)	&1.28(61)	\\
&	0.104966	&0.94(12)	&0.888(15)	&1.24(63)	&1.27(35)	\\
&	0.15745	&0.891(98)	&0.870(19)	&1.03(37)	&1.22(28)	\\
&	0.209933	&0.829(93)	&0.847(25)	&0.94(33)	&1.17(27)	\\
&	0.262416	&0.760(97)	&0.821(35)	&0.89(37)	&1.09(28)	\\
\hline
0.8&	0.0	&$-$	&0.896(14)	&$-$	&$-$	\\
&	0.0524832	&1.02(25)	&0.895(14)	&2.6(2.8)	&1.32(64)	\\
&	0.104966	&0.99(14)	&0.884(16)	&1.36(78)	&1.31(37)	\\
&	0.15745	&0.94(11)	&0.866(20)	&1.10(42)	&1.26(30)	\\
&	0.209933	&0.88(11)	&0.843(26)	&0.99(31)	&1.20(29)	\\
&	0.262416	&0.80(11)	&0.817(36)	&0.93(30)	&1.13(30)	\\
\hline
\end{tabular}
\end{table}

\begin{table}
\centering
\caption{Lattice form factor results for set 3. $ak$ here is the value of the $x$ and $y$ components of the lattice momentum for the $D_s^*$. $ak$ is calculated from the corresponding twist in Table \ref{twists}.  \label{set3}}
\begin{tabular}{ c c c c c c }
\hline
$am_h$& $ak$ & $A_0$& $A_1$& $A_2$& $V$\\\hline
0.5&	0.0	&$-$	&0.8810(73)	&$-$	&$-$	\\
&	0.0585768	&0.963(62)	&0.8705(79)	&1.00(54)	&1.42(19)	\\
&	0.117154	&0.907(40)	&0.839(11)	&0.83(18)	&1.31(13)	\\
&	0.17573	&0.818(43)	&0.792(18)	&0.75(14)	&1.20(14)	\\
&	0.234307	&0.714(59)	&0.740(32)	&0.66(21)	&1.08(18)	\\
&	0.292884	&0.620(87)	&0.676(68)	&0.45(45)	&0.97(30)	\\
\hline
0.65&	0.0	&$-$	&0.8684(75)	&$-$	&$-$	\\
&	0.0585768	&1.034(73)	&0.8584(82)	&1.00(74)	&1.47(20)	\\
&	0.117154	&0.976(48)	&0.828(11)	&0.86(24)	&1.36(14)	\\
&	0.17573	&0.881(51)	&0.782(18)	&0.79(16)	&1.24(14)	\\
&	0.234307	&0.767(69)	&0.731(33)	&0.72(17)	&1.11(20)	\\
&	0.292884	&0.66(10)	&0.668(70)	&0.59(27)	&1.00(33)	\\
\hline
0.8&	0.0	&$-$	&0.8580(77)	&$-$	&$-$	\\
&	0.0585768	&1.109(84)	&0.8485(84)	&0.96(92)	&1.54(21)	\\
&	0.117154	&1.048(55)	&0.819(11)	&0.88(30)	&1.41(14)	\\
&	0.17573	&0.948(59)	&0.774(19)	&0.83(20)	&1.28(15)	\\
&	0.234307	&0.823(81)	&0.723(33)	&0.78(19)	&1.15(22)	\\
&	0.292884	&0.70(12)	&0.660(72)	&0.67(26)	&1.04(36)	\\
\hline
\end{tabular}
\end{table}

\begin{table}
\centering
\caption{Lattice form factor results for set 4. $ak$ here is the value of the $x$ and $y$ components of the lattice momentum for the $D_s^*$. $ak$ is calculated from the corresponding twist in Table \ref{twists}.  \label{set4}}
\begin{tabular}{ c c c c c c }
\hline
$am_h$& $ak$ & $A_0$& $A_1$& $A_2$& $V$\\\hline
0.5&	0.0	&$-$	&0.9244(90)	&$-$	&$-$	\\
&	0.0356761	&0.87(25)	&0.9240(92)	&8(35)	&1.37(93)	\\
&	0.0713522	&0.88(12)	&0.9213(98)	&6(25)	&1.27(44)	\\
&	0.107028	&0.874(90)	&0.917(11)	&20(65)	&1.27(33)	\\
&	0.142704	&0.859(76)	&0.911(12)	&-7(22)	&1.25(27)	\\
&	0.178381	&0.840(69)	&0.904(13)	&-1.9(7.4)	&1.22(24)	\\
\hline
0.65&	0.0	&$-$	&0.9272(98)	&$-$	&$-$	\\
&	0.0356761	&0.90(29)	&0.927(10)	&1.7(4.0)	&1.4(1.0)	\\
&	0.0713522	&0.91(14)	&0.924(11)	&1.2(1.5)	&1.28(47)	\\
&	0.107028	&0.90(10)	&0.920(12)	&1.2(1.1)	&1.28(35)	\\
&	0.142704	&0.884(88)	&0.914(13)	&1.14(97)	&1.26(29)	\\
&	0.178381	&0.865(80)	&0.906(15)	&1.13(97)	&1.23(26)	\\
\hline
0.8&	0.0	&$-$	&0.931(10)	&$-$	&$-$	\\
&	0.0356761	&0.93(33)	&0.931(11)	&1.7(5.6)	&1.5(1.1)	\\
&	0.0713522	&0.95(16)	&0.928(12)	&1.2(1.6)	&1.31(50)	\\
&	0.107028	&0.94(12)	&0.923(13)	&1.12(84)	&1.31(37)	\\
&	0.142704	&0.92(10)	&0.917(14)	&1.07(61)	&1.28(30)	\\
&	0.178381	&0.900(92)	&0.909(16)	&1.03(53)	&1.25(27)	\\
\hline
\end{tabular}
\end{table}

\section{Reconstructing the Fit}
\label{anfullcov}

Our parameterisation of the form factors for $B_s \to D_s^*$ in the continuum limit 
is given by Eq.~(\ref{contphysfitfunc}). It consists of a pole factor with no uncertainty 
and a polynomial in $z$ for which the coefficients with their uncertainties are given 
in Table~\ref{zexpcoefficients}. 
In this section we give the correlations between the $z$-expansion coefficients which are necessary for reconstructing our results explicitly, as well as instructions for using the included ancillary files to load the $z$-expansion parameters together with their correlations automatically into \textbf{python}~\cite{python3}. 

The correlation between two coefficients is defined in the usual way as
\begin{equation}
\mathrm{Corr}(X,Y) = \frac{\langle (\bar{X}-X)(\bar{Y}-Y) \rangle }{\sqrt{\sigma^2(X)\sigma^2(Y)}}
\end{equation}
where $\sigma^2(X)$ is the variance of $X$ and $\bar{X}$ is the mean of $X$. 
The values are tabulated in \cref{corr_A0,corr_A0_A1,corr_A0_A2,corr_A0_V,corr_A1,corr_A1_A2,corr_A1_V,corr_A2,corr_A2_V,corr_V}. 

In this calculation and in the ancillary files we use the \textbf{gvar} python package to track and propagate correlations. Included in the ancillary files are two text files; \textbf{CORRELATIONS.txt} contains a dictionary including the means and variances of the $z$-expansion parameters on the first line and a dictionary detailing the correlations between these parameters on the second line, \textbf{CHECKS.txt} contains arrays of $q^2$ values and form factor mean and standard deviation values at the corresponding values of $q^2$. This file is used by the python script \textbf{load\_fit.py} as a simple check that the fit has been loaded correctly. Running \textbf{python load\_fit.py} will load the parameters from \textbf{CORRELATIONS.txt} and compare values computed at hard coded intervals in $q^2$ to those in \textbf{CHECKS.txt} which were computed as part of this work. Running \textbf{python load\_fit.py} will also produce some simple plots of the form factors across the full $q^2$ range.  We have tested \textbf{load\_fit.py} using \textbf{Python 3.7.5}~\cite{python3}, \textbf{gvar 9.2.1}~\cite{gvar}, \textbf{numpy 1.18.2}~\cite{numpy} and \textbf{matplotlib 3.1.2}~\cite{matplotlib}.

\begin{table}
\caption{\label{corr_A0} Correlation matrix for $z$-expansion coefficients of $A0$.}
\begin{tabular}{ c | c c c c }\hline
$\sigma^2 $& $a_0^{A0}$	& $a_1^{A0}$	& $a_2^{A0}$	& $a_3^{A0}$	\\\hline
$a_0^{A0}$&	1.0&	-0.3781&	-0.007023&	-0.004275\\
$a_1^{A0}$&	-0.3781&	1.0&	-0.1681&	0.00345\\
$a_2^{A0}$&	-0.007023&	-0.1681&	1.0&	-0.002318\\
$a_3^{A0}$&	-0.004275&	0.00345&	-0.002318&	1.0\\
\hline
\end{tabular}
\end{table}\begin{table}
\caption{\label{corr_A0_A1} Correlation matrix for $z$-expansion coefficients of $A0$ and $A1$.}
\begin{tabular}{ c | c c c c }\hline
$\sigma^2 $& $a_0^{A1}$	& $a_1^{A1}$	& $a_2^{A1}$	& $a_3^{A1}$	\\\hline
$a_0^{A0}$&	0.2776&	-0.02225&	0.1044&	0.01322\\
$a_1^{A0}$&	0.04791&	0.0006896&	-0.2142&	-0.02751\\
$a_2^{A0}$&	-0.01723&	0.03742&	0.0506&	-0.003768\\
$a_3^{A0}$&	0.003371&	-0.008893&	0.0279&	0.006599\\
\hline
\end{tabular}
\end{table}\begin{table}
\caption{\label{corr_A0_A2} Correlation matrix for $z$-expansion coefficients of $A0$ and $A2$.}
\begin{tabular}{ c | c c c c }\hline
$\sigma^2 $& $a_0^{A2}$	& $a_1^{A2}$	& $a_2^{A2}$	& $a_3^{A2}$	\\\hline
$a_0^{A0}$&	-0.3816&	0.2939&	-0.09207&	-0.006694\\
$a_1^{A0}$&	0.03317&	-0.4783&	0.1869&	0.01545\\
$a_2^{A0}$&	0.1246&	-0.2087&	-0.01677&	0.004254\\
$a_3^{A0}$&	0.005469&	-0.001225&	-0.03466&	-0.004052\\
\hline
\end{tabular}
\end{table}\begin{table}
\caption{\label{corr_A0_V} Correlation matrix for $z$-expansion coefficients of $A0$ and $V$.}
\begin{tabular}{ c | c c c c }\hline
$\sigma^2 $& $a_0^{V}$	& $a_1^{V}$	& $a_2^{V}$	& $a_3^{V}$	\\\hline
$a_0^{A0}$&	-0.002394&	-0.01266&	-0.000435&	-2.199e-05\\
$a_1^{A0}$&	0.003344&	0.01268&	0.000509&	2.598e-05\\
$a_2^{A0}$&	-0.001888&	0.003397&	0.0003099&	1.767e-05\\
$a_3^{A0}$&	0.0001465&	0.0001454&	2.376e-05&	1.374e-06\\
\hline
\end{tabular}
\end{table}\begin{table}
\caption{\label{corr_A1} Correlation matrix for $z$-expansion coefficients of $A1$.}
\begin{tabular}{ c | c c c c }\hline
$\sigma^2 $& $a_0^{A1}$	& $a_1^{A1}$	& $a_2^{A1}$	& $a_3^{A1}$	\\\hline
$a_0^{A1}$&	1.0&	-0.03043&	-0.01583&	-0.01588\\
$a_1^{A1}$&	-0.03043&	1.0&	-0.3144&	0.02958\\
$a_2^{A1}$&	-0.01583&	-0.3144&	1.0&	-0.09184\\
$a_3^{A1}$&	-0.01588&	0.02958&	-0.09184&	1.0\\
\hline
\end{tabular}
\end{table}\begin{table}
\caption{\label{corr_A1_A2} Correlation matrix for $z$-expansion coefficients of $A1$ and $A2$.}
\begin{tabular}{ c | c c c c }\hline
$\sigma^2 $& $a_0^{A2}$	& $a_1^{A2}$	& $a_2^{A2}$	& $a_3^{A2}$	\\\hline
$a_0^{A1}$&	0.3327&	-0.09657&	0.066&	0.01124\\
$a_1^{A1}$&	0.4326&	0.04303&	-0.1541&	-0.02324\\
$a_2^{A1}$&	-0.3016&	0.5087&	0.485&	0.05027\\
$a_3^{A1}$&	-0.005564&	-0.007639&	0.1149&	0.01492\\
\hline
\end{tabular}
\end{table}\begin{table}
\caption{\label{corr_A1_V} Correlation matrix for $z$-expansion coefficients of $A1$ and $V$.}
\begin{tabular}{ c | c c c c }\hline
$\sigma^2 $& $a_0^{V}$	& $a_1^{V}$	& $a_2^{V}$	& $a_3^{V}$	\\\hline
$a_0^{A1}$&	0.008341&	0.009546&	0.0007997&	4.367e-05\\
$a_1^{A1}$&	0.01089&	0.02048&	0.0008922&	4.38e-05\\
$a_2^{A1}$&	-0.000702&	0.01654&	0.0008034&	4.176e-05\\
$a_3^{A1}$&	-0.001079&	0.0005229&	1.918e-06&	-2.31e-08\\
\hline
\end{tabular}
\end{table}\begin{table}
\caption{\label{corr_A2} Correlation matrix for $z$-expansion coefficients of $A2$.}
\begin{tabular}{ c | c c c c }\hline
$\sigma^2 $& $a_0^{A2}$	& $a_1^{A2}$	& $a_2^{A2}$	& $a_3^{A2}$	\\\hline
$a_0^{A2}$&	1.0&	-0.6033&	0.1237&	-0.001915\\
$a_1^{A2}$&	-0.6033&	1.0&	-0.2094&	0.01039\\
$a_2^{A2}$&	0.1237&	-0.2094&	1.0&	-0.07082\\
$a_3^{A2}$&	-0.001915&	0.01039&	-0.07082&	1.0\\
\hline
\end{tabular}
\end{table}\begin{table}
\caption{\label{corr_A2_V} Correlation matrix for $z$-expansion coefficients of $A2$ and $V$.}
\begin{tabular}{ c | c c c c }\hline
$\sigma^2 $& $a_0^{V}$	& $a_1^{V}$	& $a_2^{V}$	& $a_3^{V}$	\\\hline
$a_0^{A2}$&	0.01155&	-0.025&	-0.000755&	-3.602e-05\\
$a_1^{A2}$&	-0.002784&	0.04492&	0.001724&	8.523e-05\\
$a_2^{A2}$&	0.003861&	-0.002808&	0.0001025&	7.013e-06\\
$a_3^{A2}$&	0.0008794&	4.2e-05&	2.166e-05&	1.227e-06\\
\hline
\end{tabular}
\end{table}\begin{table}
\caption{\label{corr_V} Correlation matrix for $z$-expansion coefficients of $V$.}
\begin{tabular}{ c | c c c c }\hline
$\sigma^2 $& $a_0^{V}$	& $a_1^{V}$	& $a_2^{V}$	& $a_3^{V}$	\\\hline
$a_0^{V}$&	1.0&	-0.4502&	0.0155&	0.0007512\\
$a_1^{V}$&	-0.4502&	1.0&	-0.1043&	-0.003208\\
$a_2^{V}$&	0.0155&	-0.1043&	1.0&	-0.0001705\\
$a_3^{V}$&	0.0007512&	-0.003208&	-0.0001705&	1.0\\
\hline
\end{tabular}
\end{table}

\section{BGL Parameters and Unitarity Check}
\subsection{Conversion to BGL Scheme}
\label{fullBGL}
We may also use our results to determine the parameters entering the BGL parameterisation. This allows us to check the unitarity constraints $\sum (a^\mathrm{BGL}_n)^2\leq 1$ To do this we convert our form factors to the helicity basis, given by~\cite{Cohen:2018dgz}:
\begin{align}\label{bglffactors}
g=&\frac{2}{M_{B_s}+M_{D_s^*}}V \nonumber\\
f=&(M_{B_s}+M_{D_s^*})A_1 \nonumber\\
F_1=&\frac{M_{B_s}+M_{D_s^*}}{M_{D_s^*}}\Big[ - \frac{2M_{B_s}^2 |{\vec{p}}'|^2}{(M_{B_s}+M_{D_s^*})^2}A_2 \nonumber\\
&- \frac{1}{2}(t-M_{B_s}^2+M_{D_s^*}^2)A_1\Big] \nonumber\\
F_2=&2A_0
\end{align}
where ${\vec{p}}'$ is the $D_s^*$ spatial momentum in the $B_s$ rest frame. The BGL scheme then parameterises these form factors using the expansion in $z$ space,
\begin{equation}
F(t)=\frac{1}{P(t)\phi(t,t_0)}\sum_{n=0}^\infty a^\mathrm{BGL}_{n}z(t,t_0)^n\label{bglfit}
\end{equation}
where the pole function $P_i$ is the same as we have defined in Eq.~(\ref{poleformeq}) and the outer functions, $\phi$, are defined in~\cite{Cohen:2018dgz}. We use the resonance masses given in~\cite{Cohen:2018dgz} in the pole functions $P_i$. In order to compute the outer functions we use the values of $\chi_{L(T)}(\pm u)$ computed in~\cite{Harrison:2020gvo}.
We use our results to output form factor values in the helicity basis at a large number of $q^2$ values, which we subsequently fit using Eq.~(\ref{bglfit}) truncated at $n=3$. Our results, expressed in terms of the BGL parameterisation, are given in Table~\ref{bglparams}, together with the unitarity bounds which we find to be far from saturation.

\begin{table}
\caption{\label{bglparams}BGL parameters computed by converting the physical continuum results computed using Eq.~(\ref{fitfunctionequation}) to the BGL scheme, Eq.~(\ref{bglfit}).}
\begin{tabular}{ c | c c c c c}\hline
& $a^\mathrm{BGL}_0$	& $a^\mathrm{BGL}_1$	& $a^\mathrm{BGL}_2$	& $a^\mathrm{BGL}_3$	&$\sum (a^\mathrm{BGL}_n)^2$ \\\hline
$F_1$&0.002402(90)&-0.0018(44)&-0.041(97)&0.04(82)&0.003(74)\\
$F_2$&0.0384(21)&-0.077(45)&-0.25(42)&0.0(1.0)&0.07(21)\\
$f$&0.01420(53)&-0.019(14)&-0.0(0.2)&0.0(1.0)&0.00055(84)\\
$g$&0.0300(33)&0.0(0.08)&-0.04(54)&0.0(1.0)&0.002(41)\\\hline
\end{tabular}
\end{table}

\subsection{Comparison to $B\to D^*$ BGL Coefficients}

As well as checking that the unitarity constraints are satisfied by the BGL parameters for $B_s\to D_s^*$ it is also useful and interesting to reparameterise our results using the BGL scheme for $B\to D^*$. This allows us to compare our results to those for $B\to D^*$, and also allows for our results to be more readily incorporated into analyses which assume $SU(3)_\mathrm{flav}$ symmetry.

The BGL coefficients are sensitive to the meson masses through the outer functions and through the definition of $z(q^2,t_0,t_+)$. In order to reparameterise our results using the BGL scheme for $B\to D^*$ we first convert our results to the helicity basis, using the $B_s$ and $D_s^*$ masses (as in Eq.~(\ref{bglffactors})), and then fit these to the BGL parameterisation for $B\to D^*$, in which $z(q^2,t_0,t_+)$ and the outer functions are computed using the $B$ and $D^*$ masses. In order to compare to the BGL coefficients for $B\to D^*$ determined in~\cite{Bigi:2017jbd,Bigi:2017njr} we use the masses and values of $\chi_{L(T)}(\pm u)$ given in Tables~IV and~III of~\cite{Bigi:2017jbd}. \cite{Bigi:2017jbd} and~\cite{Bigi:2017njr} both use the same definitions for the outer functions, given in~\cite{Boyd:1997kz}, which we also use here, as well as choosing $t_0=t_-$. Note that these definitions differ from those used in Appendix~\ref{fullBGL}, from~\cite{Cohen:2018dgz}, where the $t_+$ is given by the true pair production threshold in each channel (e.g. the threshold for $BD$ production for the vector channel and the threshold for $B^*D$ production for the axial-vector and pseudoscalar channels), whereas in~\cite{Boyd:1997kz} and~\cite{Bigi:2017jbd,Bigi:2017njr} it is taken to be $(M_{D^*} + M_B)^2$ for all channels. Note also that since our results for $B_s\to D_s^*$ satisfy the kinematical constraint Eq.~(\ref{eq:contconstraint}), involving the $B_s$ and $D_s^*$ masses, the equivalent constraint for $B\to D^*$ will not be satisfied and so must be imposed. We must also impose the constraint $F_1(q^2_\mathrm{max})=(M_{B}-M_{D^*})f(q^2_\mathrm{max})$ which will not otherwise be satisfied. We impose these constraints explicitly by using them to fix the zeroth-order coefficients of $F_2$ and $F_1$.

We use a BGL parameterisation including up to $z^2$ terms for $f,~F_1$ and $g$, and up to $z$ for $F_2$. The coefficients are given in Table~\ref{bglbdstarcomp}, where we see that our results are consistent with the BGL parameters for $B\to D^*$. The correlation matrix for our results in this parameterisation scheme is given in Table~\ref{bglbdstarcorr}. We conclude that there is no significant effect on the form factors evident in our results with a strange spectator quark compared to those with an up/down one.

\begin{table}
\caption{\label{bglbdstarcomp}BGL parameters for our results determined using the BGL parameterisation for $B\to D^*$, compared to the BGL parameters extracted from experimental data, lattice results and light cone sum rules for $B\to D^*$ in~\cite{Bigi:2017njr}.}
\begin{tabular}{ c | c | c }\hline
   BGL Fit: 	 & This work, $B_s\to D_s^*$ 		& $B\to D^*$\cite{Bigi:2017njr} \\\hline
$a_0^{f}$	 & 	0.01258(47)	&  0.01224(18)  \\
$a_1^{f}$	 & 	-0.008(15)	&  -0.052(+27,-15)  \\
$a_2^{f}$	 & 	-0.03(22)	&  1.0(+0,-5)  \\\hline
$a_1^{F_1}$	 & 	0.00009(402)	&  -0.0070(+54,-52)  \\
$a_2^{F_1}$	 & 	-0.066(61)	&  0.089(+96,-100)  \\\hline
$a_0^{g}$	 & 	0.0339(37)	&  0.0289(+57,-37)  \\
$a_1^{g}$	 & 	0.005(100)	&  0.08(+8,-22)  \\
$a_2^{g}$	 & 	-0.005(315)	&  -1.0(+2.0,-0)  \\\hline
$a_1^{F_2}$	 & 	0.0108(50)	&   $-$ \\\hline
\end{tabular}
\end{table}

\begin{table*}
\caption{\label{bglbdstarcorr}Correlation matrix for our BGL parameters determined using the BGL parameterisation for $B\to D^*$, given in Table~\ref{bglbdstarcomp}.}
\begin{tabular}{ c | c c c c c c c c c }\hline
		 & $a_0^{f}$	 & $a_1^{f}$	 & $a_2^{f}$	 & $a_1^{F_1}$	 & $a_2^{F_1}$	 & $a_0^{g}$	 & $a_1^{g}$	 & $a_2^{g}$	 & $a_1^{F_2}$ \\\hline
$a_0^{f}$	 & 	1.0	 & 	-0.097	 & 	0.057	 & 	-0.192	 & 	0.031	 & 	0.008	 & 	0.011	 & 	0.006	 & 	0.127\\
$a_1^{f}$	 & 		 & 	1.0	 & 	-0.546	 & 	0.191	 & 	-0.204	 & 	0.014	 & 	0.011	 & 	-0.043	 & 	-0.033\\
$a_2^{f}$	 & 		 & 		 & 	1.0	 & 	-0.051	 & 	0.007	 & 	-0.009	 & 	0.024	 & 	0.081	 & 	-0.059\\
$a_1^{F_1}$	 & 		 & 		 & 		 & 	1.0	 & 	-0.737	 & 	-0.002	 & 	0.036	 & 	-0.016	 & 	0.177\\
$a_2^{F_1}$	 & 		 & 		 & 		 & 		 & 	1.0	 & 	-0.001	 & 	-0.043	 & 	0.019	 & 	0.414\\
$a_0^{g}$	 & 		 & 		 & 		 & 		 & 		 & 	1.0	 & 	-0.382	 & 	0.006	 & 	0.002\\
$a_1^{g}$	 & 		 & 		 & 		 & 		 & 		 & 		 & 	1.0	 & 	-0.087	 & 	0.014\\
$a_2^{g}$	 & 		 & 		 & 		 & 		 & 		 & 		 & 		 & 	1.0	 & 	0.007\\
$a_1^{F_2}$	 & 		 & 		 & 		 & 		 & 		 & 		 & 		 & 		 & 	1.0\\\hline		
\end{tabular}
\end{table*}

\section{Stability of our Form Factors Under Variations of the Fit}
\label{variations}

\subsection{Variations of Correlator Fits}
\label{qsqstabappendix}
\begin{figure}
\centering
\includegraphics[scale=0.4]{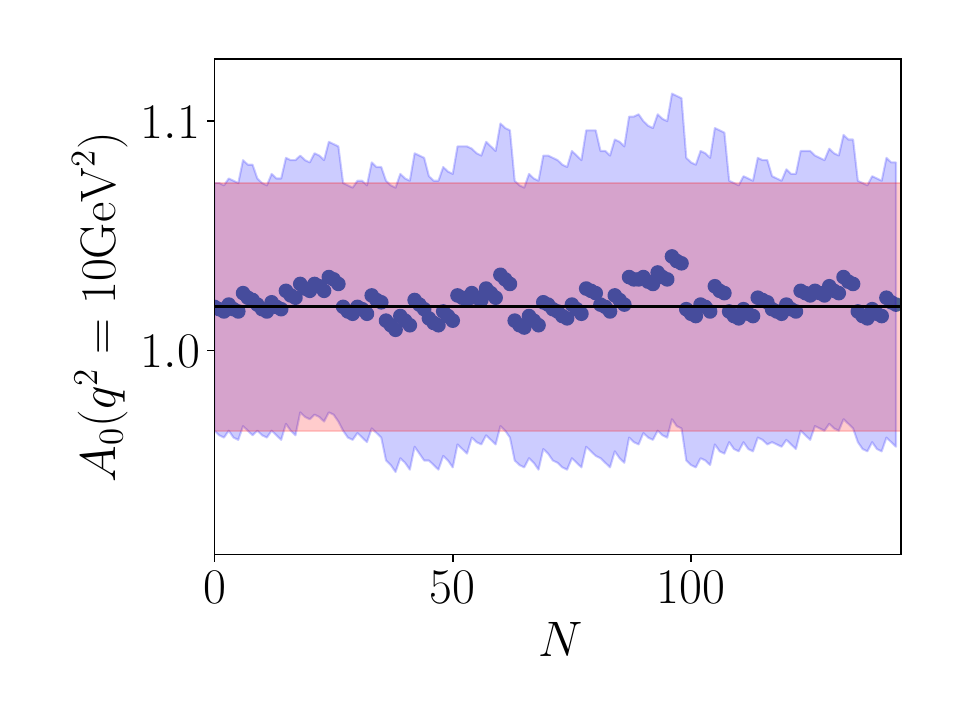}
\includegraphics[scale=0.4]{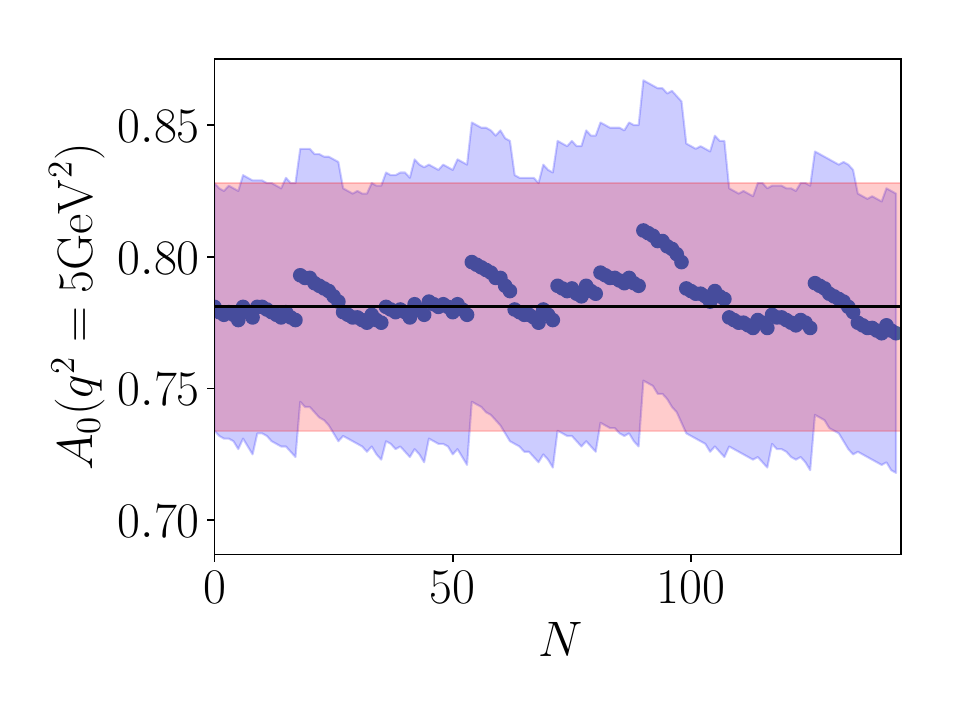}
\includegraphics[scale=0.4]{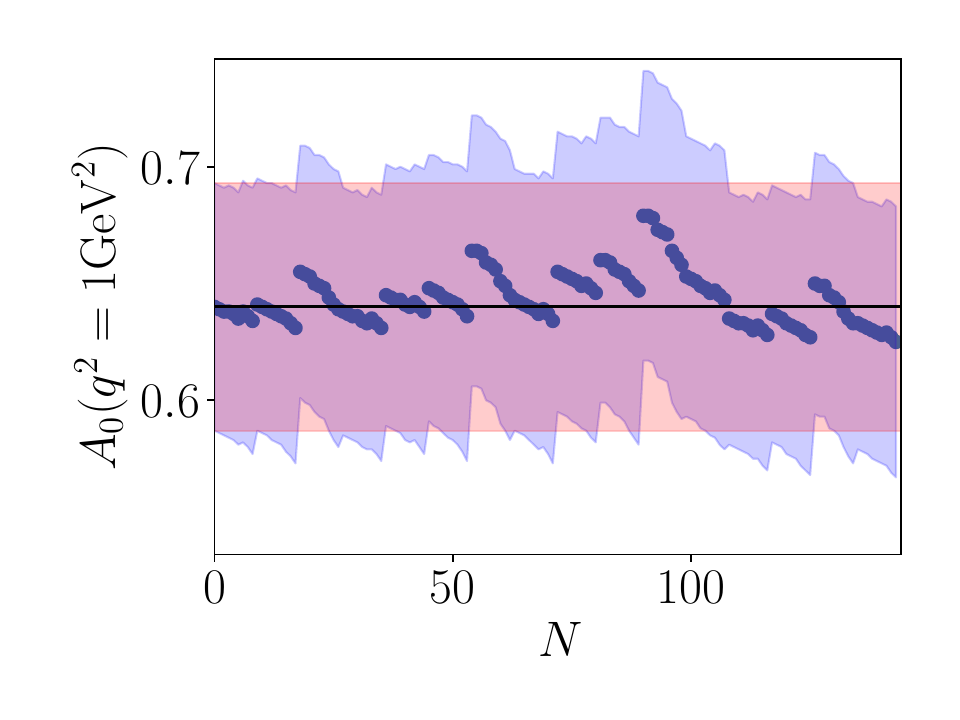}
\caption{\label{A0qsqpoints} 
As for Figure~\ref{gammastabplot} showing the stability of the 
form factor $A_0$ evaluated at $q^2=1~\mathrm{GeV}^2,5~\mathrm{GeV}^2,10~\mathrm{GeV}^2$.}
\end{figure}

\begin{figure}
\centering
\includegraphics[scale=0.4]{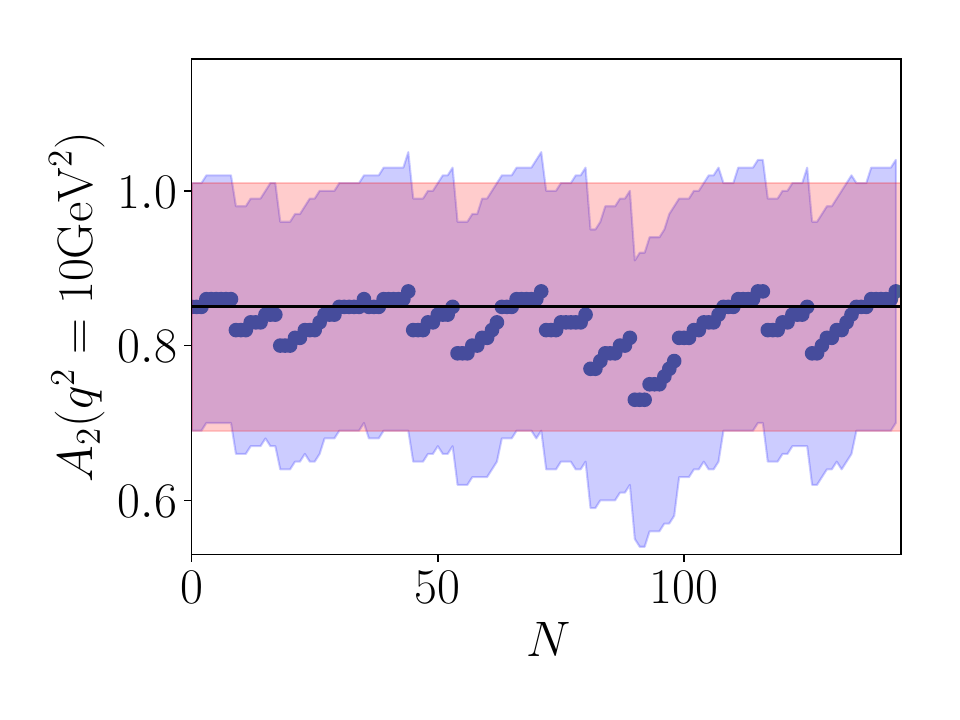}
\includegraphics[scale=0.4]{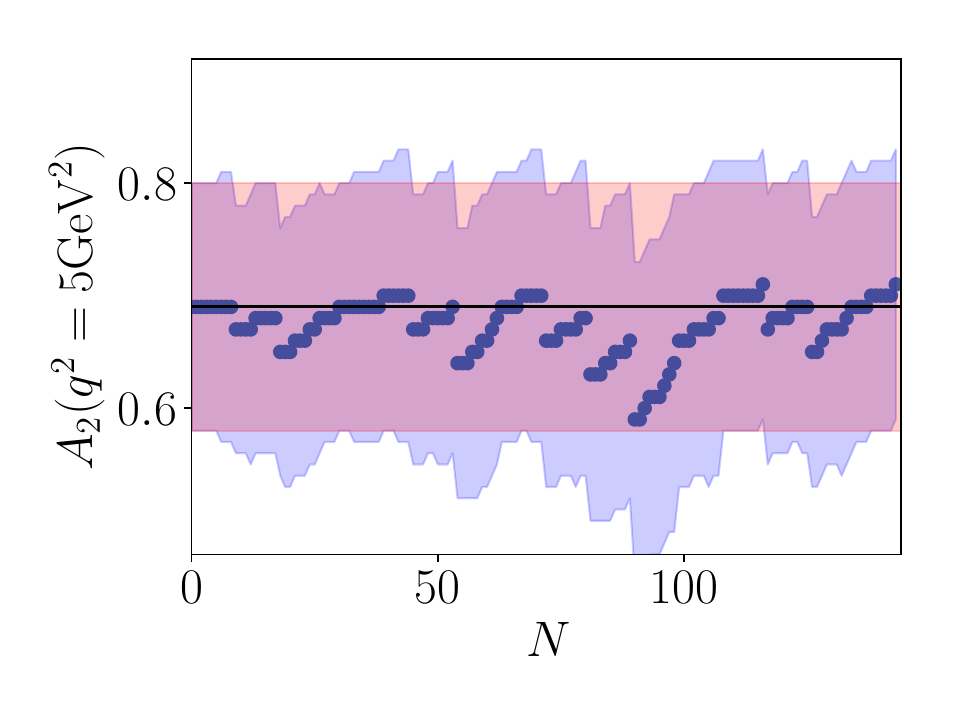}
\includegraphics[scale=0.4]{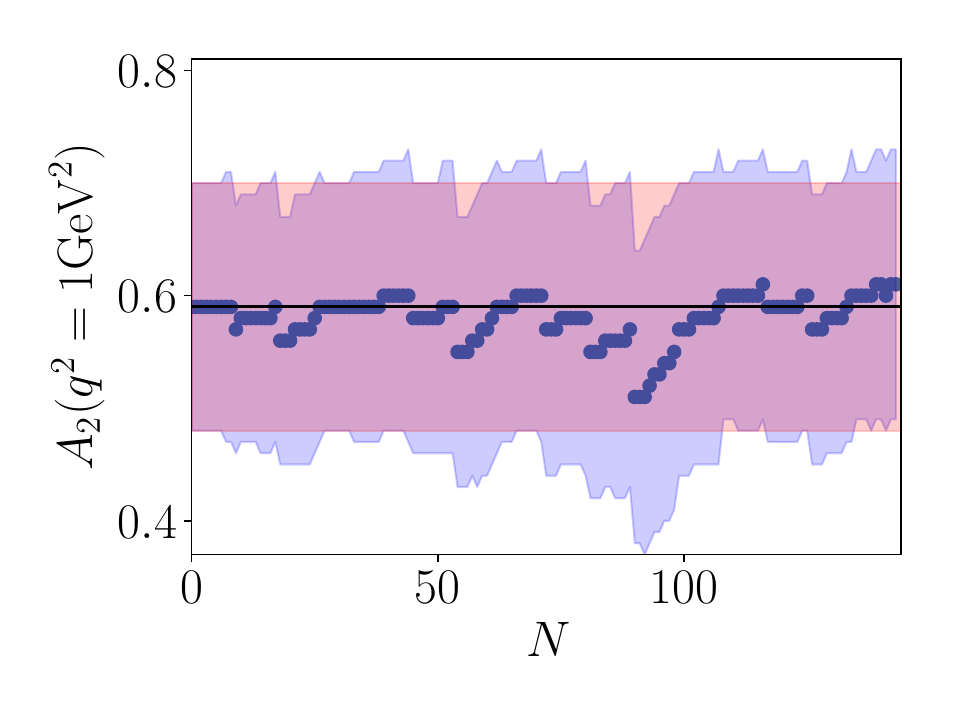}
\caption{\label{A2qsqpoints} 
As for Figure~\ref{gammastabplot} showing the stability of the 
form factor $A_2$ evaluated at $q^2=1~\mathrm{GeV}^2,5~\mathrm{GeV}^2,10~\mathrm{GeV}^2$.}
\end{figure}

In Figure~\ref{A0qsqpoints} and Figure~\ref{A2qsqpoints} we show the variation of the form factors $A_0$ and $A_2$, evaluated at $q^2=1~\mathrm{GeV}^2,5~\mathrm{GeV}^2,10~\mathrm{GeV}^2$, as in Figure~\ref{gammastabplot}. In these plots we see that these form factors do not change significantly as a result of varying the correlator fit inputs.

\subsection{Order of Expansion}

\begin{figure}
\centering
\includegraphics[scale=0.275]{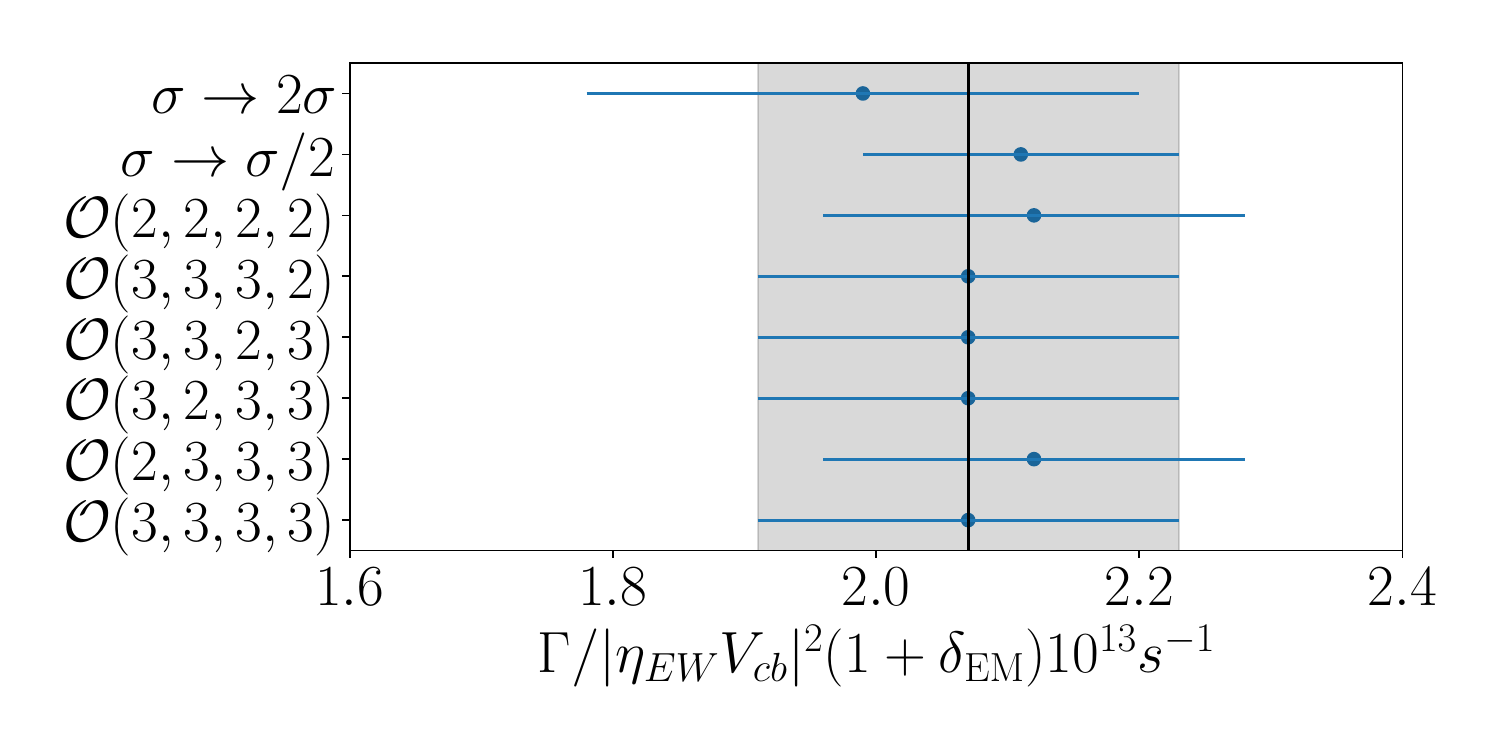}
\includegraphics[scale=0.275]{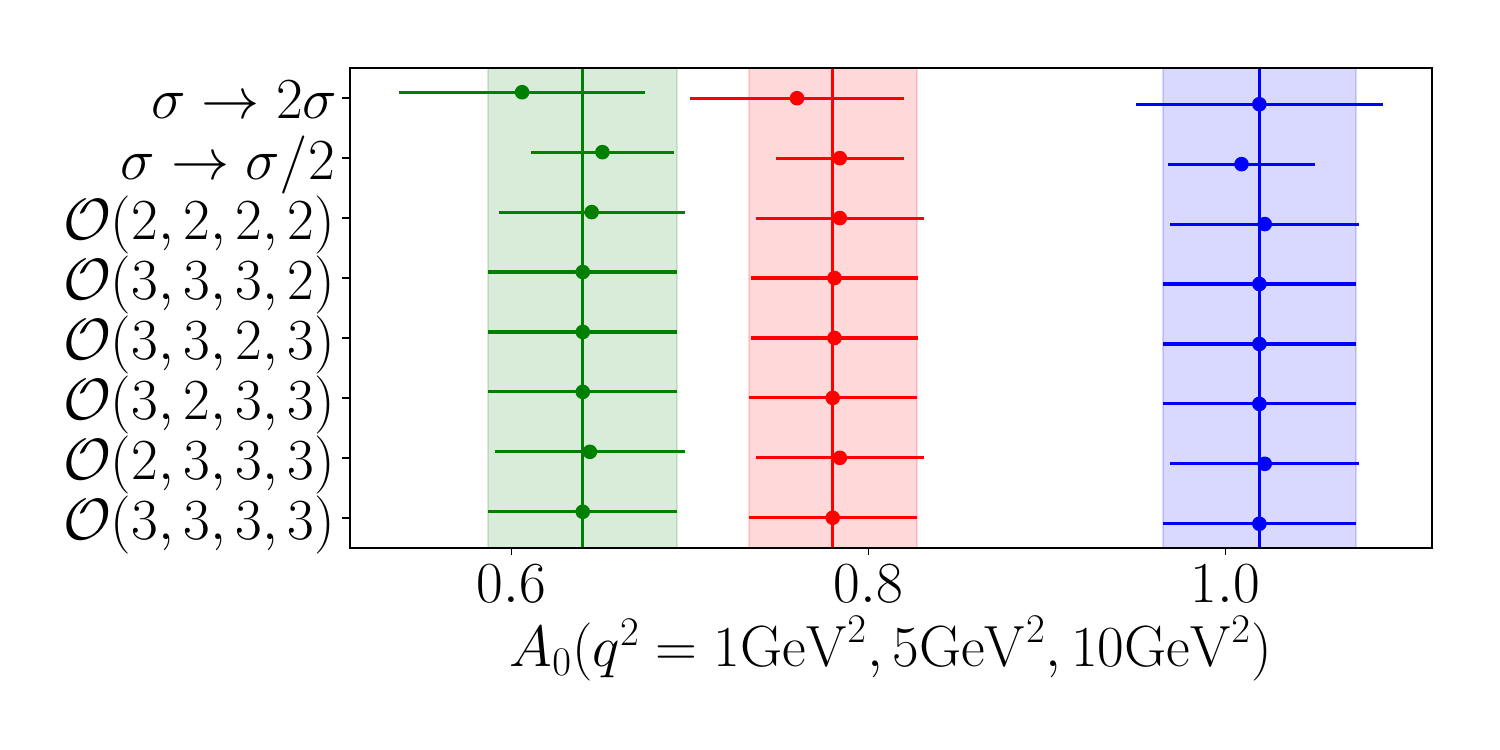}
\includegraphics[scale=0.275]{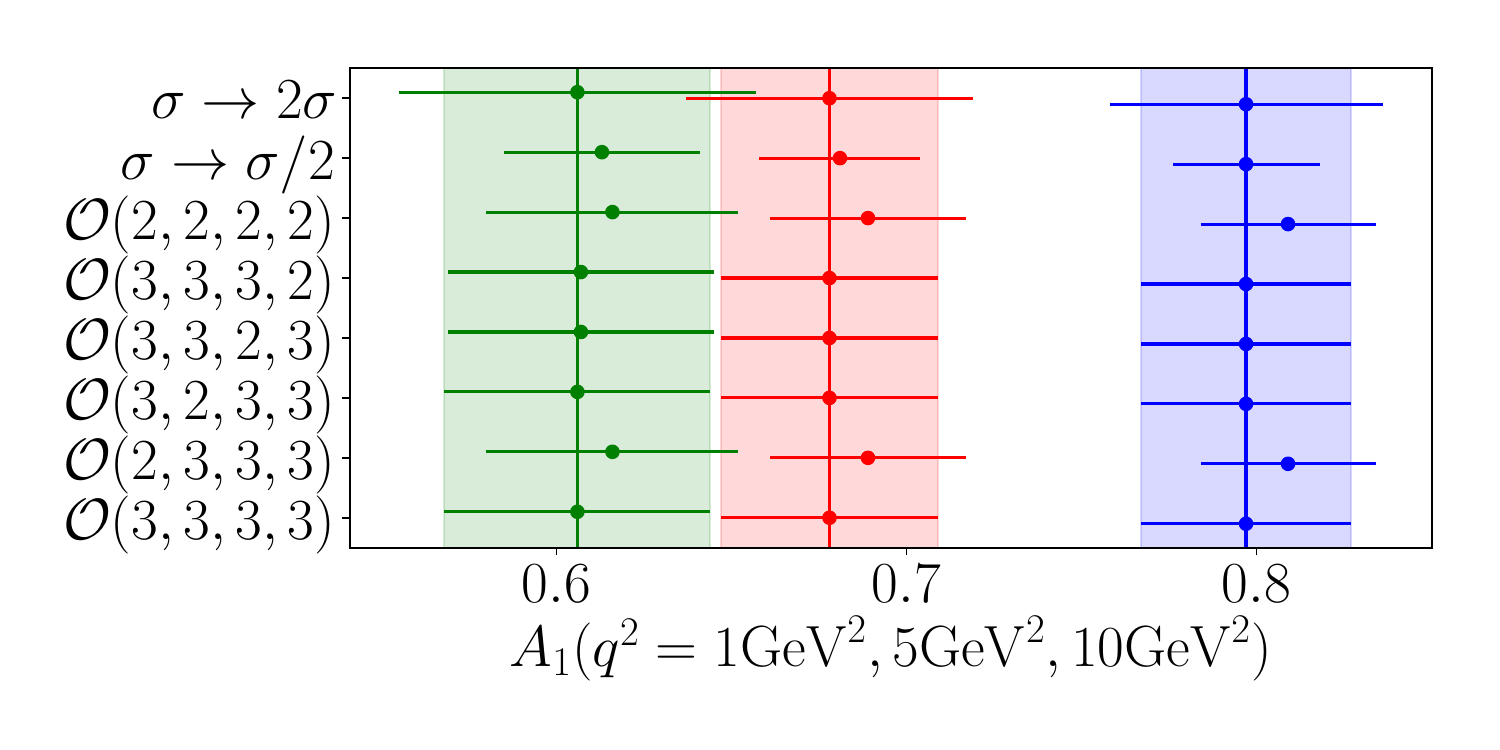}
\includegraphics[scale=0.275]{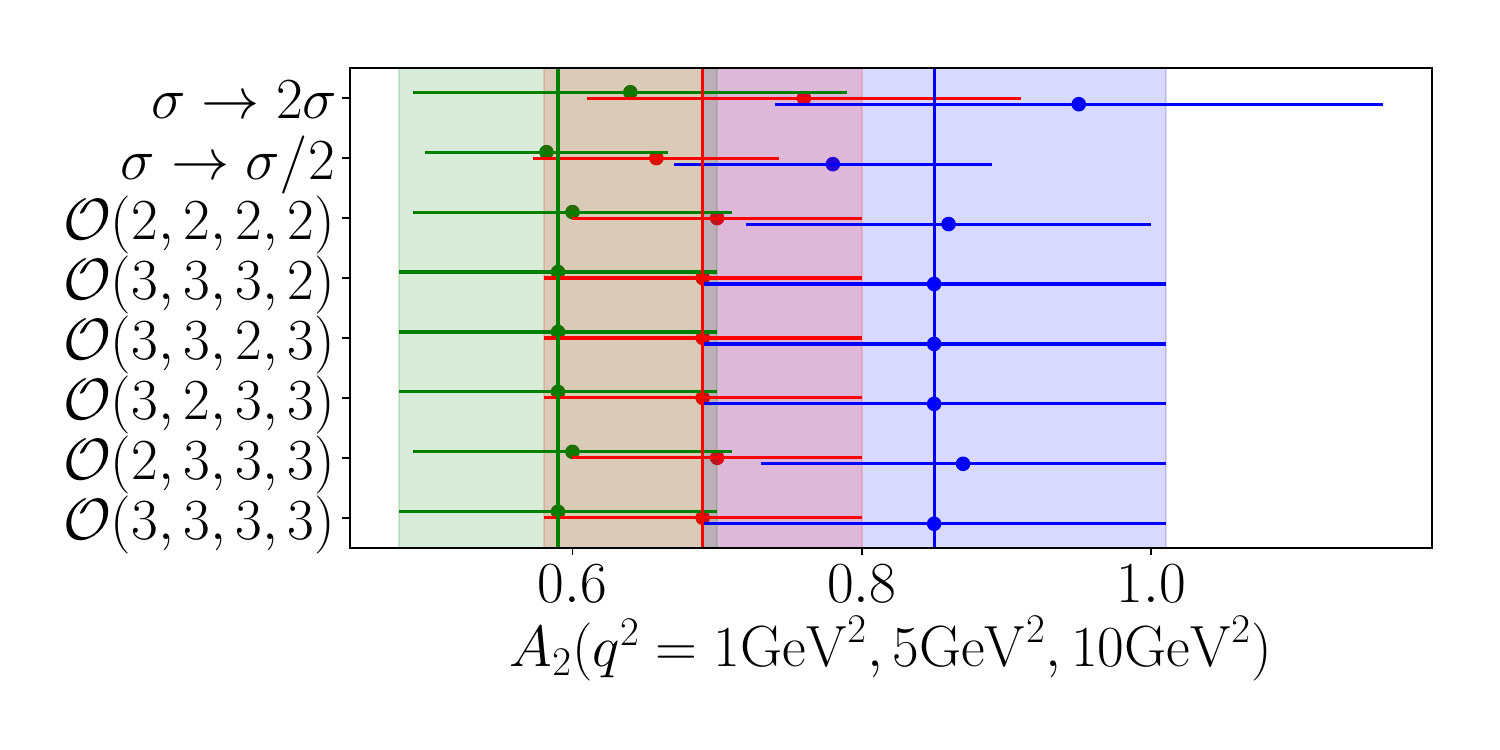}
\includegraphics[scale=0.275]{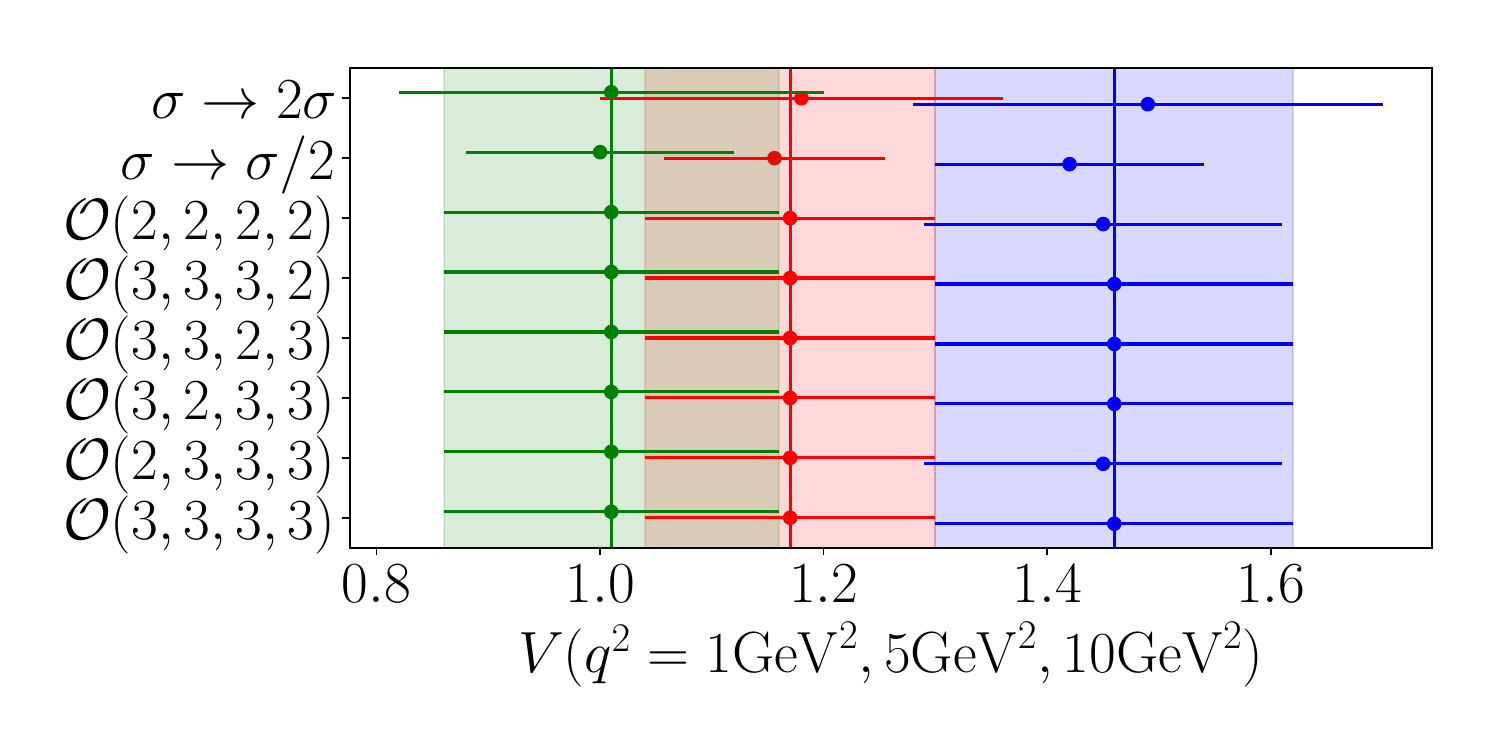}
\caption{\label{ntermstabplot} Plot showing the stability of the total rate 
for $B_s^0\to D_s^{*-}\mu^+{\nu}_\mu$ considering lower order truncations of $z$-expansion, discretisation and heavy mass dependent terms in Eq.~(\ref{fitfunctionequation}) and Eq.~(\ref{eq:anfitform}). $\mathcal{O}({n_1},{n_2},{n_3},{n_4})$ corresponds to the result including terms of highest order $\mathcal{O}( (2\Lambda/M_{\eta_h})^{n_1}, (am_c)^{2n_2}, (am_h)^{2n_3}, z^{n_4})$. The vertical black line is our final result, corresponding to $\mathcal{O}({3},{3},{3},{3})$, and the grey band is its uncertainty. We also include variations in which we multiply our prior widths either by a factor of 2 or 0.5, labelled as $0(\sigma)\rightarrow 0(\sigma\times 2)$ and $0(\sigma)\rightarrow 0(\sigma/ 2)$ respectively. Our result for the total rate is very stable to these variations. Note that here we do not include the contribution of $\delta_\mathrm{EM}$ to the uncertainty. We also include similar plots for the form factors, evaluated at $q^2=1~\mathrm{GeV}^2,5~\mathrm{GeV}^2,10~\mathrm{GeV}^2$ plotted in green, red and blue respectively.}
\end{figure}

We also investigate the effects of including fewer $z$ terms in Eq.~(\ref{fitfunctionequation}) as well as fewer $am_c$, $am_h$ and $2\Lambda_{QCD}/M_{\eta_h}$ terms in Eq.~(\ref{eq:anfitform}). Figure~\ref{ntermstabplot} gives the total width, as well as values of the form factors at several values of $q^2$, obtained using these variations, where we see that our results are insensitive to the removal of the highest order terms. We also investigate the effect of increasing or decreasing the prior widths of the parameters $b_n^{ijk}$ in Eq.~(\ref{eq:anfitform}) by a factor of 2. These results are also shown in Figure~\ref{ntermstabplot} where, as in~\cite{Harrison:2020gvo}, we see only a very small effect on the central values of our results. .

\subsection{Variation of Pole Term}

\begin{figure}
\centering
\includegraphics[scale=0.3]{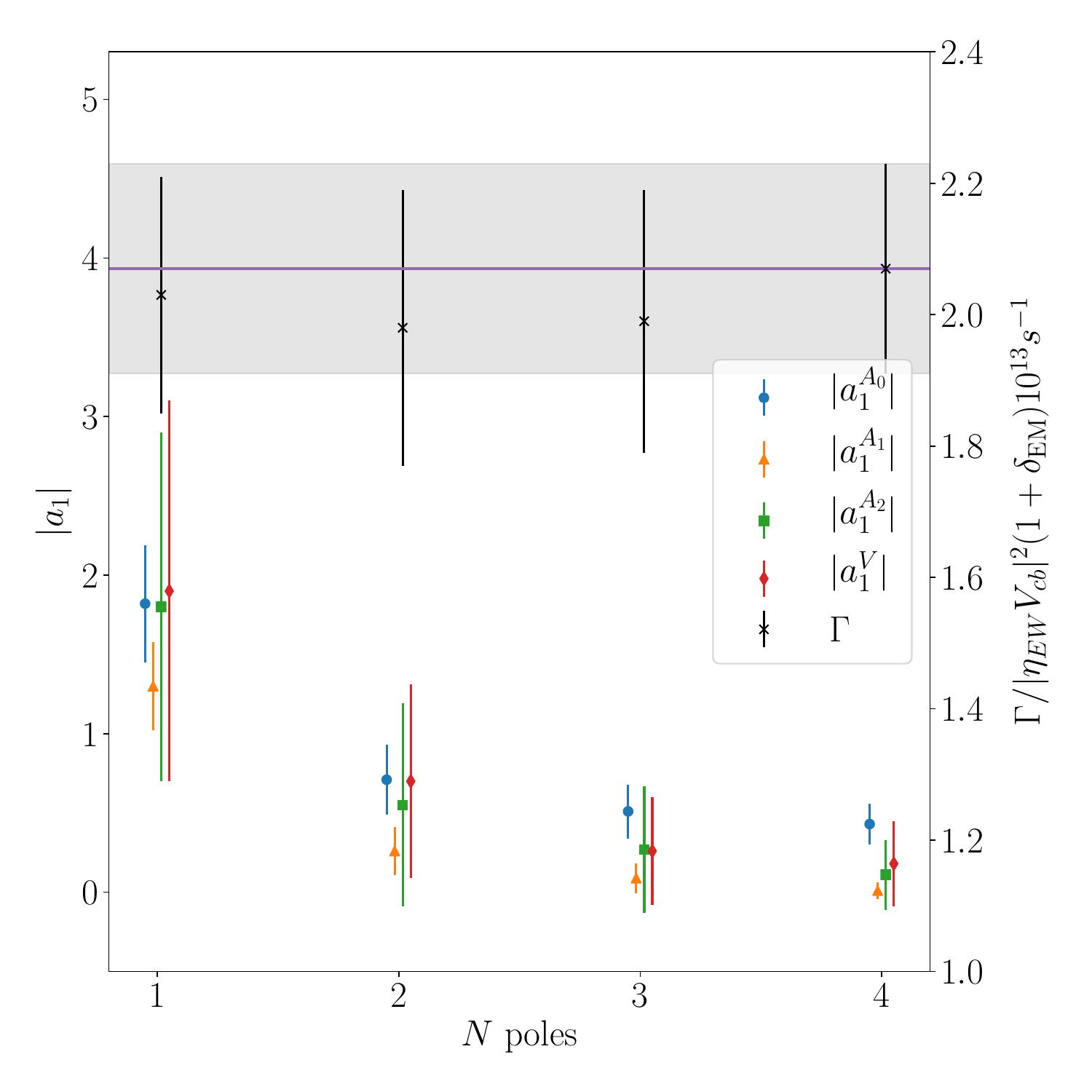}
\caption{\label{NPOLESPLOT} Magnitude of the $\mathcal{O}(z)$ coefficient, $a_1$, for each 
form factor plotted against the number of poles included in Eq.~(\ref{poleformeq}). 
The prior widths on the $b_n^{ijk}$ are scaled according to the number of 
poles, see text.  Note that the maximum number of poles included for $A_0$ is 3. 
The black crosses and error bars give the total width for the $\ell=\mu$ case, $\Gamma/|\eta_{\mathrm{EW}}V_{cb}|^2$, 
determined from that fit, using the right-hand $y$ axis. 
The grey band corresponds to our final result for the total width 
using $N_\mathrm{poles}=4$, and prior values for $b_n^{ijk}$ of $0(1)$. This 
shows how the different coefficients as a function of $N_\mathrm{poles}$ give 
a very stable result for the total width. Note that here we do not include the contribution of $\delta_\mathrm{EM}$ to the uncertainty in $\Gamma/|\eta V_{cb}|^2$.}
\end{figure}

The pole function Eq.~(\ref{poleformeq}) includes the effects of subthreshold $b\bar{c}$ resonances in $q^2$. These begin at the square of the $B_c$ mass, $(6.275\mathrm{GeV})^2$, significantly above the maximum physical value of $q^2_\mathrm{max}\approx (3.25\mathrm{GeV})^2$.  As such we do not expect the exact positions or number of poles to have a large effect on the fits, although the choice of the number of poles to include will act as a normalisation, changing the magnitude of the coefficients $a_n$ appearing in Eq.~(\ref{fitfunctionequation}).
Here we investigate the effect of including fewer poles in Eq.~(\ref{poleformeq}) by repeating our analysis including only the first $N_\mathrm{poles}$ resonances listed in Table~\ref{poletab}. 

We take a prior width on the $z$-expansion 
coefficients of $5.0 - N_\mathrm{poles}$. 
We are able to obtain a good fit, with $\chi^2/\mathrm{dof}\approx 0.1$ in all cases.  
Since there are only 3 poles for $A_0$ expected below $t_+$, we include only 
3 poles for that form factor even in the $N_\mathrm{poles}$ case. 

Figure~\ref{NPOLESPLOT} shows these results, plotting against the left-hand $y$-axis 
the magnitude of the coefficient corresponding 
to the order $z$ term, $a_1$, coming from the fits as a function of the number of poles 
included.  Results are given for each form factor.
We see that as we include fewer poles, increasingly large $z$-expansion 
coefficients are needed partly in order to account for the 
removal of physical $q^2$ dependence from missing poles but also because of 
the normalisation change. 

\subsection{Inclusion of Outer Functions}

\begin{table}
\caption{\label{bglparamsextrapolated}BGL parameters computed using the full BGL scheme including outer functions for the physical continuum extrapolation using Eq.~(\ref{fitfuncbglfullmh}). Note that these values are in good agreement with those computed by converting the physical continuum results computed using Eq.~(\ref{fitfunctionequation}) to the BGL scheme, given in Table~\ref{bglparams}}
\begin{tabular}{ c | c c c c c}\hline
& $a^\mathrm{BGL'}_0$	& $a^\mathrm{BGL'}_1$	& $a^\mathrm{BGL'}_2$	& $a^\mathrm{BGL'}_3$	&$\sum (a^\mathrm{BGL'}_n)^2$ \\\hline
$F_1$&0.00235(10)&-0.0058(63)&-0.04(13)&-0.31(91)&0.10(56)\\
$F_2$&0.0357(33)&-0.122(79)&-0.20(87)&0.0(1.0)&0.06(34)\\
$f$&0.01388(62)&-0.017(26)&-0.10(54)&0.0(1.0)&0.01(11)\\
$g$&0.0329(57)&-0.03(11)&0.01(99)&0.0(1.0)&0.002(21)\\\hline
\end{tabular}
\end{table}

In order to investigate the effect of excluding the outer functions from Eq.~(\ref{fitfunctionequation}) we consider using the BGL parameterisation described in Appendix~\ref{fullBGL} for the extrapolation to the physical point of Section~\ref{sec:physextrap}. In order to do this, we first convert our lattice data for the form factors to the helicity basis given in Eq.~(\ref{bglffactors}). To evaluate the outer functions $\phi$, defined in~\cite{Cohen:2018dgz}, as we vary the heavy mass we must evaluate the function $\chi_{L(T)}(\pm u_h)$, defined in ~\cite{Boyd:1997kz}, at different values of $u_h$. 
As in~\cite{Harrison:2020gvo} we take $u_b=0.33$, $m_b^\mathrm{pole}=4.78$ and $\alpha_s=0.22$. We use $u_h=u_b\times M_{\eta_b}/M_{\eta_h}$ to approximate the heavy mass dependence of $u_h$, as well as using $m_h^\mathrm{pole} = m_b^\mathrm{pole}\times M_{\eta_h}/ M_{\eta_b^\mathrm{phys}}$. Our BGL fit function then has the form
\begin{equation}\label{fitfuncbglfullmh}
F(t)=\frac{1}{P(t)\phi(t,t_0)}\sum_{n=0}^\infty a^\mathrm{BGL'}_{n}z(t,t_0)^n
\end{equation}
with
\begin{equation}
a^\mathrm{BGL'}_{n} = \sum_{j,k,l=0}^3 b_{n}^{'jkl}\Delta_{h}^{(j)} \left(\frac{am_c^\mathrm{val}}{\pi}\right)^{2k} \left(\frac{am_h^\mathrm{val}}{\pi}\right)^{2l}\mathcal{N}_{n}.
\end{equation}
where $\mathcal{N}$ and $\Delta_{h}^{(j)}$ have the same definitions as in the main analysis discussed in Section~\ref{sec:physextrap}. We take $t_0=t_-=q^2_\mathrm{max}$, and use the same approximate form as in Section~\ref{sec:physextrap} for the variation of the pole masses with the heavy quark mass. We take the same priors as in Section~\ref{sec:physextrap} for the coefficients $b_{n}^{'jkl}$ and for those entering $\mathcal{N}_{n}$. In this fit we also impose the kinematical constraint at $q^2=0$, which here takes the form $2 F_1(0)-F_2(0)(M_{B_s}^2-M_{D_s^*}^2)=0$. Note that this is equivalent to the constraint $2M_{D_s^*}A_0(0) = (M_{D_s^*}+M_{H_s})A_1(0) + (M_{D_s^*}-M_{H_s})A_2(0)$ from the definitions of the helicity basis form factors given in Eq.~(\ref{bglffactors}). We also have, from the definitions Eq.~(\ref{bglffactors}), the additional condition that $F_1(q^2_\mathrm{max})=(M_{B_s}-M_{D_s^*})f(q^2_\mathrm{max})$. We impose these conditions both at the physical point and on each lattice including a nuisance term as in Section~\ref{sec:physextrap}. Note that here we neglect the running of $\alpha_s$ with heavy mass, allowing for this to be taken up elsewhere in the fit. The results of fitting our lattice data to Eq.~(\ref{fitfuncbglfullmh}) are given in Table~\ref{bglparamsextrapolated}, where we see that the BGL parameters from this fit are very close to those given in Table~\ref{bglparams} which were computed from the continuum form factors extrapolated to the physical continuum using Eq.~(\ref{fitfunctionequation}) without including outer functions. In both cases the unitarity bounds are far from saturation without the need to impose this constraint in our fits. This shows that the approach adopted here and in~\cite{Harrison:2020gvo}, of excluding the outer functions from the physical continuum extrapolation, is consistent with including them.

\begin{figure*}
\centering
\includegraphics[scale=0.215]{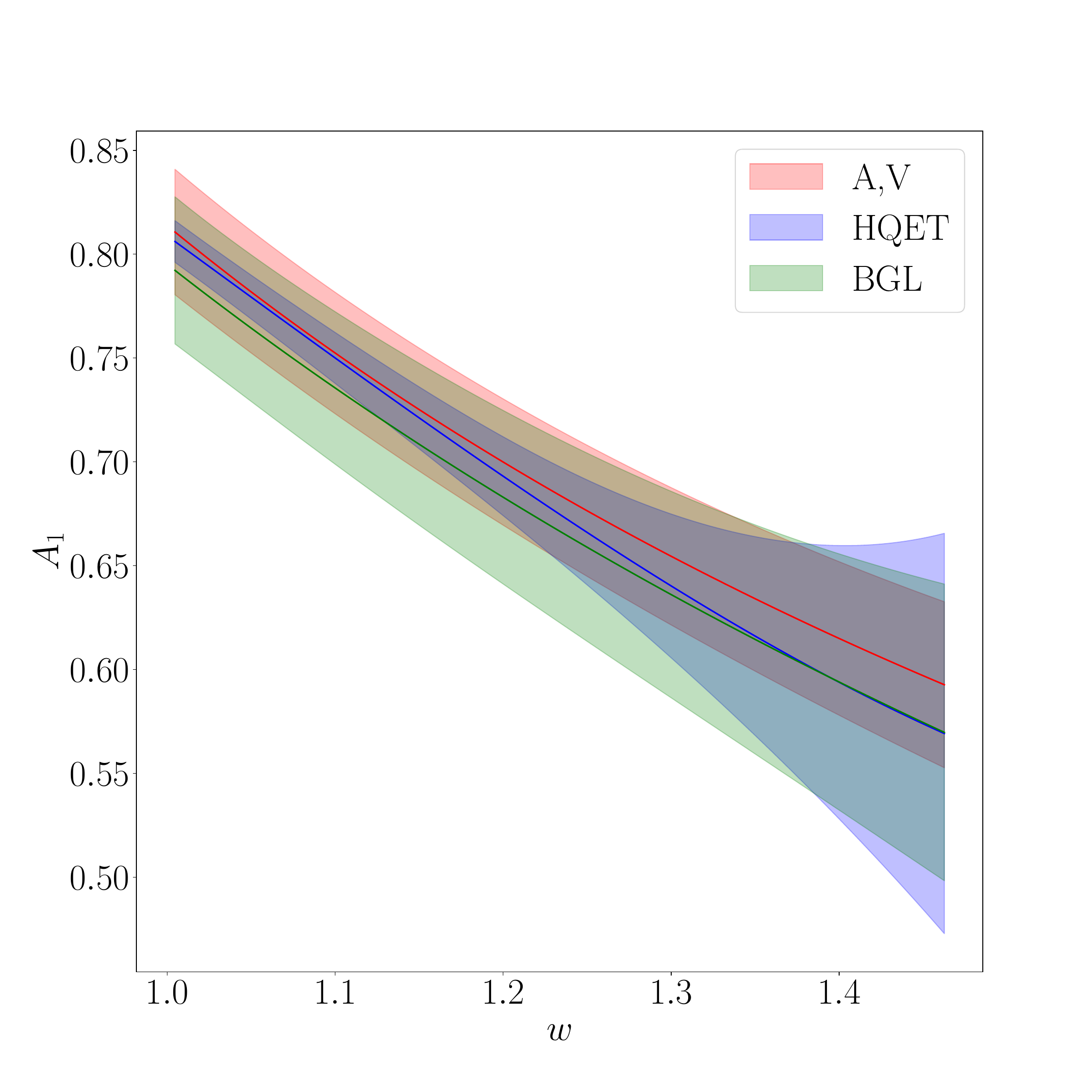}
\includegraphics[scale=0.215]{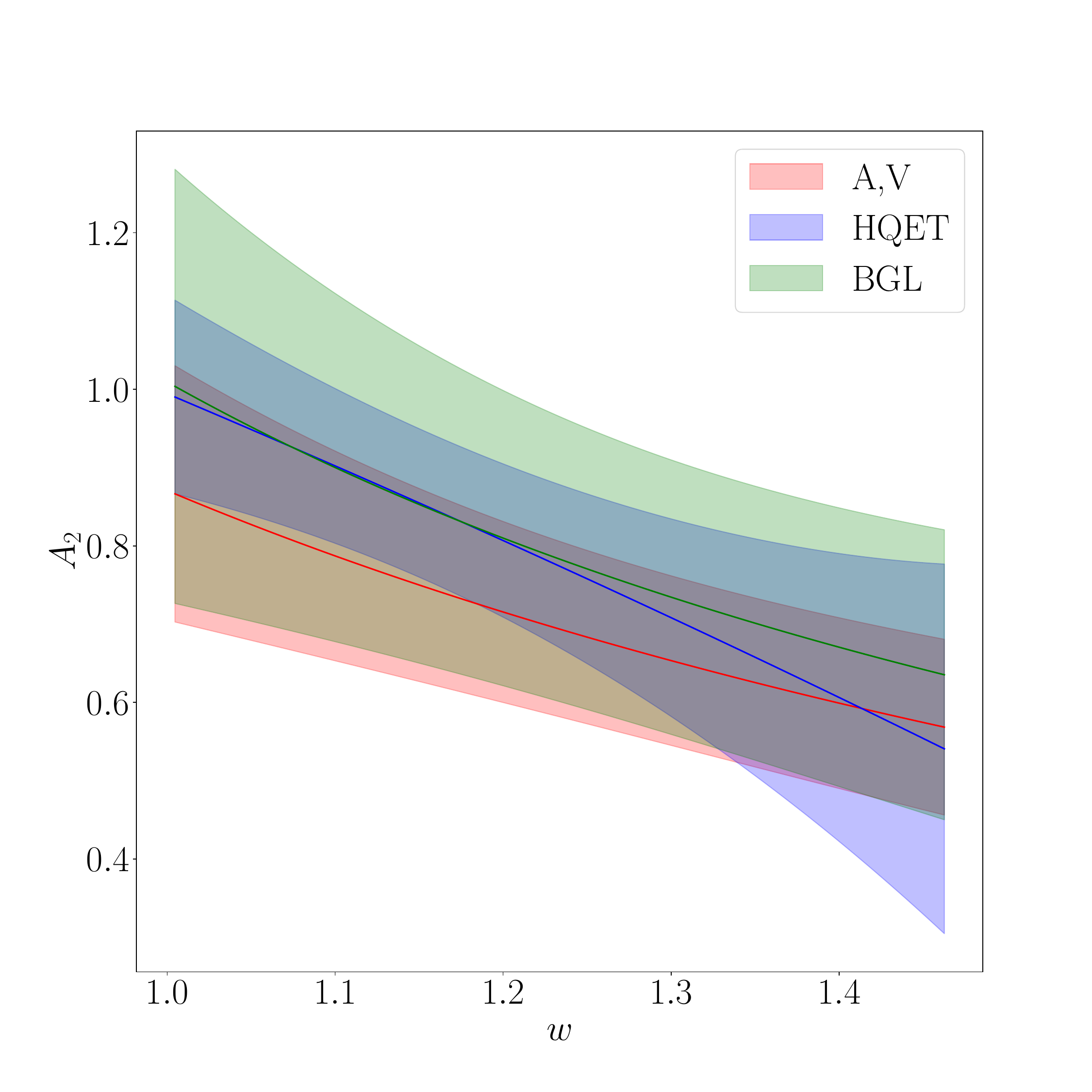}\\
\includegraphics[scale=0.215]{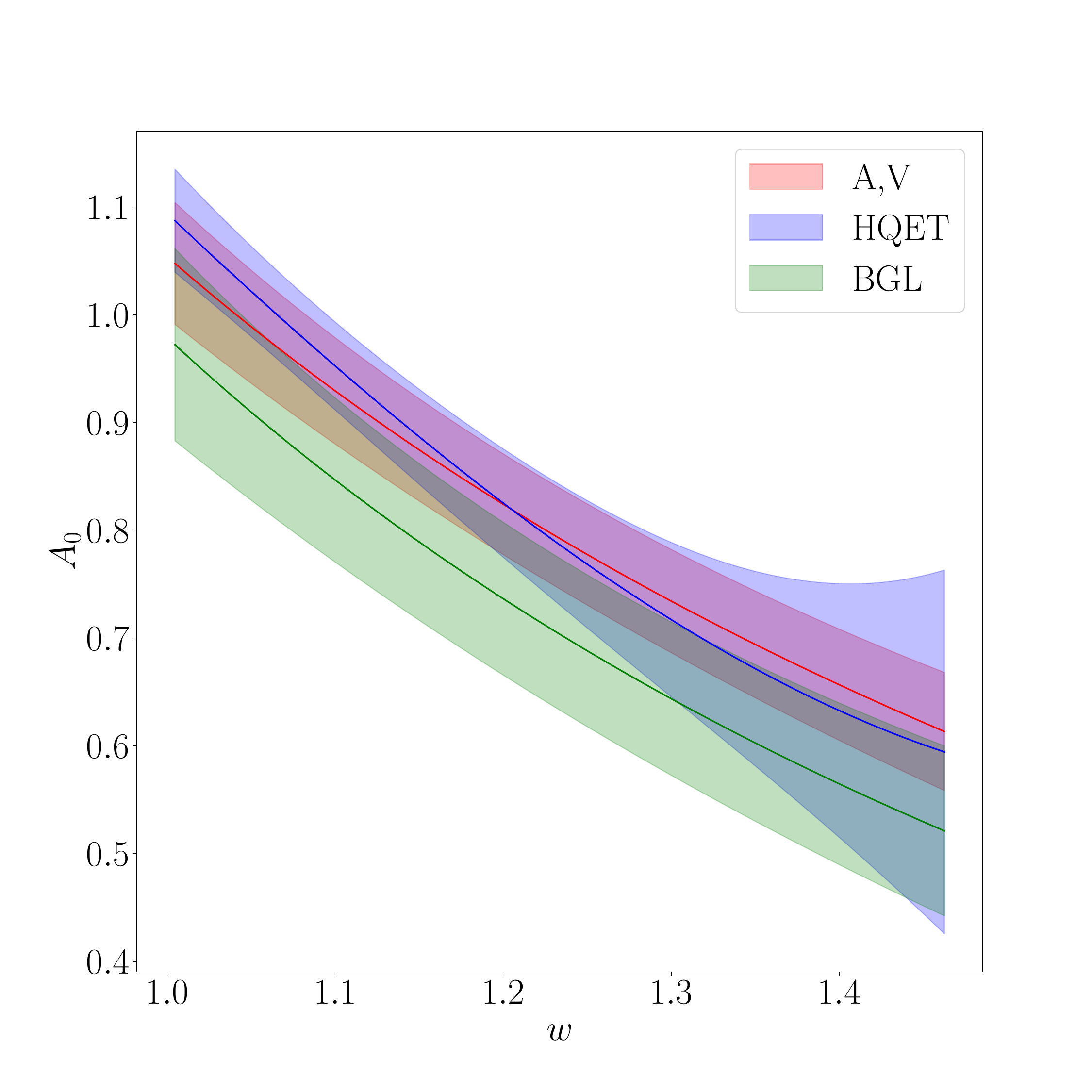}
\includegraphics[scale=0.215]{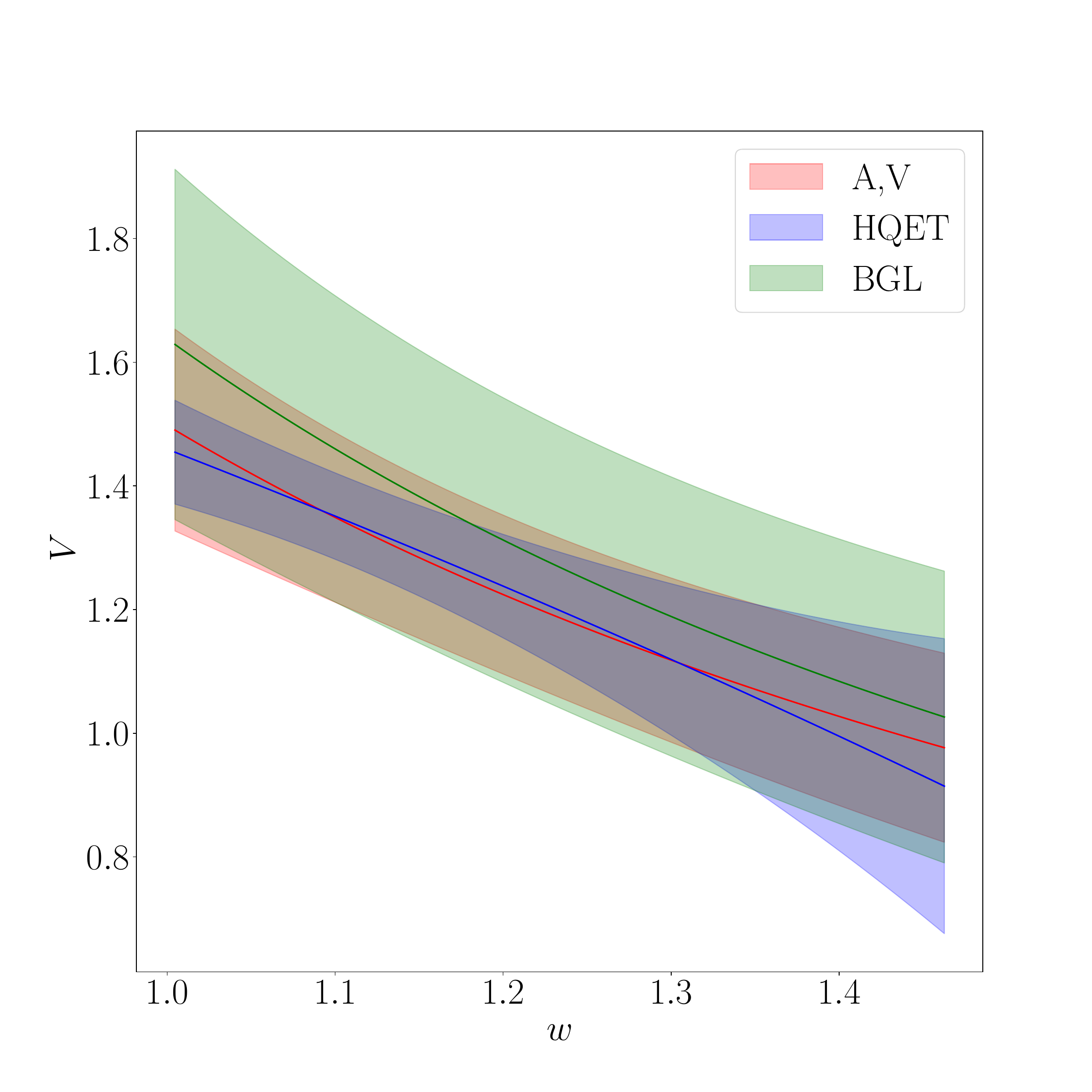}
\caption{\label{ffextrapcomparison} 
The form factors as a function of the recoil $w$. The red curves, denoted `A,V' in the legend, are the result of this work, computed using Eq.~(\ref{fitfunctionequation}) to fit the lattice form factor data without the use of outer functions. The green curves use the full BGL parameterisation Eq.~(\ref{fitfuncbglfullmh}), including the outer functions $\phi(t,t_0)$ and their heavy mass dependence, to fit the lattice form factor data. The blue curves were computed using Eq.~(\ref{fitfunchqetmh}), fitting the $q^2$ dependence using powers of $w-1$ without including any pole terms or outer functions. Here we see that the three methods produce consistent results for the form factors.}
\end{figure*}

The physical continuum form factors computed using Eq.~(\ref{fitfuncbglfullmh}) (converted to the $A_0, A_1, A_2, V$ basis) are given in Figure~\ref{ffextrapcomparison}, together with those computed using Eq.~(\ref{fitfunctionequation}). The total rate computed using the outer functions is, for the $\ell=\mu$ case, $\Gamma^\mathrm{BGL'}=1.88(20)_\mathrm{latt}(2)_\mathrm{EM}\times 10^{13} ~\mathrm{s}^{-1}$, compared to $\Gamma=\GAMMAmuTOTAL$ computed in Section~\ref{sec:discussion}, again demonstrating that the exclusion of the outer functions during the extrapolation to the physical continuum point does not have any significant effect on our results. 

\begin{figure}
\centering
\includegraphics[scale=0.215]{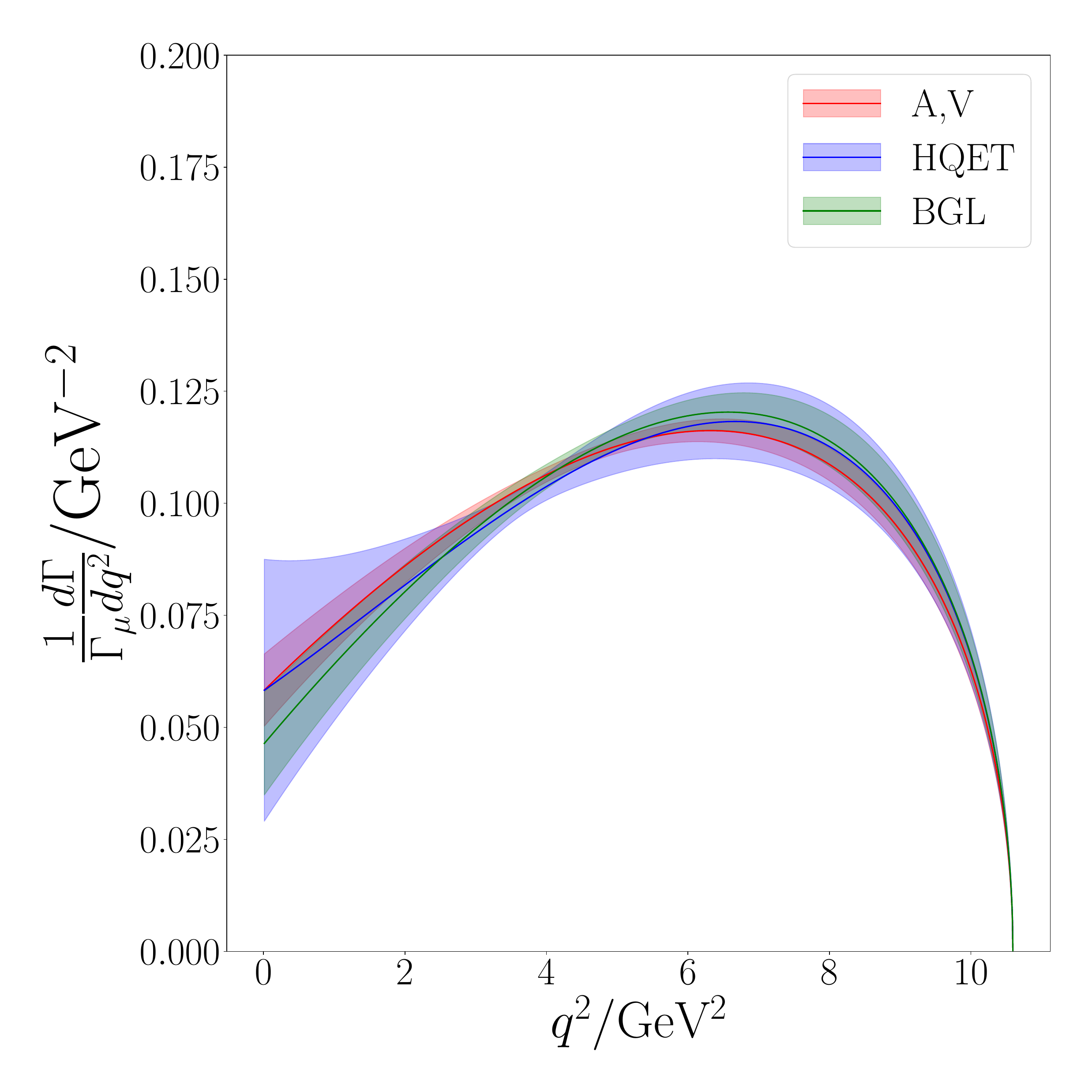}
\caption{\label{dgammadqsq_pseudo_outer_hqet_comparison} 
The normalised differential rate $\frac{1}{\Gamma}\frac{d\Gamma}{dq^2}$ resulting from fit variations. The red curve, denoted `A,V' in the legend, is the result of this work, computed using Eq.~(\ref{fitfunctionequation}) to fit the lattice form factor data without the use of outer functions. The green curve uses the full BGL parameterisation Eq.~(\ref{fitfuncbglfullmh}), including the outer functions $\phi(t,t_0)$ and their heavy mass dependence, to fit the lattice form factor data. The blue curve was computed using Eq.~(\ref{fitfunchqetmh}), fitting the $q^2$ dependence using powers of $w-1$ without including any pole terms or outer functions. Here we see that the three methods produce consistent results for the differential rate.}
\end{figure}

In Figure~\ref{ffextrapcomparison} we also plot form factors resulting from performing the extrapolation in the HQET basis. The form factors in this basis are related to the form factors in the helicity basis by
\begin{align}
g=&\frac{h_V}{M_{B_s}\sqrt{r}}\nonumber\\
f=&\sqrt{r}(1+w)h_{A_1}\nonumber\\
F_1=&M_{B_s}^2\sqrt{r}(1+w)\left((w-r)h_{A_1}-(w-1)(rh_{A_2}+h_{A_3}) \right)\nonumber\\
F_2=&\frac{1}{\sqrt{r}}\left( (1+w)h_{A_1} + (rw-1)h_{A_2} + (r-w)h_{A_3}\right),
\end{align}
and for this extrapolation we use a simple form in powers of $w-1$, which does not include any information about the pole terms:
\begin{equation}\label{fitfunchqetmh}
F(t)=\sum_{n=0}^\infty a^\mathrm{HQET}_{n}(w-1)^n.
\end{equation}
Here
\begin{equation}
a^\mathrm{HQET}_{n} = \sum_{j,k,l=0}^3 b_{n}^{''jkl}\Delta_{h}^{(j)} \left(\frac{am_c^\mathrm{val}}{\pi}\right)^{2k} \left(\frac{am_h^\mathrm{val}}{\pi}\right)^{2l}\mathcal{N}_{n}.
\end{equation}
We see in Figure~\ref{ffextrapcomparison} that our final results for the continuum form factors, extracted by fitting lattice data to Eq.~(\ref{fitfunctionequation}), are broadly consistent with the two alternative fits discussed in this section: the HQET-like fit to simple powers of $w-1$, Eq.~(\ref{fitfunchqetmh}), and the full BGL expression, Eq.~(\ref{fitfuncbglfullmh}). We also show in Figure~\ref{dgammadqsq_pseudo_outer_hqet_comparison} that the shape of the differential decay rate resulting from each of these fits is consistent. The total rate for the semimuonic mode resulting from the $w-1$ fit is $1.95(19)\times 10^13 s^{-1}$, which is consistent with our final result, $\Gamma=\GAMMAmuTOTAL$, computed using Eq.~(\ref{fitfunctionequation}) to perform the extrapolation.

%Here we vary $u_h$, the value of $u$ for a heavy quark mass $m_h$, using the simple approximation $u_h=u_b(M_{\eta_b^\mathrm{phys}}/M_{\eta_h})$, as well as also varying the $b$ quark pole mass according to $m_h^\mathrm{pole} = m_b^\mathrm{pole}/(M_{\eta_b^\mathrm{phys}}/M_{\eta_h})$. 
%As in~\cite{Harrison:2020gvo} we take $u_b=0.33$, $m_b^\mathrm{pole}=4.78$ and $\alpha_s=0.22$. 
%We include heavy mass dependence, discretisation effects and mistuning effects in the same way as in Eq.~\ref{fitfunctionequation}. Our BGL fit function then has
%\begin{equation}
%a^\mathrm{BGL}_n = \sum_{j,k,l=0}^3 b_n^{'jkl}\Delta_{h}^{(j)} \left(\frac{am_c^\mathrm{val}}{\pi}\right)^{2k} \left(\frac{am_h^\mathrm{val}}{\pi}\right)^{2l}\mathcal{N}_n.
%\end{equation}
%where $\mathcal{N}$ and $\Delta_{h}^{(j)}$ have the same definitions as in the main analysis, discussed in Section~\ref{sec:physextrap}.

%===========================================================================8

%===========================================================================8

\end{appendix}

%===========================================================================8
\bibliographystyle{apsrev4-1}
\bibliography{BsDsstar}
%===========================================================================8

\end{document}